\documentclass[runningheads,envcountsame,orivec]{llncs}

\usepackage[T1]{fontenc}
\usepackage{microtype}

\usepackage{float,caption,subcaption}
\usepackage{enumerate}
\newenvironment*{alphaenumerate}%
    {\begin{enumerate}[\upshape(a)]}%
    {\end{enumerate}}
\makeatletter
    \newenvironment*{alphaenumerate*}{%
        \let\orig@listi\@listi%
        \def\@listi{\orig@listi\topsep=0pt}%
        \begin{alphaenumerate}
    }{
        \end{alphaenumerate}%
        \def\@listi{\orig@listi}%
    }
\makeatother

\usepackage{hyperref,xurl}
\hypersetup{hidelinks,hypertexnames=false}
\newcommand*{\autoappref}[1]{\hyperref[#1]{Appendix~\ref*{#1}}}
\newcommand*{\autoeqref}[1]{\hyperref[#1]{Equation~(\ref*{#1})}}

\usepackage[UKenglish]{babel}
\addto\extrasUKenglish{%
}

\usepackage{amssymb,mathtools,ifthen,xspace}
\mathtoolsset{centercolon}

\DeclareMathOperator*{\ngl}{atan2}
\newcommand*{\dt}{\: \mathrm{d}t}
\newcommand*{\dtau}{\: \mathrm{d}\tau}
\newcommand*{\tnorm}[1][]{%
    \ifthenelse{\equal{#1}{}}%
        {\mbox{$\| \cdot \|$}}%
        {\mbox{$\| \cdot \|_{#1}$}}%
}

\newcommand*{\acmq}{{\normalfont\sffamily ACM}$\mathbb{Q}$\xspace}
\newcommand*{\settheoremcounter}[1]{%
    \setcounterref{theorem}{#1}%
    \addtocounter{theorem}{-1}%
}

\DeclareMathSymbol{\Gamma}{\mathalpha}{operators}{"00}
\DeclareMathSymbol{\Delta}{\mathalpha}{operators}{"01}
\DeclareMathSymbol{\Theta}{\mathalpha}{operators}{"02}
\DeclareMathSymbol{\Lambda}{\mathalpha}{operators}{"03}
\DeclareMathSymbol{\Xi}{\mathalpha}{operators}{"04}
\DeclareMathSymbol{\Pi}{\mathalpha}{operators}{"05}
\DeclareMathSymbol{\Sigma}{\mathalpha}{operators}{"06}
\DeclareMathSymbol{\Upsilon}{\mathalpha}{operators}{"07}
\DeclareMathSymbol{\Phi}{\mathalpha}{operators}{"08}
\DeclareMathSymbol{\Psi}{\mathalpha}{operators}{"09}
\DeclareMathSymbol{\Omega}{\mathalpha}{operators}{"0A}

\usepackage{pseudo,booktabs,tabularx,tcolorbox}
\tcbuselibrary{skins}

\newlength{\origparindent}
\pseudoset{
    font=\small,
    kwfont=\ttfamily,
    indent-length=.5\origparindent,
    compact=true,
    label=\footnotesize\arabic*.,
    ctfont=\ttfamily\color{black!60!white},
    ct-left=\hfill{\ttfamily\color{black!60!white}// },
    ct-right=,
    begin-tabular=\tabularx{\linewidth}[\pseudopos]{\pseudopreamble},
    end-tabular=\endtabularx,
    preamble={>{\pseudohpad} \pseudolabelalign >{\pseudosetup} X <{\pseudohpad}},
    hd-preamble={>{\pseudohpad} p{\linewidth-2\tabcolsep} <{\pseudohpad}},
}

\newcommand*{\pseudoinputoutput}[2]{%
    \\*%
    \midrule%
    \multicolumn{2}{>{\pseudosetup} \pseudohdpreamble}{\texttt{Input:} #1}%
    \\*%
    \multicolumn{2}{>{\pseudosetup} \pseudohdpreamble}{\texttt{Output:} #2}%
    \\*%
    \midrule%
    \pseudoprefix%
}

\newtcolorbox[auto counter]{algobox}[1]{
    empty,
    notitle,
    boxrule=0pt,
    leftrule=2pt,
    borderline west={2pt}{0pt}{black},
    boxsep=1pt,
    left=5pt,
    right=0pt,
    top=1pt,
    bottom=0pt,
    sharp corners,
    colback=white,
    label={alg:#1},
    nameref={\pseudopr{#1}},
}

\usepackage{tikz}
\usetikzlibrary{
    arrows.meta,backgrounds,calc,
    decorations.markings,intersections,
    positioning,shapes.geometric,
}

\tikzset{
    every picture/.style={font=\footnotesize},
    every label/.style={inner sep=0.6pt},
    level set/.style={util grey,fill=none},
    sublevel set/.style={draw=UmiBlue,very thick,fill=UmiBlue!10!white},
    sublevel name/.style={midway,above left,inner sep=0pt,outer sep=0pt,xshift=0.5pt,font=\footnotesize\color{UmiBlue}},
    origin/.style={draw,semithick,fill,circle,inner sep=0pt,minimum size=4.5pt},
    origin labeled/.style={origin,label={[anchor=north east]below:$\mathbf{0}$}},
    point/.style={draw,semithick,fill,diamond,inner sep=0pt,minimum size=5.7pt},
    vector/.style={draw,thick,-{Triangle[fill=white,round,width=4pt,length=6pt]}},
    util dash/.style={draw,semithick,dash pattern=on 1.75pt off 1.75pt},
    util grey/.style={draw=black!35!white,semithick,fill=black!10!white},
    main line/.style={draw,very thick},
    curve/.style={main line,miter limit=10,every node/.style={#1}},
    matching/.style={draw=UmiSkyblue,very thick},
    matching path/.style={draw,semithick,double=UmiSkyblue,double distance=1.6pt},
    matching path shape/.style={matching path,black!60!white},
    matching point/.style={point,fill=UmiSkyblue},
    valley/.style={draw=UmiOrange,ultra thick},
    diagonal/.style={draw=UmiYellow,ultra thick},
    grid line/.style={main line,line width=0.9pt},
    valley split line/.style={draw,semithick,double=UmiOrange,double distance=0.9pt},
    diagonal split line/.style={draw,semithick,double=UmiYellow,double distance=0.9pt},
    valley point/.style={point,fill=UmiOrange},
    diagonal point/.style={point,fill=UmiYellow},
    diagonal intersection/.style={draw,semithick,fill=UmiYellow,rectangle,inner sep=0pt,minimum size=3.2pt},
    diagonal indicator/.style={diagonal intersection,minimum size=2pt},
    origin indicator/.style={diagonal intersection,circle,minimum size=2.2pt},
    piece/.style={main line,-{Triangle[line width=0.6pt,width=5pt,length=4pt,fill={#1}]}},
    axis/.style={
        draw,black!60!white,semithick,-{Straight Barb},
        every node/.style={fill=white,inner xsep=3pt,inner ysep=0pt},
    },
}

\definecolor{UmiGreen}{rgb}{0.00,0.72,0.62}
\definecolor{UmiOrange}{rgb}{0.95,0.55,0.00}
\definecolor{UmiBlue}{rgb}{0.00,0.45,0.70}
\definecolor{UmiSkyblue}{rgb}{0.35,0.70,0.90}
\definecolor{UmiYellow}{rgb}{0.95,0.90,0.20}
\definecolor{UmiVermillion}{rgb}{0.80,0.40,0.00}

\newcommand{\examplecurves}{
    \coordinate (p1) at (0,1.25);
    \coordinate (p2) at (2,2.5);
    \coordinate (p3) at (4,1.75);
    \coordinate (p4) at (10,2.5);
    
    \coordinate (q1) at (0,0);
    \coordinate (q2) at (6,0.75);
    \coordinate (q3) at (8,0);
    \coordinate (q4) at (10,1.25);
    
    \path[curve={origin}] (p1) node{} -- (p2) node{} -- (p3) node{} -- (p4) node{};
    \path[curve={origin,rectangle}] (q1) node{} -- (q2) node{} -- (q3) node{} -- (q4) node{};
}

\newcommand{\borderaxes}{
    \path[axis,yshift=-11pt] (0,0) -- node {$\mathrm{dom}(\mathcal{A})$} (1,0);
    \path[axis,xshift=11pt] (1,0) -- node[rotate=90] {$\mathrm{dom}(\mathcal{B})$} (1,1);
}

\usepackage[c5paper,centering,text={\textwidth,\textheight}]{geometry}
\hypersetup{
    bookmarksdepth=2,
    pdftitle={Fundamentals of Computing Continuous Dynamic Time Warping in 2D under Different Norms},
    pdfauthor={Kevin Buchin, Maike Buchin, Jan Erik Swiadek, Sampson Wong},
}

\usepackage{multibib}
\newcites{appendix}{Additional References}
\bibliographystyleappendix{plainurl}

\begin{document}
    \title{%
        Fundamentals of Computing Continuous Dynamic Time Warping in 2D under Different Norms%
        \texorpdfstring{%
            \thanks{%
                This preprint corresponds to the full submitted version of a WALCOM 2026 paper.
                The proceedings version can be found at \href{https://doi.org/10.1007/978-981-95-7127-7_31}{\texttt{doi:10.1007/978-981-95-7127-7\_31}}.
            }%
        }{}%
    }
    \author{
        Kevin Buchin\inst{1}
        \and
        Maike Buchin\inst{2}
        \and
        Jan Erik Swiadek\inst{2}
        \and
        Sampson Wong\inst{3}
    }
    \institute{
        Technical University Dortmund, Germany\\ \email{kevin.buchin@tu-dortmund.de}
        \and
        Ruhr University Bochum, Germany\\ \email{\{maike.buchin,jan.swiadek\}@rub.de}
        \and
        University of Copenhagen, Denmark\\ \email{sampson.wong123@gmail.com}
    }
    \titlerunning{Fundamentals of Computing CDTW in 2D under Different Norms}
    \authorrunning{K. Buchin, M. Buchin, J. E. Swiadek, S. Wong}
    \maketitle
    
    \begin{abstract}
        Continuous Dynamic Time Warping (CDTW) measures the similarity of polygonal curves robustly to outliers and to sampling rates, but the design and analysis of CDTW algorithms face multiple challenges.
        We show that CDTW cannot be computed exactly under the Euclidean $2$-norm using only algebraic operations, and we give an exact algorithm~for CDTW under norms approximating the $2$-norm.
        The latter result relies on technical fundamentals that we establish, and which generalise to~any norm and to related measures such as the partial Fréchet similarity.
        
    \end{abstract}
    
    \section{Introduction}

The similarity of two curves can be measured in various ways, where the choice of technique is a trade-off depending on the application \cite{SuLZZZ2020,TaoBSBSPLPTD2021}.
Discrete similarity measures consider only the vertices of polygonal curves, often given as sampling points of some motion.
In~particular, Dynamic Time Warping (DTW) matches the vertices of the two input curves monotonically such that the sum of their distances is minimised \cite{Vints1968}.
The appeal of DTW lies in its simplicity, and in the fact that the curves can be detected as similar even if their underlying motions have different speeds.
Its main disadvantage is that it may yield poor results if the curves' sampling rates are too different or not sufficiently high \cite{TaoBSBSPLPTD2021,BuchiNW2022}.

\begin{figure}[H]%
    \centering%
    \begin{subfigure}{0.45\linewidth}
        \centering
        \begin{tikzpicture}[x={0.95\linewidth / 10},y={0.95\linewidth / 10}]
            \examplecurves
            \def\s{9}
            
            \scoped[on background layer]
                \path[matching]
                    \foreach \i in {0,...,\s} {
                        ($(p1)!{\i/\s}!(p2)$) -- ($(q1)!{\i/(3*\s)}!(q2)$)
                    }
                    \foreach \i in {1,...,\s} {
                        ($(p2)!{\i/\s}!(p3)$) -- ($(q1)!{(\i+\s)/(3*\s)}!(q2)$)
                    }
                    \foreach \i in {1,...,\s} {
                        ($(p3)!{\i/(3*\s)}!(p4)$) -- ($(q1)!{(\i+2*\s)/(3*\s)}!(q2)$)
                    }
                    \foreach \i in {1,...,\s} {
                        ($(p3)!{(\i+\s)/(3*\s}!(p4)$) -- ($(q2)!{\i/\s}!(q3)$)
                    }
                    \foreach \i in {1,...,\s} {
                        ($(p3)!{(\i+2*\s)/(3*\s}!(p4)$) -- ($(q3)!{\i/\s}!(q4)$)
                    };
        \end{tikzpicture}
        \caption{Optimal matching under the $1$-norm}
        \label{subfig:cdtw-1-norm}
    \end{subfigure}%
    \hspace*{0.05\linewidth}%
    \begin{subfigure}{0.45\linewidth}
        \centering
        \begin{tikzpicture}[x={0.95\linewidth / 10},y={0.95\linewidth / 10}]
            \examplecurves
            \def\sa{2}
            \def\sb{8}
            \def\sc{\numexpr\sb-1\relax}
            
            \scoped[on background layer]
                \path[matching]
                    \foreach \i in {0,...,\sa} {
                        (p1) -- ($(q1)!{0.071*\i/\sa}!(q2)$)
                        ($(p3)!{1-0.071*\i/\sa}!(p4)$) -- (q4)
                    }
                    \foreach \i in {1,...,\sb} {
                        ($(p1)!{0.779*\i/\sb}!(p2)$) -- ($(q1)!{0.071+(0.375-0.071)*\i/\sb}!(q2)$)
                        ($(p3)!{1-(0.071+(0.375-0.071)*\i/\sb)}!(p4)$) -- ($(q3)!{1-0.779*\i/\sb}!(q4)$)
                    }
                    \foreach \i in {1,...,\sa} {
                        ($(p1)!{0.779+(1-0.779)*\i/\sa}!(p2)$) -- ($(q1)!{0.375}!(q2)$)
                        ($(p3)!{1-0.375}!(p4)$) --
                        ($(q3)!{1-(0.779+(1-0.779)*\i/\sa)}!(q4)$)
                    }
                    \foreach \i in {1,...,\sa} {
                        ($(p2)!{0.245*\i/\sa}!(p3)$) -- ($(q1)!{0.375}!(q2)$)
                        ($(p3)!{1-0.375}!(p4)$) -- ($(q2)!{1-0.245*\i/\sa}!(q3)$)
                    }
                    \foreach \i in {1,...,\sc} {
                        ($(p2)!{0.245+(1-0.245)*\i/\sc}!(p3)$) -- ($(q1)!{0.375+(0.642-0.375)*\i/\sc}!(q2)$)
                        ($(p3)!{1-(0.375+(0.642-0.375)*\i/\sc)}!(p4)$) -- ($(q2)!{1-(0.245+(1-0.245)*\i/\sc)}!(q3)$)
                    }
                    \foreach \i in {1,...,\sa} {
                        (p3) -- ($(q1)!{0.642+(0.692-0.642)*\i/\sa}!(q2)$)
                        ($(p3)!{1-(0.642+(0.692-0.642)*\i/\sa)}!(p4)$) -- (q2)
                    }
                    \foreach \i in {1,...,\sc} {
                        ($(p3)!{(1-0.692)*\i/\sb}!(p4)$) -- ($(q1)!{0.692+(1-0.692)*\i/\sb}!(q2)$)
                    };
        \end{tikzpicture}
        \caption{Optimal matching under the $2$-norm}
    \end{subfigure}%
    \caption{Simple example of optimal matchings for our CDTW variant}%
    \label{fig:cdtw}%
\end{figure}
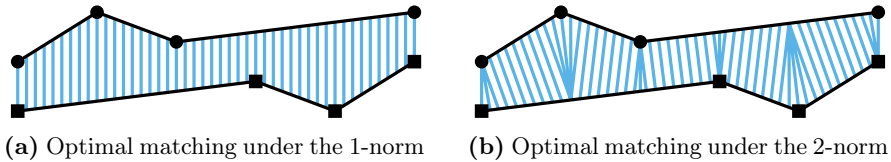

This is due to the fact that DTW disregards information by not interpreting the polygonal curves as continuous objects, given by the linear interpolations of their vertex sequences.
Therefore, several continuous versions of DTW have been proposed in the literature \cite{SerraB1994,SerraB1995,MunicP1999,BrakaPSW2005,EfratFV2007,Buchi2007,Har-PRR2025}, some of which are equivalent or nearly equivalent.
(See \autoref{subsec:related} for a brief overview.)
Each of these Continuous Dynamic Time Warping (CDTW) variants in some way utilises continuous and monotone matchings of arbitrary curve points, as illustrated in \autoref{fig:cdtw}.

Efficiently computing any CDTW variant exactly or with strong approximation guarantees has turned out to be challenging.
The strongest results so far are a pseudo-polynomial-time $(1 + \varepsilon)$-approximation for 2D curves under the Euclidean $2$-norm, due to Maheshwari, Sack and Scheffer~\cite{MahesSS2018}, and a polynomial-time exact algorithm in 1D, first described by Klaren~\cite{Klare2020}, then analysed by Buchin, Nusser and Wong~\cite{BuchiNW2022}.
The difficulty of computing CDTW has multiple reasons, which we address in this paper.
Our goals are to advance the understanding of related computational challenges and to establish approaches for handling these.

First of all, the fundamental properties of CDTW are still not well understood, especially in 2D and under different norms.
Like in the above works \cite{MahesSS2018,Klare2020,BuchiNW2022}, our formulation is based on a definition introduced by Buchin~\cite{Buchi2007} as a summed version of the Fréchet distance, which minimises the maximum distance of continuously and monotonically matched curve points~\cite{Frech1906,AltG1995}.
The CDTW definition at hand replaces the outlier-sensitive maximum with a path integral, aiming to overcome the drawbacks of the Fréchet distance and of DTW \cite{BrankBKNPW2020,BuchiNW2022}.
As our first contribution, presented in \autoref{sec:geometry}, we show that our path integral formulation admits locally optimal matchings that are advantageous from an algorithmic perspective.

Another challenge for integral-based measures is the integration itself, which typically complicates exact computations.
Many existing approaches for CDTW resort to heuristics or approximations by discretising the input curves \cite{BrakaPSW2005,EfratFV2007,MahesSS2018,BrankBKNPW2020,Har-PRR2025}.
This confines the integration to individual steps, but causes the solution quality or the running time to be dependent on the discretisation's resolution.
In contrast, the exact 1D algorithm utilises the 1D setting's piecewise linear integrands for a dynamic program that propagates piecewise quadratic functions \cite{Klare2020,BuchiNW2022}.

Brankovic~\cite{Brank2022} describes a 2D generalisation of this algorithm under the $1$-norm, which gives piecewise linear integrands too.
The $2$-norm, however, gives integrands with a square root and thus seems unsuited to the function propagation approach.
We give a new and deeper insight into the difficulty of CDTW under the $2$-norm, in that the numbers involved may be transcendental and hence not computable exactly using only algebraic operations.
This result is presented in \autoref{sec:unsolvability} and further motivates the usage of approximations for the $2$-norm \cite{Bajaj1988,DeCarGMSS2014,DeCarGMOS2014}.

In \autoref{sec:exact-algorithm} we then develop a $(1 + \varepsilon)$-approximation without discretising the curves by generalising the exact algorithm to a large class of norms with~polygons as level sets.
The final remaining challenge is the running time analysis, i.e.\ the combinatorial problem of bounding the number of propagated pieces \cite{SerraB1994,MunicP1999,BuchiNW2022}.
A technical analysis of the exact 1D algorithm has shown an $O(n^5)$ running time for curves with $n$ segments \cite{BuchiNW2022}.
Its 2D generalisation has not been analysed yet.
While the techniques of the 1D analysis do not readily facilitate a polynomial-time result in 2D, we highlight the main issue that would need to be resolved for such a result, and we prove technical properties that we deem helpful for this.

\subsection{Related Work}
\label{subsec:related}

Standard algorithms compute DTW in $O(n^2)$ time \cite{Vints1968} and the Fréchet distance in $O(n^2 \log(n))$ time \cite{AltG1995}.
There also exist improved upper bounds \cite{GoldS2018,BuchiBMM2017,ChengH2025}.
The~first continuous version of DTW was due to Serra and Berthod~\cite{SerraB1994}.
It matches curve points continuously, but sums over a finite quantity instead of using integration.
An equivalent definition was later analysed by Munich and Perona~\cite{MunicP1999}.

Serra and Berthod~\cite{SerraB1995} further introduced an integral-based measure that considers the change of distance vectors instead of the distances themselves, so it is a translation-invariant relative of CDTW.
Efrat, Fan and Venkatasubramanian~\cite{EfratFV2007} proposed a more general integral-based variant of CDTW, which is rather similar to our definition; remaining differences are explained in~\cite[Section~2.3]{MahesSS2018}.

CDTW has applications in signature verification~\cite{EfratFV2007}, map matching~\cite{BrakaPSW2005} and clustering~\cite{BrankBKNPW2020}.
Several practical works \cite{BrakaPSW2005,BrankBKNPW2020,Har-PRR2025} use CDTW formulations equivalent to some definition by Buchin \cite{Buchi2007}, which are moreover related to the lexicographic Fréchet distance \cite{Rote2014} and the partial Fréchet similarity \cite{BuchiBW2009}:
When restricted to~one pair of curve segments, optimal matchings for all these measures are realised via local optimality \cite[Section~5]{MahesSS2018}, upon which our results in \autoref{sec:geometry} build.

\subsection{Preliminaries}

A polygonal curve $P$ in $\mathbb{R}^2$ is composed of $n \in \mathbb{N}$ consecutive line segments and represented by its vertex sequence $\langle p_0,\dotsc,p_n \rangle$ with $p_{i} \neq p_{i-1}$ for $i \in \{1,\dotsc,n\}$.
The arc length of $P$ under a norm $\tnorm$ on $\mathbb{R}^2$ is $\| P \| := \sum_{i=1}^n \| p_i - p_{i-1} \|$, and we denote the corresponding arc length parametrisation with constant speed~$1$ under $\tnorm$ by $P_{\| \cdot \|} \colon [0, \| P \|] \to \mathbb{R}^2$.
Moreover, we define the set $\Pi(P)$ containing all functions $f \colon [0,1] \to \mathbb{R}^2$ that parametrise $P$ monotonically and are piecewise continuously differentiable.
Note that $\int_{0}^{1} \| f'(t) \| \dt = \| P \|$ for all $f \in \Pi(P)$.

\begin{definition}[{\cite[Chapter~6]{Buchi2007}}]
    \label{def:cdtw}
    Let $P,Q$ be polygonal curves, let $\tnorm$ be~a~norm and let $d \colon \mathbb{R}^2 \times \mathbb{R}^2 \to \mathbb{R}_{\geq 0}$ be a continuous function.
    The measure \emph{Continuous Dynamic Time Warping (CDTW)} of $P,Q$ in $(\mathbb{R}^2, \tnorm)$ under $d$ is defined by
    \[
        \mathrm{cdtw}_{\| \cdot \|,d}(P,Q)
        :=
        \inf_{(f,g) \in \Pi(P) \times \Pi(Q)}
        \int_0^1
        d(f(t), g(t)) \cdot
        \left\| \begin{pmatrix}
            \| f'(t) \| \\
            \| g'(t) \|
        \end{pmatrix} \right\|_1
        \dt
        \text{.}
    \]
    We set $\mathrm{cdtw}_{\| \cdot \|}(P,Q) := \mathrm{cdtw}_{\| \cdot \|, (p,q) \mapsto \| p - q \|}(P,Q)$ for CDTW of $P,Q$ under $\tnorm$.
\end{definition}

So far, only the vector space $(\mathbb{R}^2, \tnorm[2])$ equipped with the Euclidean \mbox{$2$-norm} has been used for 2D CDTW, where the $\tilde{p}\:\!$-norm $\tnorm[\tilde{p}]$ on $\mathbb{R}^2$ is, as usual, given by $\| x \|_{\tilde{p}} = \| (x_1,x_2)^\mathsf{T} \|_{\tilde{p}} := (|x_1|^{\tilde{p}} + |x_2|^{\tilde{p}})^{1/\tilde{p}}$ for $\tilde{p} \geq 1$.
Brankovic~\cite[Section~5.2]{Brank2022} considered CDTW in $(\mathbb{R}^2, \tnorm[2])$ under the distance function $(p,q) \mapsto \| p - q \|_1$ induced by the $1$-norm, which differs from the equipped norm.
Our formulation of CDTW under a norm employs the chosen norm for both arc lengths of curves and distances between curve points.
This not only yields an intuitive generalisation of other definitions (cf.~\cite[Section~6.2.2]{Buchi2007} and \cite[Section~5.5.1]{Har-PRR2025}), but also comes with useful geometric properties, as shown in \autoref{sec:geometry}.
The dependence on the chosen norm is likewise reflected by our definition of the parameter space.

\begin{definition}[{\cite[Section~2]{BuchiNW2022}}]
    Let $P = \langle p_0,\dotsc,p_n \rangle$ and $Q = \langle q_0,\dotsc,q_m \rangle$ be polygonal curves, and let $\tnorm$ be a norm.
    The \emph{parameter space} of $P,Q$ under $\tnorm$ is $[0, \| P \|] \times [0, \| Q \|] \subseteq \mathbb{R}^2$.
    Each pair $(i,j) \in \{1,\dotsc,n\} \times \{1,\dotsc,m\}$ is associated with a \emph{cell} $[\| \langle p_0,\dotsc,p_{i-1} \rangle \|, \| \langle p_0,\dotsc,p_i \rangle \|] \times [\| \langle q_0,\dotsc,q_{j-1} \rangle \|, \| \langle q_0,\dotsc,q_j \rangle \|]$.
\end{definition}

As usual, we represent monotone matchings of $P,Q$ by monotone paths in their parameter space under $\tnorm$.
Since \autoref{def:cdtw} combines parametrisation speeds by fixing the $1$-norm, the arc length of each path is $\sigma := \| P \| + \| Q \|$ \cite[Section~6.2.4]{Buchi2007}.
This gives a simple \emph{regularisation constraint}\footnote{Some constraint is needed to prevent infinite speed. Due to its simplicity, using the $1$-norm is popular \cite{MahesSS2018,BuchiNW2022,Brank2022}. Another natural choice is the $\infty$-norm \cite{Rote2014,Klare2020}.} for paths:
We let $\Gamma_{\| \cdot \|} (P,Q)$ contain all functions $\gamma \colon [0, \sigma] \to [0, \| P \|] \times [0, \| Q \|]$ such that there are $(f,g) \in \Pi(P) \times \Pi(Q)$ with $\gamma_1(s) = 1/\sigma \cdot \int_{0}^{s} \| f'(t / \sigma) \| \dt$ and $\gamma_2(s) = 1/\sigma \cdot \int_{0}^{s} \| g'(t / \sigma) \| \dt$, subject to the constraint $\| \gamma'(s) \|_1 = 1/\sigma \cdot (\| f'(s / \sigma) \| + \| g'(s / \sigma) \|) = 1$, for all $s \in [0, \sigma]$.

\begin{definition}[{\cite[Definition~8]{BuchiNW2022}}]
    \label{def:path-and-cost}
    Let $P,Q$ be polygonal curves, let $\tnorm$ be a norm, and let $x,y \in [0, \| P \|] \times [0, \| Q \|]$ with $(x_1 \leq y_1) \land (x_2 \leq y_2)$ be two points in parameter space.
    An \emph{$(x,y)$-path} is a restriction $\widehat{\gamma}$ of any $\gamma \in \Gamma_{\| \cdot \|} (P,Q)$ that goes through~$x$ and~$y$, i.e.\ $\gamma(\| x \|_1) = x$ and $\gamma(\| y \|_1) = y$, to the domain $[\| x \|_1, \| y \|_1]$.
    We further say that an $(x,y)$-path $\widehat{\gamma}$ is \emph{optimal} for $P,Q$ under $\tnorm$ if its assigned \emph{cost} $\int_{\| x \|_1}^{\| y \|_1} \| P_{\| \cdot \|}(\widehat{\gamma}_1(t)) - Q_{\| \cdot \|}(\widehat{\gamma}_2(t)) \| \dt$ is minimum among all $(x,y)$-paths.
\end{definition}

Thus, the cost of an optimal $(\mathbf{0}, (\| P \|, \| Q \|)^{\mathsf{T}})$-path for $P,Q$ under $\tnorm$, which goes from the bottom left $\mathbf{0} := (0,0)^{\mathsf{T}}$ to the top right of the parameter space, is equal to the value $\mathrm{cdtw}_{\| \cdot \|}(P,Q)$.
This is easy to verify (cf.~\cite[Lemma~1]{BuchiNW2022}).

\section{Geometry of Parameter Space Cells}
\label{sec:geometry}

To construct a dynamic program propagating optimal path costs through the grid of parameter space cells (see~\autoref{sec:exact-algorithm}), we need to characterise optimal $(x,y)$-paths within a single cell.
For now, we represent each cell $(i,j)$ by corresponding \emph{polygonal segments}, i.e.\ single-segment curves $\overline{P} := \langle p_{i-1}, p_i \rangle$ and $\overline{Q} := \langle q_{j-1}, q_j \rangle$ whose arc length parametrisations $\overline{P}_{\| \cdot \|},\overline{Q}_{\| \cdot \|}$ under a norm $\tnorm$ have domains extended to $\mathbb{R}$.
\autoref{subsec:optimal-paths} states some known basics.
In \autoref{subsec:valleys} we establish a robust characterisation of optimal $(x,y)$-paths for $\overline{P},\overline{Q}$ under arbitrary $\tnorm$.
Note that this is also useful for the lexicographic Fréchet distance \cite{Rote2014} and the partial Fréchet similarity \cite{BuchiBW2009}.
Omitted proofs are given in \autoappref{app:geometry-proofs}.

\subsection{Optimal Paths through Cell Terrain}
\label{subsec:optimal-paths}

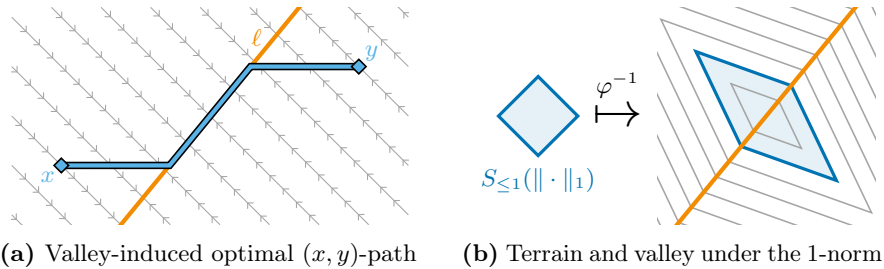
\begin{figure}[H]%
    \centering%
    \begin{subfigure}{0.45\linewidth}
        \centering
        \begin{tikzpicture}[
            x=.95\linewidth,y=.95\linewidth,
            decoration={
                markings,mark=between positions 3.5pt and 1-3.5pt step 7pt with {
                    \arrowreversed{Straight Barb[round,width=3pt,length=1.5pt]}
                },
            },
        ]
            \coordinate (bl) at (0,0);
            \coordinate (tr) at (1,0.55);
            \coordinate (m) at ($(bl)!0.5!(tr)$);
            \clip[name path=clip] (bl) rectangle (tr);
            
            \path[valley,name path=vall,shift={(m)}] ($-2*(0.63,0.77)$) -- ($2*(0.63,0.77)$);
            \path[name intersections={of=vall and clip,by={a,b},sort by=vall}];
            \path (a) -- node[pos=0.86,left=1.5pt] {\color{UmiOrange}$\ell$} (b);
            
            \begin{scope}[on background layer]
                \clip (bl) rectangle (tr);
                
                \foreach \i in {-2,...,10} {
                    \coordinate (z) at ($(a)!{\i/8}!(b)$);
                    \path[name path=slope] ($(z) - 2*(1,-1)$) -- ($(z) + 2*(1,-1)$);
                    
                    \path[name intersections={of=slope and clip,sort by=slope}];
                    \path[util grey,thin,postaction={decorate}] (z) -- (intersection-1);
                    \path[util grey,thin,postaction={decorate}] (z) -- (intersection-2);
                };
            \end{scope}
            
            \coordinate (x) at ($(a) + (-0.15,0.15)$);
            \coordinate (y) at ($(b) + (0.15,-0.15)$);
            \path[name path=rect] (x) rectangle (y);
            
            \path[name intersections={of=vall and rect,sort by=vall}];
            \path[matching path] (x) -- (intersection-1) -- (intersection-2) -- (y);
            
            \node[matching point,label={below left:\color{UmiSkyblue}$x$}] at (x) {};
            \node[matching point,label={above right:\color{UmiSkyblue}$y$}] at (y) {};
        \end{tikzpicture}
        \caption{Valley-induced optimal $(x,y)$-path}
        \label{subfig:valley-definition}
    \end{subfigure}%
    \hspace*{0.05\linewidth}%
    \begin{subfigure}{0.45\linewidth}
        \centering
        \begin{tikzpicture}[x={0.95\linewidth / 10},y={0.95\linewidth / 10}]
            \coordinate (bl) at (0,0);
            \coordinate (tr) at (10,5.5);
            \coordinate (m) at ($(bl)!0.5!(tr)$);
            
            \begin{scope}[shift={($(m) + (-3.5,0)$)}]
                \path[sublevel set] (1,0) -- (0,1) -- (-1,0) -- (0,-1) -- cycle;
                \node[anchor=north,below=1.25pt] at (0,-1) {\color{UmiBlue}$S_{\leq 1}(\tnorm[1])$};
            \end{scope}
            
            \begin{scope}[shift={($(m) + (-1.5,0)$)}]
                \node[anchor=mid,scale=2] at (0,0) {$\mapsto$};
                \node[anchor=south,above=3.5pt] at (0,0) {$\varphi^{-1}$};
            \end{scope}
            
            \begin{scope}[shift={($(m) + (2.25,0)$)}]
                \clip (-2.75,-2.75) rectangle (2.75,2.75);
                
                \path[sublevel set] (0.63,0.77) -- (-1.77,1.63) -- (-0.63,-0.77) -- (1.77,-1.63) -- cycle;
                
                \foreach \i in {0.5,1.5,2,...,5} {
                    \path[level set] ($\i*(0.63,0.77)$) -- ($\i*(-1.77,1.63)$) -- ($-\i*(0.63,0.77)$) -- ($-\i*(-1.77,1.63)$) -- cycle;
                };
                
                \path[valley] ($-5*(0.63,0.77)$) -- ($5*(0.63,0.77)$);
            \end{scope}
        \end{tikzpicture}
        \caption{Terrain and valley under the $1$-norm}
        \label{subfig:valley-example}
    \end{subfigure}%
    \caption{Connection between optimal paths through cell terrain \\ and geometric shape of cell terrain, as provided by valleys}%
\end{figure}

As we regularise path speeds to the value $1$ via the $1$-norm, all $\gamma \in \Gamma_{\| \cdot \|}(\overline{P},\overline{Q})$ and $s \in [0,\sigma]$ satisfy $\| \gamma(s) \|_1 = \gamma_1(s) + \gamma_2(s) = s$.
Intuitively, the points in parameter space reachable after time exactly $s$ are those on the line of slope~$-1$ through $\gamma(s)$.
Hence, terrain minima on lines of slope $-1$ may induce optimal $(x,y)$-paths by means of steepest monotone descent \cite{Rote2014}.
If such minima are themselves attained on a line, known as a \emph{valley} \cite{BuchiNW2022}, there is an optimal $(x,y)$-path travelling as long as possible on the valley or as close as possible to the valley, see \autoref{subfig:valley-definition}.

\begin{definition}[{\cite[Definition~2]{Brank2022}}]
    \label{def:valley}
    A \emph{valley} for polygonal segments $\overline{P},\overline{Q}$ under a norm $\tnorm$ is a line $\ell \subseteq \mathbb{R}^2$ not of slope~$-1$ such that every point $z \in \ell$ is a~\emph{sink}, i.e.\ the function $t \mapsto \| \overline{P}_{\| \cdot \|}(z_1 + t) - \overline{Q}_{\| \cdot \|}(z_2 - t) \|$ that evaluates along the line of slope~$-1$ through $z$ is non-increasing on $\mathbb{R}_{\leq 0}$ and non-decreasing on $\mathbb{R}_{\geq 0}$.
\end{definition}

\begin{theorem}[{\cite[Section~5.5]{Brank2022}}]
    \label{thm:optimal-paths}
    Let $\overline{P},\overline{Q}$ be polygonal segments with a valley $\ell$ of positive slope under a norm $\tnorm$, and let $x,y \in \mathbb{R}^2$ with $\Xi := [x_1,y_1] \times [x_2,y_2] \neq \varnothing$ be two points in parameter space.
    If $\ell \cap \Xi \neq \varnothing$, then the $(x,y)$-path tracing line segments from $x$ to $\widehat{x}$ to $\widehat{y}$ to $y$ is optimal for $\overline{P},\overline{Q}$, where $\widehat{x},\widehat{y} \in \ell \cap \Xi$ share a coordinate with $x,y$ respectively.
    Else, the $(x,y)$-path tracing line segments from $x$ to $\xi$ to $y$ is optimal, where $\xi \in \{ (x_1,y_2)^{\mathsf{T}}, (y_1,x_2)^{\mathsf{T}} \}$ is closest to $\ell$ in $\Xi$.
\end{theorem}

The terrain that paths travel through in a cell consists of affinely transformed versions of the (sub)level sets of $\tnorm$, see \autoref{subfig:valley-example}.
For $\mu \in \mathbb{R}$ the \emph{$\mu$-sublevel set} of a function $h \colon \mathrm{dom}(h) \to \mathbb{R}$ is the set $S_{\leq \mu}(h) := \{ z \in \mathrm{dom}(h) \mid h(z) \leq \mu \}$.

\begin{lemma}[{\cite[Lemma~3]{AltG1995}}]
    \label{thm:affine}
    Let $\overline{P},\overline{Q}$ be polygonal segments, let $\tnorm$ be a norm, and let $\varphi \colon \mathbb{R}^2 \to \mathbb{R}^2$ be the affine map defined through $\varphi(z) := \overline{P}_{\| \cdot \|}(z_1) - \overline{Q}_{\| \cdot \|}(z_2)$.
    If \smash{$\overline{P},\overline{Q}$} are parallel, then $\varphi$ is constant either on every line of slope $1$ or on every line of slope $-1$ in the extended parameter space $\mathbb{R}^2$.
    Else, $\varphi$ has an affine inverse map $\varphi^{-1} \colon \mathbb{R}^2 \to \mathbb{R}^2$ satisfying $S_{\leq \mu}(\tnorm \circ \varphi) = \varphi^{-1}(S_{\leq \mu}(\tnorm))$ for all $\mu \geq 0$.
\end{lemma}

\subsection{Existence and Computation of Valleys}
\label{subsec:valleys}

Every cell has a valley under the $2$-norm~\cite[Lemma~4]{MahesSS2018} and $1$-norm~\cite[Lemma~24]{Brank2022}, but for CDTW in $(\mathbb{R}^2,\tnorm[2])$ under $(p,q) \mapsto \| p - q \|_1$ there are cells whose single valley has negative slope.
Because monotone paths cannot travel on such valleys, computing optimal $(x,y)$-paths then deviates from \autoref{thm:optimal-paths} (see~\cite[Lemma~27]{Brank2022}), and handling their costs algorithmically becomes more difficult.
We generalise and improve upon previous results by showing that every norm~$\tnorm$ guarantees valleys of positive slope for our formulation of CDTW under $\tnorm$ from \autoref{def:cdtw}.

To this end, we employ the geometric definition of norms:
Given an absorbing set $K \subseteq \mathbb{R}^2$, i.e.\ every $z \in \mathbb{R}^2$ satisfies $z \in \lambda K$ for all $\lambda \in \mathbb{R}$ with large enough $|\lambda|$, let its \emph{gauge} be denoted by $\mathcal{G}_K \colon \mathbb{R}^2 \to \mathbb{R}_{\geq 0}$, where $\mathcal{G}_K(z) := \inf \{ \lambda \geq 0 \mid z \in \lambda K \}$.
If $K$ is balanced, i.e.\ $\lambda K \subseteq K$ for all $\lambda \in [-1,1]$, as well as closed and bounded, then the sublevel sets of $\mathcal{G}_K$ are $S_{\leq \mu}(\mathcal{G}_K) = \mu K$ for all $\mu \geq 0$.
If $K$ is also convex, i.e.\ $\lambda z + (1 - \lambda) z' \in K$ for all $z,z' \in K$ and $\lambda \in [0,1]$, then its gauge $\mathcal{G}_K$ is a norm on $\mathbb{R}^2$.
Conversely, any norm $\tnorm$ is equal to the gauge of its absorbing, balanced, closed, bounded and convex $1$-sublevel set $S_{\leq 1}(\tnorm)$.
(See~\cite[pp.~39--40]{SchaeW1999}.)

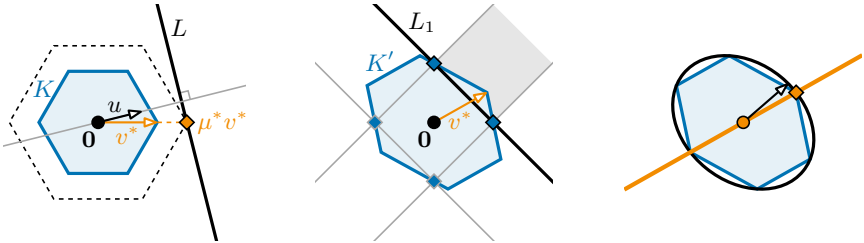
\begin{figure}[H]%
    \centering%
    \begin{subfigure}{0.3\linewidth}
        \centering
        \begin{tikzpicture}[x=0.85\linewidth,y=0.85\linewidth]
            \path[clip,name path=clip] (0,0) rectangle (1.025,1);
            
            \begin{scope}[shift={(0.4,0.5)},scale=0.25]
                \path[sublevel set] \foreach \i in {1,...,6} {
                    \ifnum\i>1 -- \fi
                    ({cos(\i * 60)},{sin(\i * 60)})
                    \ifnum\i=3 node[sublevel name] {$K$} \fi
                } -- cycle;
                
                \path[util dash] \foreach \i in {1,...,6} {
                    \ifnum\i>1 -- \fi
                    ($1.5*({cos(\i * 60)},{sin(\i * 60)})$)
                } -- cycle;
                
                \path[main line,name path=line] ($(1.5,0) - (1,-4)$) -- ($(1.5,0) + (1,-4)$);
                \path[name intersections={of=line and clip,sort by=line}]
                    (intersection-1) -- (intersection-2) node[pos=0.1,right,inner sep=2.6pt] {$L$};
                
                \path[util grey,name path=perp] (-4,-1) -- (4,1);
                \scoped[on background layer]
                    \path[util grey,fill=none,rotate=atan(1/4),name intersections={of=line and perp}]
                        ($(intersection-1) + 0.15*(1,0)$) |- ($(intersection-1) + 0.15*(0,1)$);
                
                \path[vector,UmiOrange] (0,0) -- node[below,inner sep=1.2pt] {$v^*$} (1,0);
                \path[util dash,UmiOrange] (1,0) -- (1.5,0);
                \node[point,fill=UmiOrange,label={right:\color{UmiOrange}$\mu^* v^*$}] at (1.5,0) {};
                
                \path[vector] (0,0) -- node[pos=0.375,above,inner sep=1.8pt] {$u$} ($0.185*(4,1)$);
                \node[origin labeled] at (0,0) {};
            \end{scope}
        \end{tikzpicture}
        \caption{Minimising the norm}
        \label{subfig:valley-minimisation}
    \end{subfigure}%
    \hspace*{0.1\linewidth/3}%
    \begin{subfigure}{0.3\linewidth}
        \centering
        \begin{tikzpicture}[x=0.85\linewidth,y=0.85\linewidth]
            \path[clip,name path=clip] (0,0) rectangle (1,1);
            
            \begin{scope}[shift={(0.5,0.5)},scale=0.25]
                \path[sublevel set] \foreach \i in {1,...,6} {
                    \ifnum\i>1 -- \fi
                    ({-0.894*cos(\i * 60) + 0.789*sin(\i * 60)},{-0.5064*cos(\i * 60) - 1.0128*sin(\i * 60)})
                    coordinate (v\i)
                    \ifnum\i=5 node[sublevel name] {$K'$} \fi
                } -- cycle;
                
                \path[main line,name path=line] ($(0.5,0.5) - 4*(1,-1)$) -- ($(0.5,0.5) + 4*(1,-1)$);
                \path[name intersections={of=line and clip,sort by=line}]
                    (intersection-1) -- (intersection-2) node[pos=0.1,right=2.5pt] {$L_1$};
                
                \path[util grey] ($(-0.5,-0.5) - 4*(1,-1)$) -- ($(-0.5,-0.5) + 4*(1,-1)$);
                \path[util grey] ($(0.5,-0.5) - 4*(1,1)$) -- ($(0.5,-0.5) + 4*(1,1)$);
                \path[util grey,name path=perp] ($(-0.5,0.5) - 4*(1,1)$) -- ($(-0.5,0.5) + 4*(1,1)$);
                
                \scoped[on background layer]
                    \path[util grey,draw=none,name intersections={of=perp and clip,sort by=perp}]
                        (1,0) -- (0,1) -- (intersection-2) -- ($(intersection-2) + (1,-1)$) -- cycle;
                
                \path[vector,UmiOrange] (0,0) -- node[below,inner sep=1.2pt] {$v^*$} (v3) {};
                \node[origin labeled] at (0,0) {};
                
                \node[point,fill=UmiBlue] at (1,0) {};
                \node[point,fill=UmiBlue] at (0,1) {};
                \node[point,util grey,fill=UmiBlue] at (-1,0) {};
                \node[point,util grey,fill=UmiBlue] at (0,-1) {};
            \end{scope}
        \end{tikzpicture}
        \caption{Finding positive slope}
        \label{subfig:valley-positive-slope}
    \end{subfigure}%
    \hspace*{0.1\linewidth/3}%
    \begin{subfigure}{0.3\linewidth}
        \centering
        \begin{tikzpicture}[x=0.85\linewidth,y=0.85\linewidth]
            \path[clip] (0,0) rectangle (1,1);
            
            \begin{scope}[shift={(0.5,0.5)},scale=0.25]
                \path[sublevel set] \foreach \i in {1,...,6} {
                    \ifnum\i>1 -- \fi
                    ({-0.894*cos(\i * 60) + 0.789*sin(\i * 60)},{-0.5064*cos(\i * 60) - 1.0128*sin(\i * 60)})
                    coordinate (v\i)
                } -- cycle;
                
                \path[main line] (0,0)
                    ellipse[x radius=1.3059,y radius=0.9993,rotate=-39.2476];
                \path[valley] ($2.25*(v6)$) -- ($2.25*(v3)$);
                
                \path[vector] (0,0) -- (0.7584,0.6599) {};
                \node[origin,fill=UmiOrange] at (0,0) {};
                \node[point,fill=UmiOrange] at (v3) {};
            \end{scope}
        \end{tikzpicture}
        \caption{Computing a valley}
        \label{subfig:valley-computation}
    \end{subfigure}%
    \caption{Valley characterisation under a norm with regular hexagons as sublevel sets}%
\end{figure}

To find sinks for valleys, we minimise a norm $\mathcal{G}_K$ on a line $L$ by duality (cf.~\cite[Section~5.1.6]{BoydV2009}).
The intuition is:
We scale $K$ by $\mu^* \geq 0$ such that $L$ is a \emph{supporting line} of $\mu^* K$, i.e.\ $L \cap \mu^* K \neq \varnothing$ and $\mu^* K$ lies within a closed half-plane bounded by $L$.
The minimum of $\mathcal{G}_K$ on $L$ is then attained on $L \cap \mu^* K$, see~\autoref{subfig:valley-minimisation}.

\begin{lemma}
    \label{thm:duality}
    Let $L = \{ z \in \mathbb{R}^2 \mid u^{\mathsf{T}} \cdot z = t \}$ be a line with $u \in \mathbb{R}^2 \setminus \{ \mathbf{0} \}$ and $t \in \mathbb{R}$.
    The gauge $\mathcal{G}_K$ of an absorbing, balanced, closed and bounded~$K \subseteq \mathbb{R}^2$ attains its minimum on $L$ at $(t/t^*) v^*$, where $v^* \in \arg \max \{ u^\mathsf{T} \cdot v \mid v \in K \}$ and $t^* := u^\mathsf{T} \cdot v^*$.
\end{lemma}

\begin{proof}
    We have $(t/t^*) v^* \in L$ due to $u^{\mathsf{T}} \cdot (t/t^*) v^* = t$ by definition of $t^*$.
    If $t = 0$, then $(t/t^*) v^* = \mathbf{0}$ minimises the gauge $\mathcal{G}_K$.
    Otherwise, it is $\mathbf{0} \notin L$, so $\mathcal{G}_K(z) > 0$ and $z / \mathcal{G}_K(z) \in K$ hold for all $z \in L$, as $K$ is bounded and closed.
    This yields
    \[
        \mathcal{G}_K(z)
        = \frac{|t|}{|t|} \cdot \mathcal{G}_K(z)
        = \frac{|t|}{|u^{\mathsf{T}} \cdot z|} \cdot \mathcal{G}_K(z)
        = \frac{|t|}{|u^{\mathsf{T}} \cdot (z / \mathcal{G}_K(z))|}
        \geq \frac{|t|}{|u^{\mathsf{T}} \cdot v^*|}
        = \frac{|t|}{|t^*|}
    \]
    for all $z \in L$, where the inequality follows from $v^*$ maximising $v \mapsto u^{\mathsf{T}} \cdot v$ on $K$ and thereby also maximising $v \mapsto |u^{\mathsf{T}} \cdot v|$ on $K$, as $K$ is balanced.
    On top of that, $v^* \in K$ and $K$ being balanced imply $|t/t^*| \geq |t/t^*| \mathcal{G}_K(v^*) = \mathcal{G}_K((t/t^*) v^*)$.
    \qed
\end{proof}

We next prove the main result of this section, which additionally uses the affine map $\varphi$ from \autoref{thm:affine}.
Note that polygonal segments $\overline{P}, \overline{Q}$ give $\varphi(\mathbf{0}) = p_{i-1} - q_{j-1}$, while the translational part of an analogously defined affine map for a parameter space cell $(i,j)$ of curves $P,Q$ may depend on all of $p_0,\dotsc,p_i$ and $q_0,\dotsc,q_j$.

\begin{theorem}
    \label{thm:valleys}
    Let $\overline{P} = \langle p_{i-1}, p_i \rangle$ and $\overline{Q} = \langle q_{j-1}, q_j \rangle$ be polygonal segments, and let $\tnorm$ be a norm.
    There exists a valley $\ell$ of positive slope for $\overline{P},\overline{Q}$ under $\tnorm$.
    Computing such a valley reduces to finding a sink via \autoref{thm:duality}, where we use $K := S_{\leq 1}(\tnorm)$ or $K' := \varphi^{-1}(K) - \varphi^{-1}(\mathbf{0})$ with $\varphi \colon \mathbb{R}^2 \to \mathbb{R}^2$ from \autoref{thm:affine}.
\end{theorem}

\begin{proof}
    By definition of $K$ and $\varphi$, we have $\mathcal{G}_K(\varphi(z)) = \| \overline{P}_{\| \cdot \|}(z_1) - \overline{Q}_{\| \cdot \|}(z_2) \|$ for all $z \in \mathbb{R}^2$.
    Moreover, $K$ is absorbing, balanced, closed, bounded and convex.
    
    Assume first that $\overline{P},\overline{Q}$ are parallel, so \autoref{thm:affine} says that $\varphi$ and $\mathcal{G}_K \circ \varphi$ are constant either on every line of slope $1$ or $-1$.
    In the latter case every line not of slope~$-1$ is a valley by \autoref{def:valley}.
    In the former case it suffices to find one sink.
    For this, consider $L := \{ \varphi(s, -s) \mid s \in \mathbb{R} \}$, where $\varphi(s, -s) = 2s \cdot \frac{p_i - p_{i-1}}{\| p_i - p_{i-1} \|} + \varphi(\mathbf{0})$ holds for all $s \in \mathbb{R}$.
    We obtain an $s^* \in \mathbb{R}$ such that $\mathcal{G}_K$ attains its minimum on $L$ at $\varphi(s^*, -s^*)$ by applying \autoref{thm:duality} for any $u \in \mathbb{R}^2 \setminus \{\mathbf{0}\}$ with $u^{\mathsf{T}} \cdot (p_i - p_{i-1}) = 0$ and $t := u^{\mathsf{T}} \cdot \varphi(\mathbf{0})$.
    Also, the properties of~$K$ imply that $L \cap S_{\leq \mu}(\mathcal{G}_K) = L \cap \mu K$ is a convex subset of $L$ for each $\mu \geq 0$.
    It follows that $(s^*, -s^*)^{\mathsf{T}}$ is a sink and that the line $\ell := \{ (\lambda + s^*, \lambda - s^*)^{\mathsf{T}} \mid \lambda \in \mathbb{R} \}$ is a valley of slope $1$ for $\overline{P},\overline{Q}$.
    
    Now assume that $\overline{P},\overline{Q}$ are not parallel.
    Then \autoref{thm:affine} says $S_{\leq \mu}(\mathcal{G}_K \circ \varphi) = \varphi^{-1}(S_{\leq \mu}(\mathcal{G}_K)) = \varphi^{-1}(\mu K)$ for all $\mu \geq 0$.
    Because $\varphi^{-1}$ is a bijective affine map, the set $K' = \varphi^{-1}(K) - \varphi^{-1}(\mathbf{0})$ inherits the properties of $K$ and further satisfies $\mu K' = \varphi^{-1}(\mu K) - \varphi^{-1}(\mathbf{0})$ for all $\mu \geq 0$. Thus, $\mathcal{G}_{K'} = (\mathcal{G}_K \circ \varphi) \circ (z \mapsto z + \varphi^{-1}(\mathbf{0}))$ is fulfilled.
    Given some line $L_t := \{ z \in \mathbb{R}^2 \mid z_1 + z_2 = t \}$ of slope~$-1$ with $t \in \mathbb{R}$, we apply \autoref{thm:duality} for $K'$ and $u' = (1,1)^{\mathsf{T}}$ to see that $\mathcal{G}_{K'}$ attains its minimum on $L_t$ at a $(t/t^*) v^*$, and hence that $(t/t^*) v^* + \varphi^{-1}(\mathbf{0})$ is a sink similar to before.
    By fixing any $v^* \in \arg \max \{ (1,1) \cdot v \mid v \in K' \}$ and $t^* := (1,1) \cdot v^*$, we have that each point on $\ell := \{ \lambda v^* + \varphi^{-1}(\mathbf{0}) \mid \lambda \in \mathbb{R} \}$ is a sink, so $\ell$ is a valley for $\overline{P},\overline{Q}$.
    
    It remains to show that there is such a valley $\ell$ of positive slope by establishing that an appropriate $v^*$ with positive components exists.
    To this end, observe that $\varphi$ gives unit vectors $\varphi(\pm 1, 0) - \varphi(\mathbf{0}) = \pm \frac{p_i - p_{i-1}}{\| p_i - p_{i-1} \|}$ and $\varphi(0, \pm 1) - \varphi(\mathbf{0}) = \mp \frac{q_j - q_{j-1}}{\| q_j - q_{j-1} \|}$ that lie in the boundary of the unit sublevel set $K = S_{\leq 1}(\tnorm)$.
    In consequence, the boundary of the linearly transformed set $K' = \varphi^{-1}(K) - \varphi^{-1}(\mathbf{0})$ contains~the inverse elements $(\pm 1, 0)^{\mathsf{T}}$ and $(0, \pm 1)^{\mathsf{T}}$.
    Because $K'$ is closed and convex, it is
    \[
        L_1 \cap \mathbb{R}_{\geq 0}^2 \subseteq \{ v \in K' \mid (1,1) \cdot v \geq 1 \} \subseteq L_1 \cup (L_1 \cap \mathbb{R}_{\geq 0}^2 + \{ (\lambda, \lambda)^{\mathsf{T}} \mid \lambda > 0 \})
    \]
    with $L_1 = \{ z \in \mathbb{R}^2 \mid z_1 + z_2 = 1 \}$.
    These inclusions ensure a~$v^* \in \mathbb{R}_{> 0}^2$ as desired, see \autoref{subfig:valley-positive-slope}.
    The latter inclusion holds since $K'$ has supporting lines through all of its boundary points due to convexity (see~\cite[Section~2.5.2]{BoydV2009}) and each such line through $(1,0)^{\mathsf{T}} \in L_1$ or $(0,1)^{\mathsf{T}} \in L_1$ is constrained by $(0, \pm 1)^{\mathsf{T}}, (\pm 1,0)^{\mathsf{T}} \in K'$.
    \qed
\end{proof}

To approximate the $2$-norm, we can use norms with polygons as sublevel sets.
Given a suitable $k$-gon $K$, in particular we have that $K$ is convex and $k$ is even, \autoref{thm:valleys} allows computing valleys in $O(\log(k))$ time via binary~search on the vertices of $K$ or $K'$ (cf.~\cite[Section~2.2]{ChazeD1987}).
If $K$ is regular, we achieve $O(1)$ time by maximising $v \mapsto (1,1) \cdot v$ on the ellipse circumscribing $K'$, see \autoref{subfig:valley-computation}.
More generally, this works if $K = \psi(R_k)$ is a linearly transformed regular $k$-gon.

\begin{corollary}
    \label{thm:valleys-for-polygonal-norms}
    Let $\tnorm$ be a norm with $S_{\leq 1}(\tnorm) = \psi(R_k)$ for a regular $k$-gon~$R_k$ and a linear map $\psi \colon \mathbb{R}^2 \to \mathbb{R}^2$.
    Computing a valley $\ell$ of positive slope under $\tnorm$ for arbitrary $\overline{P},\overline{Q}$ is possible in $O(1)$ time, independent of $k \in \{4,6,\dotsc\}$.
\end{corollary}

\section{Unsolvability within \texorpdfstring{\acmq}{ACMQ} under the 2-Norm}
\label{sec:unsolvability}

It has been shown that the partial Fréchet similarity, a measure related to CDTW, cannot be computed exactly under the $2$-norm by radicals over $\mathbb{Q}$ \cite{DeCarGMSS2014}.
Thus, its computational problem is unsolvable within \acmq, the \emph{Algebraic Computation Model over $\mathbb{Q}$} that only allows consecutive applications of the four basic arithmetic operations and the extraction of roots $\sqrt[c]{\parbox{0.375em}{\centering\ensuremath{\cdot}}}$ for all~$c \in \mathbb{N}$.
Such algebraic operations are sufficient for many problems from computational geometry, e.g.\ the Fréchet distance can be computed using only basic arithmetic and square roots \cite{AltG1995}.

The abovementioned result and others of its kind \cite{DeCarGMOS2014,Bajaj1988} utilise Galois theory.
They provide a constraint for potential exact algorithms and hence a motivation for approximation algorithms.
Note that, from a numerical perspective, radicals are also approximations, but they enable symbolic computations~\cite[Remark~1]{DeCarGMOS2014}.
Moreover, there is a recent development \cite{WildbR2025} that solves general polynomials of arbitrary degree by power series, which might start to compete with radicals.

That does not affect our \acmq unsolvability result for CDTW under $\tnorm[2]$, as we show that the numbers involved may not even be algebraic, i.e.\ they do not solve any polynomial equation over $\mathbb{Q}$.
This stronger result uses the following consequence of Baker's Theorem from transcendental number theory.

\begin{lemma}[{\cite[Theorem~2.2]{Baker1990}}]
    \label{thm:baker}
    Let $\alpha_1,\dotsc,\alpha_r \in \mathbb{R}_{> 0}$ and $\beta_0,\beta_1,\dotsc,\beta_r \in \mathbb{R}$ be algebraic numbers.
    The equality $\beta_0 + \beta_1 \ln(\alpha_1) + \dotsb + \beta_r \ln(\alpha_r) = 0$ holds if and only if we have $\beta_0 = 0 = \beta_1 \ln(\alpha_1) + \dotsb + \beta_r \ln(\alpha_r)$.
\end{lemma}

A single example that gives a transcendental CDTW value under the $2$-norm already establishes the unsolvability of the general case.
This of course does not rule out practical numerical computations, but we demonstrate another obstacle:
Optimal paths can involve transcendental numbers too, so that approximations of their costs could introduce cumulative errors when propagated.

\begin{theorem}
    \label{thm:transcendence}
    CDTW under $\tnorm[2]$ is unsolvable within \acmq because polygonal curves $P,Q$ may exhibit the following, even with integer vertices.
    \begin{alphaenumerate*}
        \item 
        The number $\mathrm{cdtw}_{\| \cdot \|_2}(P,Q)$ is transcendental.
        
        \item 
        Every optimal $(\mathbf{0}, (\| P \|_2, \| Q \|_2)^{\mathsf{T}})$-path for $P,Q$ under $\tnorm[2]$ has  bending points whose coordinates are transcendental numbers.
    \end{alphaenumerate*}
\end{theorem}

\begin{proof}[Sketch]
    Showing (b) requires more effort than showing (a).
    For the sake of clarity, we defer some details from the proof of (b) to \autoappref{app:transcendence-details}.
    \begin{alphaenumerate}
        \item
        Let $P := \langle (1,2)^{\mathsf{T}}, (1,-4)^{\mathsf{T}} \rangle$ and $Q := \langle (0,0)^{\mathsf{T}}, (5,0)^{\mathsf{T}} \rangle$, so that $\| P \|_2 + \| Q \|_2 = 6 + 5 = 11$ holds and the parameter space has only one cell, which corresponds to the right cell in \autoref{fig:transcendence}.
        The map $z \mapsto z_1 \cdot (0,-1)^{\mathsf{T}} - z_2 \cdot (1,0)^{\mathsf{T}} + (1,2)^{\mathsf{T}}$ realises $\varphi$ from \autoref{thm:affine} and attains $\mathbf{0}$ at $(2,1)^{\mathsf{T}}$.
        The line $\ell := \{ z \in \mathbb{R}^2 \mid z_1 - z_2 = 1 \}$ through that point is a valley of the cell since the $2$-norm always gives valleys of slope $1$ along ellipse axes in cell terrain (see~\cite{MahesSS2018,Rote2014}).
        
        By \autoref{thm:optimal-paths}, the valley $\ell$ induces an optimal path $\gamma^* \in \Gamma_{\| \cdot \|_2}(P,Q)$ tracing line segments from $\mathbf{0}$ to $\smash{(1,0)^{\mathsf{T}}}$ to $\smash{(6,5)^{\mathsf{T}}}$.
        Evaluating its cost yields
        \begin{align*}
            \mathrm{cdtw}_{\| \cdot \|_2}(P,Q) = &\textstyle\int_{0}^{1} \left\| \varphi(t,0) \right\|_2 \dt + \int_{1}^{11} \left\| \varphi(\tfrac{t+1}{2}, \tfrac{t-1}{2}) \right\|_2 \dt \\[5pt]
            = &\textstyle\int_{0}^{1} \sqrt{t^2 - 4t + 5} \dt + \int_{1}^{11} \tfrac{|t - 3|}{\sqrt{2}} \dt \\[2pt]
            = &\left[ \tfrac{1}{2} \cdot \ln(\alpha_1) - \tfrac{1}{\;\sqrt{2}\;} - \tfrac{1}{2} \cdot \ln(\alpha_2) + \sqrt{5} \right] + 17 \cdot \sqrt{2}
        \end{align*}
        with $\alpha_1 := \sqrt{2} - 1$ and $\alpha_2 := \sqrt{5} - 2$.
        The left integral makes use of $\int h(t) \dt = \frac{4 \lambda_0 - \lambda_1^2}{8} \cdot \ln(|2 \cdot h(s) + 2s + \lambda_1|) + \frac{2s + \lambda_1}{4} \cdot h(s) + C$ for $h(s) := \sqrt{s^2 + \lambda_1 s + \lambda_0}$ with $\lambda_0,\lambda_1 \in \mathbb{R}$ (see~\cite[Equation~4.3.4.1.5]{Jeffr2008}).
        It is $\tfrac{1}{2} \cdot \ln(\alpha_1) - \tfrac{1}{2} \cdot \ln(\alpha_2) \neq 0$ due to $\alpha_1 \neq \alpha_2$.
        Thus, the numbers $\tfrac{1}{2} \cdot \ln(\alpha_1) - \tfrac{1}{2} \cdot \ln(\alpha_2)$ and $\mathrm{cdtw}_{\| \cdot \|_2}(P,Q)$ are transcendental, as assuming otherwise contradicts \autoref{thm:baker}.
        
        \begin{figure}[H]%
            \centering%
            \begin{tikzpicture}[x={0.825\linewidth / 11},y={0.825\linewidth / 11}]
                \begin{scope}
                    \clip (0,0) rectangle (5,5);
                    
                    \foreach \i in {1,...,5} {
                        \path[level set] (2.5,2.5)
                            ellipse[x radius={1/sqrt(8/5)},y radius={1/sqrt(2/5)},rotate=45,scale=\i];
                    };
                \end{scope}
                
                \begin{scope}
                    \clip (5,0) rectangle (11,5);
                    
                    \foreach \i in {1,...,5} {
                        \path[level set] (7,1) circle[radius=\i];
                    };
                \end{scope}
                
                \path[valley] (0,0) -- (5,5);
                \path[valley] (6,0) -- (11,5);
                
                \path[main line] (5,0) -- (5,5);
                \path[main line] (0,0) rectangle (11,5);
                
                \foreach \i in {0,5,11} {
                    \path[main line] (\i,0) -- (\i,-5pt);
                    \node[below=5pt] at (\i,0) {$\i$};
                };
                
                \foreach \i in {1,...,4,6,7,...,10} {
                    \path[main line] (\i,0) -- (\i,-3pt);
                };
                
                \foreach \i in {0,5} {
                    \path[main line] (0,\i) -- (-5pt,\i);
                    \node[left=5pt] at (0,\i) {$\i$};
                };
                
                \foreach \i in {1,...,4} {
                    \path[main line] (0,\i) -- (-3pt,\i);
                };
                
                \foreach \s in {2,5,8} {
                    \coordinate (x) at (\s/2,\s/2);
                    \coordinate (y) at (\s/2+6,\s/2);
                    \path[matching path] (x) -- (y);
                    
                    \node[point,fill=UmiSkyblue] (x\s) at (x) {};
                    \node[point,fill=UmiSkyblue] (y\s) at (y) {};
                    \node[inner sep=0pt,below right=3pt and 1pt] at (x) {\color{UmiSkyblue}$s = \s$};
                };
                
                \node[point,fill=UmiSkyblue] (x0) at (0,0) {};
                \node[point,fill=UmiSkyblue] (y10) at (11,5) {};
                
                \foreach \x/\y in {x0/x2,x2/x5,x5/x8,y2/y5,y5/y8,y8/y10} {
                    \path[matching path,dashed,dash expand off] (\x) -- (\y);
                };
            \end{tikzpicture}%
            \caption{Parameter space whose optimal path candidates switch valleys}%
            \label{fig:transcendence}%
        \end{figure}
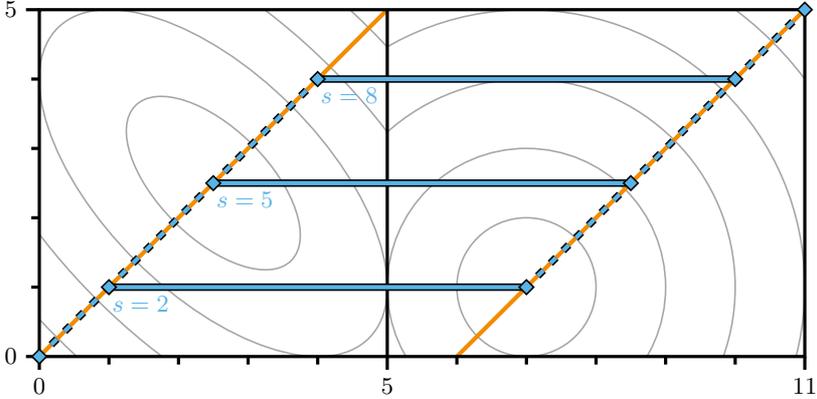
        
        \item
        Now let $P := \langle (4,-2)^{\mathsf{T}}, (1,2)^{\mathsf{T}}, (1,-4)^{\mathsf{T}} \rangle$ and $Q := \langle (0,0)^{\mathsf{T}}, (5,0)^{\mathsf{T}} \rangle$, so that this time $\| P \|_2 + \| Q \|_2 = 11 + 5 = 16$ holds and there are two cells with valleys $\ell_1 := \{ z \in \mathbb{R}^2 \mid z_1 - z_2 = 0 \}$ and $\ell_2 := \{ z \in \mathbb{R}^2 \mid z_1 - z_2 = 6 \}$~respectively.
        Optimal path candidates are all $\gamma_s \in \Gamma_{\| \cdot \|_2}(P,Q)$ tracing line segments from $\mathbf{0}$ to $(s/2, s/2)^{\mathsf{T}}$ to $(s/2 + 6, s/2)^{\mathsf{T}}$ to $(11, 5)^{\mathsf{T}}$ for an $s \in [0,10]$, where~$\gamma_s(s) = (s/2, s/2)^{\mathsf{T}}$ is the bending point of $\gamma_s$ on $\ell_1$ like in \autoref{fig:transcendence}.
		This is because any other path can be transformed into this shape to decrease its cost, by splitting it at the inter-cell boundary and using \autoref{thm:optimal-paths} in each cell.
        
        \begin{claim}[B1]
            Let $\mathcal{C} \colon [0,10] \to \mathbb{R}_{\geq 0}$ be defined by $\mathcal{C}(s)$ being the cost of the optimal path candidate $\gamma_s$ for $s \in [0,10]$.
            We have that $\mathcal{C}$ is continuous.
        \end{claim}
        
        By continuity, $\mathcal{C}$ attains its minimum on the closed interval $[0,10]$, so there is a candidate $\gamma_{s^*}$ that is optimal and has cost equal to $\mathrm{cdtw}_{\| \cdot \|_2}(P,Q)$.
        
        \begin{claim}[B2]
            The function $\mathcal{C}$ is differentiable on $[0,10]$, and its derivative at~$s$ is
            $
                \mathcal{C}'(s) =
                - \frac{1}{5} \cdot \int_0^{5/4 \cdot (2 - \kappa_s)} \frac{t}{\;\sqrt{(t + 2\kappa_s)^2 + \kappa_s^2}\;} \dt
                - \frac{1}{2} \cdot \int_{-(\nu_s + 2)}^{0} \frac{t}{\;\sqrt{(t + \nu_s)^2 + \nu_s^2}\;} \dt
            $
            for $\kappa_s := 2/5 \cdot (s - 5)$ and $\nu_s := 1/2 \cdot (s - 2)$.
            It is $\mathcal{C}'(0) < 0$ and $\mathcal{C}'(10) > 0$.
        \end{claim}
    
        The bounds for $\mathcal{C}'$ at values $0$ and $10$ imply that any $s^*$ minimising $\mathcal{C}$ lies in the open interval $(0,10)$, which means $\mathcal{C}'(s^*) = 0$ by differentiability.
        
        \begin{claim}[B3]
            Some algebraic functions $\alpha_1, \alpha_2, \beta_0, \beta_1, \beta_2 \colon [0,10] \setminus \{2,5\} \to \mathbb{R}$ satisfying $\mathcal{C}'(s) = \beta_0(s) + \beta_1(s) \cdot \ln(\alpha_1(s)) + \beta_2(s) \cdot \ln(\alpha_2(s))$ exist.
            We also have $\mathcal{C}'(2) \neq 0 \neq \mathcal{C}'(5)$, and $\beta_0(s^*_0) = 0$ is fulfilled on $(0,10)$ only by~$s^*_0 := 1/929 \cdot \allowbreak \big( 1880 \cdot \sqrt{10} - 3070 \big) + 60 \cdot \big( \sqrt{16 \cdot \sqrt{10} + 61} \big) / \big( 80 \cdot \sqrt{10} + 269 \big) \in (4,4.5)$.
        \end{claim}
        
        Due to \autoref{thm:baker}, $s^*_0$ is the only algebraic candidate that could minimise $\mathcal{C}$.
        Numerically, we obtain $s^*_0 \approx 4.31$ with $\mathcal{C}'(s^*_0) \approx 0.80$, whereas $\mathcal{C}'(s^*) = 0$ is fulfilled on $(0,10)$ only by $s^* \approx 2.08$.
        We can prove $\mathcal{C}'(s^*_0) \neq 0$ as follows.
        
        \begin{claim}[B4]
            It is $\beta_1(s) \cdot \ln(\alpha_1(s)) + \beta_2(s) \cdot \ln(\alpha_2(s)) > 0$ for all $s \in (4,4.5)$.
        \end{claim}
        
        So the minimum of $\mathcal{C}$ is not attained at an algebraic value.
        All $s^*$ minimising~$\mathcal{C}$, along with the bending points of all optimal $\gamma_{s^*}$, are thus transcendental.
        \qed
    \end{alphaenumerate}
\end{proof}

\section{An Exact Algorithm under Polygonal Norms}
\label{sec:exact-algorithm}

This section uses a parameter $k \in \{4,6,\dotsc\}$, a bijective linear map $\psi \colon \mathbb{R}^2 \to \mathbb{R}^2$, and a norm $\tnorm$ on $\mathbb{R}^2$ as follows:
Whenever we write $\tnorm$, the respective definition or result applies to every possible norm.
For the algorithm we assume a $1$-sublevel set $S_{\leq 1}(\tnorm) = \psi(R_k)$, where $R_k$ denotes the convex regular $k$-gon with vertices $v_1,\dotsc,v_k$ defined by $v_r := (\cos(\theta_r),\sin(\theta_r))^{\mathsf{T}}$ for $\theta_r := r \cdot 2\pi/k$ and $r \in \{1,\dotsc,k\}$.
In that case we use the gauge notation $\mathcal{G}_{\psi(R_k)}$, as specified in \autoref{subsec:valleys}.

This setting covers all norms with regular polygons as sublevel sets, including the $1$-norm and the $\infty$-norm, and allows to approximate the $2$-norm.
Instead of discretising the input curves like in previous approaches \cite{BrakaPSW2005,EfratFV2007,MahesSS2018,BrankBKNPW2020,Har-PRR2025}, we hence essentially discretise the norm.
Omitted proofs are given in \autoappref{app:algorithm-proofs}.

We apply a dynamic programming scheme known from other exact algorithms for CDTW and related measures \cite{SerraB1994,SerraB1995,MunicP1999,BuchiBW2009,Klare2020,BuchiNW2022,Brank2022}.
It propagates optimal path costs through the grid of parameter space cells in the form of functions.

\begin{definition}[{\cite[Definition~5]{BuchiNW2022}}]
    \label{def:borders-and-optimum-function}
    Let $P, Q$ be polygonal curves.
    For every parameter space cell $[a_1,b_1] \times [a_2,b_2] \subseteq [0,\| P \|] \times [0,\| Q \|]$ we define its \emph{borders} as parametrisations along the top/right/bottom/left side of its boundary, e.g.\ its top border is $[a_1,b_1] \to \mathbb{R}^2, t \mapsto (t,b_2)^{\mathsf{T}}$.
    A cell border~$\mathcal{B} \colon \mathrm{dom}(\mathcal{B}) \to \mathbb{R}^2$ has an associated \emph{optimum function} denoted by $\mathrm{opt}_{0,\mathcal{B}} \colon \mathrm{dom}(\mathcal{B}) \to \mathbb{R}_{\geq 0}$, where $\mathrm{opt}_{0,\mathcal{B}}(t)$ is the cost of an optimal $(\mathbf{0},\mathcal{B}(t))$-path under $\tnorm$ for each $t \in \mathrm{dom}(\mathcal{B})$.
\end{definition}

In the dynamic program's base case we compute optimum functions of borders on the parameter space's bottom and left side via costs of straight paths.

\subsection{Propagation Procedure}

Our propagation procedure builds upon \cite[Section~5.2]{Klare2020} and \cite[Section~5.3]{Brank2022}.

\begin{definition}
    \label{def:propagation-space}
    Let $\mathcal{A}$ be the bottom or left border and let $\mathcal{B}$ be the top or right border of one cell.
    The \emph{propagation space} of $\mathcal{A},\mathcal{B}$ is defined by $\Sigma_{\mathcal{A},\mathcal{B}} := \{ (s,t)^{\mathsf{T}} \in \mathrm{dom}(\mathcal{A}) \times \mathrm{dom}(\mathcal{B}) \mid (\mathcal{A}_1(s) \leq \mathcal{B}_1(t)) \land (\mathcal{A}_2(s) \leq \mathcal{B}_2(t)) \}$.
    If $\mathcal{A},\mathcal{B}$ are a pair of left and top border or a pair of bottom and right border, we call them \emph{adjoining} borders and write $\mathrm{adj}(\mathcal{B}) := \mathcal{A}$.
    Else, we call $\mathcal{A},\mathcal{B}$ \emph{opposing} and write $\mathrm{opp}(\mathcal{B}) := \mathcal{A}$.
\end{definition}

We equip $\mathrm{opt}_{\mathcal{A},\mathcal{B}} \colon \Sigma_{\mathcal{A},\mathcal{B}} \to \mathbb{R}_{\geq 0}$ on the propagation space, where $\mathrm{opt}_{\mathcal{A},\mathcal{B}}(s,t)$ is the cost of an optimal $(\mathcal{A}(s),\mathcal{B}(t))$-path under $\tnorm$ for each $(s,t)^{\mathsf{T}} \in \Sigma_{\mathcal{A},\mathcal{B}}$, so
\begin{equation}
    \tag{$\star$}\label{eq:propagation}
    \mathrm{opt}_{0,\mathcal{B}}(t) =
    \min_{\mathcal{A} \in \{\mathrm{adj}(\mathcal{B}),\mathrm{opp}(\mathcal{B})\}}
    \, \inf \big\{ \mathrm{opt}_{0,\mathcal{A}}(s) + \mathrm{opt}_{\mathcal{A},\mathcal{B}}(s,t) \bigm| (s,t)^{\mathsf{T}} \in \Sigma_{\mathcal{A},\mathcal{B}} \big\}
\end{equation}
holds for all $t \in \mathrm{dom}(\mathcal{B})$.
After determining optimum functions of $\mathrm{adj}(\mathcal{B}),\mathrm{opp}(\mathcal{B})$, we propagate costs to $\mathcal{B}$ by implementing \autoeqref{eq:propagation} with the help of \autoref{thm:optimal-paths}.
Our implementation \nameref{alg:Propagate} and its correctness are outlined below.

\begin{lemma}
    \label{thm:polygonal-norm-properties}
    \begin{alphaenumerate*}
        \item 
        We can evaluate $\mathcal{G}_{\psi(R_k)}$ in $O(1)$ time.
        Its restriction to each~cone $\psi(\{ \lambda v_{r-1} + \lambda' v_{r} \mid \lambda, \lambda' \geq 0 \})$ is linear, where $r \in \{1,\dotsc,k\}$ and $v_0 := v_k$.
        
        \item 
        Let $\Sigma_{\mathcal{A},\mathcal{B}}$ be the propagation space of borders $\mathcal{A},\mathcal{B}$ under $\mathcal{G}_{\psi(R_k)}$.
        There is an arrangement $(V,E,F)$ of $O(k)$ lines with $\Sigma_{\mathcal{A},\mathcal{B}} = \bigcup F'$ for a subset $F' \subseteq F$~of closed faces such that each restriction of $\mathrm{opt}_{\mathcal{A},\mathcal{B}}$ to an $f \in F'$ is quadratic.
        
        \item 
        The function $\mathrm{opt}_{0,\mathcal{B}}$ is piecewise quadratic for every border $\mathcal{B}$ under $\mathcal{G}_{\psi(R_k)}$.
    \end{alphaenumerate*}
\end{lemma}

\begin{algobox}{Propagate}
    \begin{pseudo}*
        \pseudohd{\nameref*{alg:Propagate}}(\overline{P},\overline{Q}, \mathcal{B}, \mathrm{opt}_{0,\mathrm{adj}(\mathcal{B})}, \mathrm{opt}_{0,\mathrm{opp}(\mathcal{B})})%
        \pseudoinputoutput%
            {Polygonal segments $\overline{P},\overline{Q}$ and a top/right border $\mathcal{B}$ of their cell under $\mathcal{G}_{\psi(R_k)}$.
            Optimum functions $\mathrm{opt}_{0,\mathcal{A}}$ of $\mathcal{A} \in \{ \mathrm{adj}(\mathcal{B}), \mathrm{opp}(\mathcal{B}) \}$ as stacks of quadratic pieces.}%
            {Optimum function $\mathrm{opt}_{0,\mathcal{B}}$ of $\mathcal{B}$ as a stack of quadratic pieces.}%
        $\mathcal{S} \gets$ empty stack; $H \gets \varnothing$; $\ell \gets$ valley of positive slope for $\overline{P},\overline{Q}$ under $\mathcal{G}_{\psi(R_k)}$ \\
        \pseudokw{foreach} $\mathcal{A}$ \pseudokw{in} $\langle \mathrm{adj}(\mathcal{B}), \mathrm{opp}(\mathcal{B}) \rangle$ \pseudokw{do} \pseudoct{start with adjoining border} \\+
            $(V,E,F) \gets$ arrangement as in \autoref{thm:polygonal-norm-properties}b, based on $\ell$ and \autoref{thm:optimal-paths} \\
            overlay $(V,E,F)$ with vertical lines arranged at breakpoints of $\mathrm{opt}_{0,\mathcal{A}}$ \\
            $\mathcal{I}$ $\gets$ ascending list of closed intervals split by vertical lines of $(V,E,F)$ \\
            \pseudokw{if} $\mathcal{A} = \mathrm{adj}(\mathcal{B})$ \pseudokw{then} reverse $\mathcal{I}$ \pseudoct{choose order depending on $\mathcal{A}$} \\
            \pseudokw{foreach} interval $I$ \pseudokw{in} $\mathcal{I}$ \pseudokw{do} \pseudoct{process intervals in order} \\+
                \pseudokw{foreach} edge $e$ \pseudokw{in} $\{ e \in E \mid e \subseteq (I \times \mathbb{R}) \cap \Sigma_{\mathcal{A},\mathcal{B}} \}$ \pseudokw{do} \\+
                    add $[t \mapsto \mathrm{opt}_{0,\mathcal{A}}(s_e(t)) + \mathrm{opt}_{\mathcal{A},\mathcal{B}}(s_e(t),t)]$ to $H$, where $(s_e(t),t)^{\mathsf{T}} \in e$ \\-
                \pseudokw{foreach} face $f$ \pseudokw{in} $\{ f \in F \mid f \subseteq (I \times \mathbb{R}) \cap \Sigma_{\mathcal{A},\mathcal{B}} \}$ \pseudokw{do} \\+
                    $e_f \gets \{ (s,t)^{\mathsf{T}} \in f \mid \partial_s [\mathrm{opt}_{0,\mathcal{A}}(s) + \mathrm{opt}_{\mathcal{A},\mathcal{B}}(s,t)] = 0 \}$ \pseudoct{extremal edge} \\
                    add $[t \mapsto \mathrm{opt}_{0,\mathcal{A}}(s_f(t)) + \mathrm{opt}_{\mathcal{A},\mathcal{B}}(s_f(t),t)]$ to $H$, where $(s_f(t),t)^{\mathsf{T}} \in e_f$ \\-
                $\mathcal{S}_I \gets$ lower envelope of $H$ as a stack of quadratic pieces; $H \gets \varnothing$ \\
                update top of $\mathcal{S}$ with suffix of smaller-valued pieces on top of $\mathcal{S}_I$ \\--
        \pseudokw{return} $\mathcal{S}$
    \end{pseudo}
\end{algobox}

Due to \autoref{thm:polygonal-norm-properties}c, \nameref{alg:Propagate} exclusively deals with quadratic pieces, and its method to compute these is based on arrangements as in \autoref{thm:polygonal-norm-properties}b.
The~use of stack data structures (see \cite[Lemma~12]{BuchiNW2022}) speeds up computations and relies on the following property, which is a stronger version of \cite[Lemma~11]{BuchiNW2022}.

\begin{lemma}
    \label{thm:propagation-order}
    Let $\mathcal{B}$ be a top/right border under $\tnorm$, let $\mathcal{A} \colon [a,b] \to \mathbb{R}^2$ be~a~parametrisation in direction $(1,0)^{\mathsf{T}}$ or $(0,1)^{\mathsf{T}}$, and let $[s\;\!\triangleright\;\!t] := \mathrm{opt}_{0,\mathcal{A}}(s) + \mathrm{opt}_{\mathcal{A},\mathcal{B}}(s,t)$.
    If $[s \triangleright t] < [s' \triangleright t]$ and $[s \triangleright t'] \geq [s' \triangleright t']$ for $t < t'$, then $s < s'$ if and only if $\mathcal{A}$ and $\mathcal{B}$ have the same direction.
    This yields a propagation order for $\mathcal{A} \in \{ \mathrm{adj}(\mathcal{B}),\mathrm{opp}(\mathcal{B}) \}$.
\end{lemma}

\begin{theorem}
    \label{thm:propagation-correctness}
    \nameref{alg:Propagate} correctly computes optimum functions under $\mathcal{G}_{\psi(R_k)}$.
\end{theorem}

\begin{proof}
    We obtain $\ell, \mathrm{opt}_{\mathcal{A},\mathcal{B}}, (V,E,F)$ via \autoref{thm:valleys-for-polygonal-norms}, \autoref{thm:optimal-paths} and \autoref{thm:polygonal-norm-properties}b.
    Lines~4~\&~5 ensure that the loops cover all of $\Sigma_{\mathcal{A},\mathcal{B}}$ and that lines~9~\&~12 deal with quadratic restrictions of $\mathrm{opt}_{0,\mathcal{A}}$ and $\mathrm{opt}_{\mathcal{A},\mathcal{B}}$.
    Minima of $s \mapsto \mathrm{opt}_{0,\mathcal{A}}(s) + \mathrm{opt}_{\mathcal{A},\mathcal{B}}(s,t)$ are attained on edges and face interiors.
    The latter minima come with derivative value $0$, which yields a linear relation of $s$ and $t$ determined in line~11.
    To realise \autoeqref{eq:propagation}, lines~13~\&~14 process lower envelopes.
    Here, \autoref{thm:propagation-order} and the~order of $\mathcal{I}$ imply that if a value on $\mathcal{S}$ is improved by $\mathcal{S}_I$, all values above it are too.
    \qed
\end{proof}

\begin{corollary}
    \label{thm:algorithm-results}
    We can compute CDTW exactly under $\mathcal{G}_{\psi(R_k)}$, yielding a $(1+\varepsilon)$-approximation for CDTW under $\tnorm[2]$ using $\psi := \mathrm{id}$ and some $k \in O(\varepsilon^{-1/2})$.
\end{corollary}

\subsection{Complexity and Properties}
\label{subsec:complexity}

We consider the ideas from the 1D running time analysis~\cite{BuchiNW2022} in our setting.

\begin{proposition}
    \label{thm:running-time}
    The running time of our algorithm for CDTW under $\mathcal{G}_{\psi(R_k)}$ is bounded by $O(N \cdot k^2 \log(k) \alpha(k))$, where $N$ denotes the total number of quadratic pieces over all optimum functions and $\alpha$ is the inverse Ackermann function.
\end{proposition}

An essential ingredient to showing $N \in O(n^5)$ in 1D is an inductive bound~on the number of distinct pairs $(\lambda_1,\lambda_2)^{\mathsf{T}} \in \mathbb{R}^2$ such that $s \mapsto \textstyle\sum_{i=0}^{2} \lambda_i s^i$ is a quadratic piece \cite[Lemma~14]{BuchiNW2022}.
It is not clear if such a bound holds in 2D, as dealing with the arrangements from \autoref{thm:polygonal-norm-properties}b poses a greater challenge in this setting.

That is, the cones from \autoref{thm:polygonal-norm-properties}a can induce lines of negative slope, which do not occur in 1D.
Even optimal paths in the order from \autoref{thm:propagation-order} may have bending points that repeatedly switch faces via these lines, see \autoref{fig:doubling}.
It is an unresolved issue whether this could cause an exponential doubling behaviour.

\begin{figure}[H]%
    \centering%
    \begin{minipage}[b]{0.45\linewidth}%
        \centering%
        \begin{tikzpicture}[x=0.95\linewidth,y={0.95\linewidth * 0.55}]
            \useasboundingbox (0,0) rectangle (1,1);
            \path[main line] (0,1) -- (1,1);
            
            \foreach \j/\c [remember=\j as \i] in {1/,0.675/white,0/black} {
                \ifthenelse{\equal{\j}{0} \OR \equal{\j}{1}}{
                    \path[main line] (\j,-5pt) -- (\j,0);
                }{
                    \path[main line] (\j,-3pt) -- (\j,3pt);
                }
                
                \ifthenelse{\equal{\c}{}}{}{
                    \path[piece=\c] (\i,0) -- (\j,0);
                }
            };
            
            \foreach \j/\c [remember=\j as \i] in {0/,1/white} {
                \path[main line] (-5pt,\j) -- (0,\j);
                
                \ifthenelse{\equal{\c}{}}{}{
                    \path[piece=\c] (0,\i) -- (0,\j);
                }
            };
            
            \foreach \j/\c [remember=\j as \i] in {0/,0.2/white,0.4/white,0.6/black,0.8/black,1/white} {
                \ifthenelse{\equal{\j}{0} \OR \equal{\j}{1}}{
                    \path[main line] (1,\j) -- ($(1,\j) + (5pt,0)$);
                }{
                    \path[main line] ($(1,\j) + (-3pt,0)$) -- ($(1,\j) + (3pt,0)$);
                }
                
                \ifthenelse{\equal{\c}{}}{}{
                    \path[piece=\c] (1,\i) -- (1,\j);
                }
            };
            
            \begin{scope}[on background layer]
                \clip (0,0) rectangle (1,1);
                
                \path[diagonal,name path=diag] (0,1) -- (1,0);
                \path[matching path,name path=path1] (0.85,0) |- (1,0.1);
                \path[matching path,name path=path2] (0.775,0) |- (1,0.3);
                \path[matching path,name path=path3] (0.575,0) |- (1,0.5);
                \path[matching path,name path=path4] (0.25,0) |- (1,0.7);
                \path[matching path,name path=path5] (0,0.9) -- (1,0.9);
                
                \foreach \i in {1,...,5} {
                    \path[name intersections={of=diag and path\i}];
                    \node[diagonal intersection] at (intersection-1) {};
                };
            \end{scope}
        \end{tikzpicture}%
        \captionof{figure}{Doubling of adjoining pieces}%
        \label{fig:doubling}%
    \end{minipage}%
    \hspace*{0.05\linewidth}%
    \begin{minipage}[b]{0.45\linewidth}%
        \centering%
        \begin{tikzpicture}[x=0.95\linewidth,y={0.95\linewidth * 0.55}]
            \path[main line] (0.4,0.35) -- (0.4,1);
            \path[matching path] (0,0) -- (0.125,0.45) -- (0.4,0.45);
            
            \node[origin] at (0,0) {};
            \node[matching point,label={right:\color{UmiSkyblue}$z$}] (z) at (0.4,0.45) {};
            \node[matching point] (t) at (0.4,0.9) {};
            \path[matching path,dashed,dash expand off] (z) -- (t);
            
            \begin{scope}[shift={(0.525,0)}]
                \coordinate (tr) at (0.4,1);
                \path[main line] (0.4,0.35) -- (tr);
                \path[matching path] (0,0) -- (0.25,0.9) -- (0.4,0.9);
                
                \node[origin] at (0,0) {};
                \node[matching point,label={[label distance=0.45pt]left:\color{UmiSkyblue}$z$}] (z) at (0.125,0.45) {};
                \node[matching point] (t) at (0.4,0.45) {};
                \node[matching point] at (0.4,0.9) {};
                \path[matching path,dashed,dash expand off] (z) -- (t);
            \end{scope}
            
            \pgfresetboundingbox
            \useasboundingbox (0,0) rectangle (tr);
        \end{tikzpicture}%
        \captionof{figure}{Prefix-based path creation}%
        \label{fig:path-creation}%
    \end{minipage}%
\end{figure}

Still, we next generalise another important property, the continuity of optimum functions \cite[Lemma~7]{BuchiNW2022}.
We use $\varlimsup$ and $\varliminf$ to denote $\limsup$ and $\liminf$.

\begin{lemma}
    \label{thm:converging-costs}
    Consider the parameter space of polygonal curves $P,Q$ under~$\tnorm$.
    Let $(\gamma_r)_{r \in \mathbb{N}}$ be a sequence of $(x_r,y_r)$-paths that converge to an $(x,y)$-path~$\gamma$, i.e.\ it is $\lim_{r \to \infty} (x_r,y_r) = (x,y)$ and $\lim_{r \to \infty} \gamma_r(s) = \gamma(s)$ for all $s \in (\| x \|_1, \| y \|_1)$.
    Then the associated sequence $(\mathrm{cost}(\gamma_r))_{r \in \mathbb{N}}$ of  path costs converges to $\mathrm{cost}(\gamma)$.
\end{lemma}

\begin{proof}
    First, assume $(x_r,y_r) = (x,y)$ for all $r \in \mathbb{N}$.
    Then the path costs are~integrals over $[\| x \|_1, \| y \|_1]$ by \autoref{def:path-and-cost}.
    Their integrands converge and are uniformly bounded by $M := \max \{ \| P_{\| \cdot \|}(z_1) - Q_{\| \cdot \|}(z_2) \| \mid z \in [0,\| P \|] \times [0,\| Q \|] \}$.
    The Dominated Convergence Theorem (see \cite{Luxem1971}) implies $\lim_{r \to \infty} \mathrm{cost}(\gamma_r) = \mathrm{cost}(\gamma)$.
    
    Now consider general paths $\gamma_r \colon [\| x_r \|_1, \| y_r \|_1] \to [0,\|P\|] \times [0,\|Q\|]$.
    We can normalise their domains to $[\| x \|_1, \| y \|_1]$ by removing excess prefix/suffix paths and adding (arbitrary) missing prefix/suffix paths.
    The costs of these prefix/suffix paths converge to $0$ because of $\lim_{r \to \infty} (x_r,y_r) = (x,y)$ and the uniform bound~$M$.
    Consequently, the normalisation does not change the limit of $(\mathrm{cost}(\gamma_r))_{r \in \mathbb{N}}$.
    \qed
\end{proof}

\begin{theorem}
    \label{thm:continuity}
    The function $\mathrm{opt}_{0,\mathcal{B}}$ is continuous for every border $\mathcal{B}$ under $\tnorm$.
\end{theorem}
    
\begin{proof}
    We prove that $\varlimsup_{t \to t_0} \mathrm{opt}_{0,\mathcal{B}}(t) \leq \mathrm{opt}_{0,\mathcal{B}}(t_0) \leq \varliminf_{t \to t_0} \mathrm{opt}_{0,\mathcal{B}}(t)$ holds for $t_0 \in \mathrm{dom}(\mathcal{B})$, which implies the result.
    Let $\gamma_0^*$ be an optimal $(\mathbf{0},\mathcal{B}(t_0))$-path, and create $(\mathbf{0},\mathcal{B}(t))$-paths $\gamma_t$ for all $t \in \mathrm{dom}(\mathcal{B})$ by concatenating the prefix path of~$\gamma_0^*$ up to its final point $z \in [0,\mathcal{B}_1(t)] \times [0,\mathcal{B}_2(t)]$ with the straight $(z,\mathcal{B}(t))$-path, as in \autoref{fig:path-creation}.
    It is $\mathrm{opt}_{0,\mathcal{B}}(t) \leq \mathrm{cost}(\gamma_t)$ for $t \in \mathrm{dom}(\mathcal{B})$, while $(\gamma_t)_t$ converges to~$\gamma_0^*$~for $t \to t_0$, so \autoref{thm:converging-costs} gives $\mathop{\smash{\varlimsup}}_{t \to t_0} \mathrm{opt}_{0,\mathcal{B}}(t) \leq \lim_{t \to t_0} \mathrm{cost}(\gamma_t) = \mathrm{opt}_{0,\mathcal{B}}(t_0)$.
    
    Now consider optimal $(\mathbf{0},\mathcal{B}(t))$-paths $\gamma_t^*$ for $t \in \mathrm{dom}(\mathcal{B})$, and create $(\mathbf{0},\mathcal{B}(t_0))$-paths $\gamma_0^t$ with $\mathrm{opt}_{0,\mathcal{B}}(t_0) \leq \mathrm{cost}(\gamma_0^t)$ as above.
    Thus, $\mathrm{opt}_{0,\mathcal{B}}(t_0) \leq \varliminf_{t \to t_0} \mathrm{cost}(\gamma_0^t)$.
    By construction, $\gamma_0^t$ and $\gamma_t^*$ share a prefix path up to some point $z_t$.
    For $t \nearrow t_0$ we have $(z_t)_t \to \mathcal{B}(t_0)$, so that the suffix costs of $(\gamma_0^t)_t$ converge to $0$, while for $t \searrow t_0$ we may assume by \autoref{thm:propagation-order} that there is a $z$ with $(z_t)_t \to z$.
    Then \autoref{thm:converging-costs} implies that the suffix costs of $(\gamma_0^t)_t$ and of $(\gamma_t^*)_t$ both converge to the cost of the straight $(z,\mathcal{B}(t_0))$-path.
	It hence follows $\varliminf_{t \to t_0} \mathrm{cost}(\gamma_0^t) = \varliminf_{t \to t_0} \mathrm{opt}_{0,\mathcal{B}}(t)$.
    \qed
\end{proof}

\section{Conclusion}

We have contributed several fundamentals for the computation of CDTW in~2D, tackling the challenges associated with an integral-based measure.
To facilitate algorithm design and analysis, our results comprise technical properties that~hold under all norms and attest to the robustness of our CDTW formulation.
Moreover, we have shown that CDTW under the $2$-norm can involve transcendental~numbers.
Finding a tight bound on our algorithm's complexity remains an open~problem.

    \bibliography{literature.bib}

\begin{thebibliography}{1}

\bibitem{ChewD1985}
L.~Paul Chew and Robert L.~(Scot) {Drysdale, III}.
\newblock Voronoi diagrams based on convex distance functions.
\newblock In {\em Proceedings of the First Annual Symposium on Computational
  Geometry}, pages 235--244. ACM, 1985.
\newblock \href {https://doi.org/10.1145/323233.323264}
  {\path{doi:10.1145/323233.323264}}.

\bibitem{deBerCKO2008}
Mark {de Berg}, Otfried Cheong, Marc {van Kreveld}, and Mark {Overmars}.
\newblock {\em Computational Geometry: Algorithms and Applications}.
\newblock Springer, 3rd edition, 2008.
\newblock \href {https://doi.org/10.1007/978-3-540-77974-2}
  {\path{doi:10.1007/978-3-540-77974-2}}.

\bibitem{Hersh1989}
John Hershberger.
\newblock Finding the upper envelope of $n$ line segments in {$O(n \log n)$}
  time.
\newblock {\em Information Processing Letters}, 33(4):169--174, 1989.
\newblock \href {https://doi.org/10.1016/0020-0190(89)90136-1}
  {\path{doi:10.1016/0020-0190(89)90136-1}}.

\bibitem{ProttM1985}
Murray~H. Protter and Charles~B. {Morrey, Jr.}
\newblock {\em Intermediate Calculus}.
\newblock Springer, 2nd edition, 1985.
\newblock \href {https://doi.org/10.1007/978-1-4612-1086-3}
  {\path{doi:10.1007/978-1-4612-1086-3}}.

\end{thebibliography}


\begin{thebibliography}{10}

\bibitem{AltG1995}
Helmut Alt and Michael Godau.
\newblock Computing the {Fréchet} distance between two polygonal curves.
\newblock {\em International Journal of Computational Geometry \&
  Applications}, 5(1--2):75--91, 1995.
\newblock \href {https://doi.org/10.1142/S0218195995000064}
  {\path{doi:10.1142/S0218195995000064}}.

\bibitem{Bajaj1988}
Chandrajit Bajaj.
\newblock The algebraic degree of geometric optimization problems.
\newblock {\em Discrete \& Computational Geometry}, 3(2):177--191, 1988.
\newblock \href {https://doi.org/10.1007/BF02187906}
  {\path{doi:10.1007/BF02187906}}.

\bibitem{Baker1990}
Alan Baker.
\newblock {\em Transcendental Number Theory}.
\newblock Cambridge University Press, 1990.
\newblock Reissue with updated material.
\newblock \href {https://doi.org/10.1017/CBO9780511565977}
  {\path{doi:10.1017/CBO9780511565977}}.

\bibitem{BoydV2009}
Stephen Boyd and Lieven Vandenberghe.
\newblock {\em Convex Optimization}.
\newblock Cambridge University Press, 7th edition, 2009.
\newblock URL: \url{https://web.stanford.edu/~boyd/cvxbook/}.

\bibitem{BrakaPSW2005}
Sotiris Brakatsoulas, Dieter Pfoser, Randall Salas, and Carola Wenk.
\newblock On map-matching vehicle tracking data.
\newblock In {\em Proceedings of the 31st International Conference on Very
  Large Data Bases}, pages 853--864. VLDB Endowment, 2005.
\newblock URL: \url{https://dl.acm.org/doi/10.5555/1083592.1083691}.

\bibitem{Brank2022}
Milutin Brankovic.
\newblock {\em Graphs and Trajectories in Practical Geometric Problems}.
\newblock PhD thesis, University of Sydney, 2022.

\bibitem{BrankBKNPW2020}
Milutin Brankovic, Kevin Buchin, Koen Klaren, André Nusser, Aleksandr Popov,
  and Sampson Wong.
\newblock $(k,l)$-medians clustering of trajectories using {Continuous Dynamic
  Time Warping}.
\newblock In {\em Proceedings of the 28th International Conference on Advances
  in Geographic Information Systems}, pages 99--110. ACM, 2020.
\newblock \href {https://doi.org/10.1145/3397536.3422245}
  {\path{doi:10.1145/3397536.3422245}}.

\bibitem{BuchiBMM2017}
Kevin Buchin, Maike Buchin, Wouter Meulemans, and Wolfgang Mulzer.
\newblock Four soviets walk the dog: Improved bounds for computing the
  {Fréchet} distance.
\newblock {\em Discrete \& Computational Geometry}, 58(1):180--216, 2017.
\newblock \href {https://doi.org/10.1007/S00454-017-9878-7}
  {\path{doi:10.1007/S00454-017-9878-7}}.

\bibitem{BuchiBW2009}
Kevin Buchin, Maike Buchin, and Yusu Wang.
\newblock Exact algorithms for partial curve matching via the {Fréchet}
  distance.
\newblock In {\em Proceedings of the Twentieth Annual ACM-SIAM Symposium on
  Discrete Algorithms}, pages 645--654. SIAM, 2009.
\newblock \href {https://doi.org/10.1137/1.9781611973068.71}
  {\path{doi:10.1137/1.9781611973068.71}}.

\bibitem{BuchiNW2022}
Kevin Buchin, André Nusser, and Sampson Wong.
\newblock Computing {Continuous Dynamic Time Warping} of time series in
  polynomial time.
\newblock In {\em 38th International Symposium on Computational Geometry},
  pages 22:1--22:16. Schloss Dagstuhl -- Leibniz-Zentrum für Informatik, 2022.
\newblock \href {https://arxiv.org/abs/2203.04531} {\path{arXiv:2203.04531}},
  \href {https://doi.org/10.4230/LIPICS.SOCG.2022.22}
  {\path{doi:10.4230/LIPICS.SOCG.2022.22}}.

\bibitem{Buchi2007}
Maike Buchin.
\newblock {\em On the Computability of the {Fréchet} Distance between
  Triangulated Surfaces}.
\newblock PhD thesis, Freie Universität Berlin, 2007.
\newblock URL: \url{https://refubium.fu-berlin.de/handle/fub188/1909}.

\bibitem{ChazeD1987}
Bernard Chazelle and David~P. Dobkin.
\newblock Intersection of convex objects in two and three dimensions.
\newblock {\em Journal of the ACM}, 34(1):1--27, 1987.
\newblock \href {https://doi.org/10.1145/7531.24036}
  {\path{doi:10.1145/7531.24036}}.

\bibitem{ChengH2025}
Siu-Wing Cheng and Haoqiang Huang.
\newblock {Fréchet} distance in subquadratic time.
\newblock In {\em Proceedings of the 2025 Annual ACM-SIAM Symposium on Discrete
  Algorithms}, pages 5100--5113. SIAM, 2025.
\newblock \href {https://doi.org/10.1137/1.9781611978322.173}
  {\path{doi:10.1137/1.9781611978322.173}}.

\bibitem{DeCarGMSS2014}
Jean-Lou {De Carufel}, Amin Gheibi, Anil Maheshwari, Jörg-Rüdiger Sack, and
  Christian Scheffer.
\newblock Similarity of polygonal curves in the presence of outliers.
\newblock {\em Computational Geometry}, 47(5):625--641, 2014.
\newblock \href {https://doi.org/10.1016/J.COMGEO.2014.01.002}
  {\path{doi:10.1016/J.COMGEO.2014.01.002}}.

\bibitem{DeCarGMOS2014}
Jean-Lou {De Carufel}, Carsten Grimm, Anil Maheshwari, Megan Owen, and Michiel
  Smid.
\newblock A note on the unsolvability of the weighted region shortest path
  problem.
\newblock {\em Computational Geometry}, 47(7):724--727, 2014.
\newblock \href {https://doi.org/10.1016/J.COMGEO.2014.02.004}
  {\path{doi:10.1016/J.COMGEO.2014.02.004}}.

\bibitem{EfratFV2007}
Alon Efrat, Quanfu Fan, and Suresh Venkatasubramanian.
\newblock Curve matching, time warping, and light fields: New algorithms for
  computing similarity between curves.
\newblock {\em Journal of Mathematical Imaging and Vision}, 27(3):203--216,
  2007.
\newblock \href {https://doi.org/10.1007/S10851-006-0647-0}
  {\path{doi:10.1007/S10851-006-0647-0}}.

\bibitem{Frech1906}
Maurice Fréchet.
\newblock Sur quelques points du calcul fonctionnel.
\newblock {\em Rendiconti del Circolo Matematico di Palermo}, 22(1):1--72,
  1906.
\newblock \href {https://doi.org/10.1007/BF03018603}
  {\path{doi:10.1007/BF03018603}}.

\bibitem{GoldS2018}
Omer Gold and Micha Sharir.
\newblock {Dynamic Time Warping} and {Geometric Edit Distance}: Breaking the
  quadratic barrier.
\newblock {\em ACM Transactions on Algorithms}, 14(4):50:1--50:17, 2018.
\newblock \href {https://doi.org/10.1145/3230734} {\path{doi:10.1145/3230734}}.

\bibitem{Har-PRR2025}
Sariel Har-Peled, Benjamin Raichel, and Eliot~W. Robson.
\newblock The {Fréchet} distance unleashed: Approximating a dog with a frog.
\newblock In {\em 41st International Symposium on Computational Geometry},
  pages 54:1--54:13. Schloss Dagstuhl -- Leibniz-Zentrum für Informatik, 2025.
\newblock \href {https://arxiv.org/abs/2407.03101} {\path{arXiv:2407.03101}},
  \href {https://doi.org/10.4230/LIPICS.SOCG.2025.54}
  {\path{doi:10.4230/LIPICS.SOCG.2025.54}}.

\bibitem{Jeffr2008}
Alan Jeffrey and Hui-Hui Dai.
\newblock {\em Handbook of Mathematical Formulas and Integrals}.
\newblock Elsevier, 4th edition, 2008.

\bibitem{Klare2020}
Koen Klaren.
\newblock {Continuous Dynamic Time Warping} for clustering curves.
\newblock Master's thesis, Eindhoven University of Technology, 2020.
\newblock URL:
  \url{https://research.tue.nl/en/studentTheses/continuous-dynamic-time-warping-for-clustering-curves}.

\bibitem{Luxem1971}
Wilhelmus A.~J. Luxemburg.
\newblock {Arzelà's Dominated Convergence Theorem} for the {Riemann} integral.
\newblock {\em The American Mathematical Monthly}, 78(9):970--979, 1971.
\newblock \href {https://doi.org/10.1080/00029890.1971.11992915}
  {\path{doi:10.1080/00029890.1971.11992915}}.

\bibitem{MahesSS2018}
Anil Maheshwari, Jörg-Rüdiger Sack, and Christian Scheffer.
\newblock Approximating the integral {Fréchet} distance.
\newblock {\em Computational Geometry}, 70--71:13--30, 2018.
\newblock \href {https://doi.org/10.1016/J.COMGEO.2018.01.001}
  {\path{doi:10.1016/J.COMGEO.2018.01.001}}.

\bibitem{MunicP1999}
Mario~E. Munich and Pietro Perona.
\newblock {Continuous Dynamic Time Warping} for translation-invariant curve
  alignment with applications to signature verification.
\newblock In {\em Proceedings of the Seventh IEEE International Conference on
  Computer Vision}, pages 108--115. IEEE, 1999.
\newblock \href {https://doi.org/10.1109/ICCV.1999.791205}
  {\path{doi:10.1109/ICCV.1999.791205}}.

\bibitem{Rote2014}
Günter Rote.
\newblock Lexicographic {Fréchet} matchings.
\newblock Extended abstract from the 30th European Workshop on Computational
  Geometry, 2014.
\newblock URL: \url{https://refubium.fu-berlin.de/handle/fub188/15658}.

\bibitem{SchaeW1999}
Helmut~H. Schaefer and Manfred~P. Wolff.
\newblock {\em Topological Vector Spaces}.
\newblock Springer, 2nd edition, 1999.
\newblock \href {https://doi.org/10.1007/978-1-4612-1468-7}
  {\path{doi:10.1007/978-1-4612-1468-7}}.

\bibitem{SerraB1994}
Bruno Serra and Marc Berthod.
\newblock Subpixel contour matching using continuous dynamic programming.
\newblock In {\em 1994 Proceedings of IEEE Conference on Computer Vision and
  Pattern Recognition}, pages 202--207. IEEE, 1994.
\newblock \href {https://doi.org/10.1109/CVPR.1994.323830}
  {\path{doi:10.1109/CVPR.1994.323830}}.

\bibitem{SerraB1995}
Bruno Serra and Marc Berthod.
\newblock Optimal subpixel matching of contour chains and segments.
\newblock In {\em Proceedings of IEEE International Conference on Computer
  Vision}, pages 402--407. IEEE, 1995.
\newblock \href {https://doi.org/10.1109/ICCV.1995.466911}
  {\path{doi:10.1109/ICCV.1995.466911}}.

\bibitem{SuLZZZ2020}
Han Su, Shuncheng Liu, Bolong Zheng, Xiaofang Zhou, and Kai Zheng.
\newblock A survey of trajectory distance measures and performance evaluation.
\newblock {\em The VLDB Journal}, 29(1):3--32, 2020.
\newblock \href {https://doi.org/10.1007/S00778-019-00574-9}
  {\path{doi:10.1007/S00778-019-00574-9}}.

\bibitem{TaoBSBSPLPTD2021}
Yaguang Tao, Alan Both, Rodrigo~I. Silveira, Kevin Buchin, Stef Sijben, Ross~S.
  Purves, Patrick Laube, Dongliang Peng, Kevin Toohey, and Matt Duckham.
\newblock A comparative analysis of trajectory similarity measures.
\newblock {\em GIScience \& Remote Sensing}, 58(5):643--669, 2021.
\newblock \href {https://doi.org/10.1080/15481603.2021.1908927}
  {\path{doi:10.1080/15481603.2021.1908927}}.

\bibitem{Vints1968}
Taras~K. Vintsyuk.
\newblock Speech discrimination by dynamic programming.
\newblock {\em Cybernetics and Systems Analysis}, 4(1):52--57, 1968.
\newblock \href {https://doi.org/10.1007/BF01074755}
  {\path{doi:10.1007/BF01074755}}.

\bibitem{WildbR2025}
Norman~J. Wildberger and Dean Rubine.
\newblock A hyper-{Catalan} series solution to polynomial equations, and the
  {Geode}.
\newblock {\em The American Mathematical Monthly}, 132(5):383--402, 2025.
\newblock \href {https://doi.org/10.1080/00029890.2025.2460966}
  {\path{doi:10.1080/00029890.2025.2460966}}.

\end{thebibliography}
    
    \clearpage
    \appendix
    
    \section{Additional Proofs for \autoref{sec:geometry}}
\label{app:geometry-proofs}

For the sake of completeness, we give proofs of \autoref{thm:optimal-paths} and \autoref{thm:affine}.

\settheoremcounter{thm:optimal-paths}
\begin{theorem}[{\cite[Section~5.5]{Brank2022}}]
    Let $\overline{P},\overline{Q}$ be polygonal segments with a valley $\ell$ of positive slope under a norm $\tnorm$, and let $x,y \in \mathbb{R}^2$ with $\Xi := [x_1,y_1] \times [x_2,y_2] \neq \varnothing$ be two points in parameter space.
    If $\ell \cap \Xi \neq \varnothing$, then the $(x,y)$-path tracing line segments from $x$ to $\widehat{x}$ to $\widehat{y}$ to $y$ is optimal for $\overline{P},\overline{Q}$, where $\widehat{x},\widehat{y} \in \ell \cap \Xi$ share a coordinate with $x,y$ respectively.
    Else, the $(x,y)$-path tracing line segments from $x$ to $\xi$ to $y$ is optimal, where $\xi \in \{ (x_1,y_2)^{\mathsf{T}}, (y_1,x_2)^{\mathsf{T}} \}$ is closest to $\ell$ in $\Xi$.
\end{theorem}

\begin{proof}
	Consider the lines $L_s := \{ z \in \mathbb{R}^2 \mid z_1 + z_2 = s \}$ for $s \in [\|x\|_1,\|y\|_1]$, each of which has slope $-1$ and contains a segment $L_s \cap \Xi = \{ \gamma(s) \mid \gamma \text{ is an } (x,y)\text{-path} \}$.
    Let $\chi(s) \in L_s \cap \ell$ be the intersection point of $L_s$ with the valley $\ell$, and let $t^*_s \in I_s$ have minimum absolute value $|t^*_s|$ on $I_s := \{ t \in \mathbb{R} \mid (\chi_1(s) + t, \chi_2(s) - t)^{\mathsf{T}} \in \Xi \}$.
    Then $\gamma^* \colon [\|x\|_1,\|y\|_1] \to \Xi$ given by $\gamma^*(s) := (\chi_1(s) + t^*_s, \chi_2(s) - t^*_s)^{\mathsf{T}}$ realises the $(x,y)$-path from the statement and is optimal, as we show in the following.
    
    First, it indeed traces the stated line segments in both cases:
    By construction, we have $\gamma^*(s) \in L_s \cap \Xi$ for all $s \in [\|x\|_1,\|y\|_1]$.
    If $t^*_s = 0$, then $\gamma^*(s)$ also lies on $\ell$.
    Else, $\gamma^*(s)$ lies on the boundary of $\Xi$ due to $t^*_s$ minimising \mbox{$| \cdot |$} on $I_s$, where each boundary point of $\Xi$ shares a coordinate with $x$ or $y$ (cf.~\autoref{subfig:valley-definition}).
    Particularly, it is $\gamma^*(\|x\|_1) = x$ because of $L_{\|x\|_1} \cap \Xi = \{ x \}$.
    Similarly, it is $\gamma^*(\|y\|_1) = y$.
        
    The remaining properties of an $(x,y)$-path hold as well:
    Monotonicity of $\gamma^*$ follows from the positive slope of $\ell$.
    We hence get
    $
        \| 1/\delta \cdot  (\gamma^*(s + \delta) - \gamma^*(s)) \|_1
        = 1/\delta \cdot (\| (\gamma^*(s + \delta) \|_1 - \| \gamma^*(s)) \|_1)
        = 1/\delta \cdot (s + \delta - s)
        = 1
    $
    for $s, s + \delta \in [\|x\|_1,\|y\|_1]$ with $\delta \neq 0$.
    This implies that $\gamma^*$ satisfies the path regularisation constraint.
        
    Regarding optimality:
    Let $\gamma \colon [\|x\|_1,\|y\|_1] \to \Xi$ be an arbitrary $(x,y)$-path and let $s \in [\|x\|_1,\|y\|_1]$.
    It is $\gamma(s) \in L_s \in \Xi$ due to the regularisation constraint and monotonicity, so there is a $t \in I_s$ with $\gamma(s) = (\chi_1(s) + t, \chi_2(s) - t)^{\mathsf{T}}$.
    By $|t^*_s| \leq |t|$ and \autoref{def:valley}, it is $\| \overline{P}_{\| \cdot \|}(\gamma^*_1(s)) - \overline{Q}_{\| \cdot \|}(\gamma^*_2(s)) \| \leq \| \overline{P}_{\| \cdot \|}(\gamma_1(s)) - \overline{Q}_{\| \cdot \|}(\gamma_2(s)) \|$.
    \autoref{def:path-and-cost} gives $\mathrm{cost}(\gamma^*) \leq \mathrm{cost}(\gamma)$ and optimality of $\gamma^*$, as $\gamma$ was arbitrary.
    \qed
\end{proof}

\settheoremcounter{thm:affine}
\begin{lemma}[{\cite[Lemma~3]{AltG1995}}]
    Let $\overline{P},\overline{Q}$ be polygonal segments, let $\tnorm$ be a norm, and let $\varphi \colon \mathbb{R}^2 \to \mathbb{R}^2$ be the affine map defined through $\varphi(z) := \overline{P}_{\| \cdot \|}(z_1) - \overline{Q}_{\| \cdot \|}(z_2)$.
    If \smash{$\overline{P},\overline{Q}$} are parallel, then $\varphi$ is constant either on every line of slope $1$ or on every line of slope $-1$ in the extended parameter space $\mathbb{R}^2$.
    Else, $\varphi$ has an affine inverse map $\varphi^{-1} \colon \mathbb{R}^2 \to \mathbb{R}^2$ satisfying $S_{\leq \mu}(\tnorm \circ \varphi) = \varphi^{-1}(S_{\leq \mu}(\tnorm))$ for all $\mu \geq 0$.
\end{lemma}

\begin{proof}
    If $\overline{P} = \langle p_{i-1}, p_i \rangle$ and $\overline{Q} = \langle q_{j-1}, q_j \rangle$ are parallel, the normalised direction vectors $w_P := \varphi(1, 0) - \varphi(\mathbf{0}) = \frac{p_i - p_{i-1}}{\| p_i - p_{i-1} \|}$ and $w_Q := \varphi(0, -1) - \varphi(\mathbf{0}) = \frac{q_j - q_{j-1}}{\| q_j - q_{j-1} \|}$ differ at most by sign.
    In particular, codirectional $\smash{\overline{P}},\smash{\overline{Q}}$ yield $w_P = w_Q$ and thus $\varphi(z) = (z_1 - z_2) \cdot w_P + \varphi(\mathbf{0})$ for all $z \in \mathbb{R}^2$, meaning that $\varphi$ is constant on every line $\{ z \in \mathbb{R}^2 \mid z_1 - z_2 = t \}$ of slope~$1$ with $t \in \mathbb{R}$.
    The case of opposite directions yields $w_P = -w_Q$, so that the result then applies to every line of slope $-1$.

    If $\overline{P},\overline{Q}$ are not parallel, then $w_P = \varphi(1, 0) - \varphi(\mathbf{0})$ and $w_Q = \varphi(0, -1) - \varphi(\mathbf{0})$ form a basis of $\mathbb{R}^2$.
    Consequently, the affine map $\varphi$ is bijective and has an affine inverse map $\varphi^{-1} \colon \mathbb{R}^2 \to \mathbb{R}^2$.
    Given some arbitrary $\mu \geq 0$ and $z \in \mathbb{R}^2$, we therefore obtain
    $
        z \in S_{\leq \mu}(\tnorm \circ \varphi)
        \iff \varphi(z) \in S_{\leq \mu}(\tnorm)
        \iff z \in \varphi^{-1}(S_{\leq \mu}(\tnorm))
    $.
    \qed
\end{proof}

We next apply \autoref{thm:valleys} to polygonal norms, as used in \autoref{sec:exact-algorithm}.

\settheoremcounter{thm:valleys-for-polygonal-norms}
\begin{corollary}
    Let $\tnorm$ be a norm with $S_{\leq 1}(\tnorm) = \psi(R_k)$ for a regular $k$-gon~$R_k$ and a linear map $\psi \colon \mathbb{R}^2 \to \mathbb{R}^2$.
    Computing a valley $\ell$ of positive slope under $\tnorm$ for arbitrary $\overline{P},\overline{Q}$ is possible in $O(1)$ time, independent of $k \in \{4,6,\dotsc\}$.
\end{corollary}

\begin{proof}
    As $\tnorm$ is a norm, we have that $R_k$ is convex and $\psi$ is bijective.
    Two~convex regular $k$-gons can differ only by rotation and scaling.
    A transformation between them is a bijective linear map that can be computed in $O(1)$ time by considering a vertex pair of each polygon.
    Thus, we may assume w.l.o.g.\ that $R_k$ is of the form from \autoref{sec:exact-algorithm}, i.e.\ $R_k$ has vertices $v_1,\dotsc,v_k$ defined by $v_r := (\cos(\theta_r),\sin(\theta_r))^{\mathsf{T}}$ for $\theta_r := r \cdot 2\pi/k$ and $r \in \{1,\dotsc,k\}$.
    The vertices $\psi(v_1),\dotsc,\psi(v_k)$ of $\psi(R_k)$ are contained~in the ellipse parametrised by $\theta \mapsto \psi(\cos(\theta), \sin(\theta))$ on $(0,2\pi]$.
    
    Given polygonal segments $\overline{P} = \langle p_{i-1},p_i \rangle$ and $\overline{Q} = \langle q_{j-1}, q_j \rangle$, we precompute the values $\| p_i - p_{i-1} \|$ and $\| q_j - q_{j-1} \|$, which is possible in $O(1)$ time by \autoref{thm:polygonal-norm-properties}a.
    Hence, we can evaluate the affine map $\varphi \colon \mathbb{R}^2 \to \mathbb{R}^2$ from \autoref{thm:affine} and, if $\overline{P},\overline{Q}$ are not parallel, its inverse $\varphi^{-1}$ in $O(1)$ time whenever needed.
    In the following, we implement \autoref{thm:valleys} by performing its stated reduction to \autoref{thm:duality}.
    
    If $\overline{P},\overline{Q}$ are parallel with opposite directions, then the proofs of \autoref{thm:affine} and \autoref{thm:valleys} say that there is nothing to do.
    In case of codirectional $\overline{P},\overline{Q}$ we need to search for a $v^* \in \arg \max \{ u^\mathsf{T} \cdot v \mid v \in \psi(R_k) \}$ using any orthogonal~$u \in \mathbb{R}^2 \setminus \{\mathbf{0}\}$ with $u^{\mathsf{T}} \cdot (p_i - p_{i-1}) = 0$.
    As $\psi(R_k)$ is a convex polygon, the maximum is attained at a vertex $\psi(v_{r^*}) = \psi(\cos(\theta_{r^*}),\sin(\theta_{r^*}))$ for one or for two $r^* \in \{1,\dotsc,k\}$.
    
    To find such an $r^*$, we consider the function $\theta \mapsto u^{\mathsf{T}} \cdot \psi(\cos(\theta), \sin(\theta))$, which is periodic and differentiable on $\mathbb{R}$.
    We can compute a maximising angle $\theta^* \in (0,2\pi]$ by solving the equation $u^\mathsf{T} \cdot \psi(-\sin(\theta), \cos(\theta)) = 0$ for $\theta$.
    One of the two vertices next to the ellipse point $\psi(\cos(\theta^*),\sin(\theta^*))$ is a suitable $v^* := \psi(v_{r^*})$ since the maximised function is increasing on $[\theta^* - \pi, \theta^*]$ and decreasing on $[\theta^*, \theta^* + \pi]$, see \autoref{subfig:valley-computation}.
    Thus, $r^+ := \lceil \theta^* / \theta_1  \rceil$ gives $r^* \in \{ r^+, r^+ - 1 \}$ when using $v_0 := v_k$.
    All computations are independent of $k$, so performing them takes $O(1)$ time.\footnote{\label{fn:rounding}Assuming that unit-cost integer rounding is allowed as long as the resulting integers have a logarithmic number of bits.}
    
    If $\overline{P},\overline{Q}$ are not parallel, the vertices of $K' := \varphi^{-1}(\psi(R_k)) - \varphi^{-1}(\mathbf{0})$ are instead now contained in the ellipse parametrised by $\theta \mapsto \varphi^{-1}(\psi(\cos(\theta), \sin(\theta))) - \varphi^{-1}(\mathbf{0})$.
    We proceed analogously to before, though there is an edge case:
    It can happen that our vertex $\varphi^{-1}(\psi(v_{r^*})) - \varphi^{-1}(\mathbf{0})$ maximising $v \mapsto (1,1) \cdot v$ on $K'$ does not have two positive components.
    However, the proof of \autoref{thm:valleys} then implies that there is an incident edge of $K'$ through $(1,0)^{\mathsf{T}}$ and $(0,1)^{\mathsf{T}}$, so the relevant maximum is attained at every point on that edge and we can choose e.g.\ $v^* := (1/2, 1/2)^{\mathsf{T}}$.
    \qed
\end{proof}

\section{Proof Details for \texorpdfstring{\autoref{thm:transcendence}b}{\autoref*{thm:transcendence}b}}
\label{app:transcendence-details}

In the following, we provide details to complete the proof sketch of \autoref{thm:transcendence}b.
When manipulating expressions, we still do not go through every individual step.
Instead, we convey the main ideas and describe how to arrive at any expressions we give, which makes it possible to verify everything by hand or software.

We start with the first claim, which does not even require calculations.

\begin{claim}[B1]
    Let $\mathcal{C} \colon [0,10] \to \mathbb{R}_{\geq 0}$ be defined by $\mathcal{C}(s)$ being the cost of the optimal path candidate $\gamma_s$ for $s \in [0,10]$.
    We have that $\mathcal{C}$ is continuous.
\end{claim}

\begin{proof}
    The continuity follows directly from \autoref{thm:converging-costs} because for every $s \in [0,10]$ we have that the optimal path candidates $\gamma_{\tilde{s}}$ converge to $\gamma_s$ for $\tilde{s} \to s$.
    \qed
\end{proof}

In order to prove the remaining claims, we nonetheless determine $\mathcal{C}(s)$ through the costs of the three line segments that $\gamma_s$ traces for $s \in [0,10]$.
The first and the third segment, which each travel on a valley, give simple integrals:
\begin{gather*}
    \int_0^s \| P_{\| \cdot \|_2}(t/2) - Q_{\| \cdot \|_2}(t/2) \|_2 \dt
    = \frac{2}{\sqrt{5}} \cdot \int_0^s |t - 5| \dt
    \text{,}\tag{I1}\label{eq:int1}\\[3pt]
    \int_{s+6}^{16} \| P_{\| \cdot \|_2}(t/2 + 3) - Q_{\| \cdot \|_2}(t/2 - 3) \|_2 \dt
    = \frac{1}{\sqrt{2}} \cdot \int_{s+6}^{16} |t - 8| \dt
    \text{.}\tag{I2}\label{eq:int2}
\end{gather*}

The second segment is horizontal and travels from one valley to the next, as illustrated in \autoref{fig:transcendence}.
Its cost is more complicated and needs to be split into a sum of two integrals since $\gamma_s$ crosses the inter-cell boundary at $(5,s/2)^{\mathsf{T}}$:
\begin{align}
    &\int_s^{s + 6} \| P_{\| \cdot \|_2}(t - s/2) - Q_{\| \cdot \|_2}(s/2) \|_2 \dt
    \nonumber\\[3pt]
    = &\int_s^{s/2 + 5} \sqrt{t^2 - (2/5 \cdot s + 8) \cdot t + (s^2 / 5 + 20)} \dt
    \tag{I3}\label{eq:int3}\\[3pt]
    + &\int_{s/2 + 5}^{s+6} \sqrt{t^2 - (s + 14) \cdot t + (s^2/2 + 6s + 50)} \dt
    \text{.}\tag{I4}\label{eq:int4}
\end{align}

Thus, $\mathcal{C}(s)$ is equal the sum of the four integrals \eqref{eq:int1} to \eqref{eq:int4}.
The derivative~$\mathcal{C}'$ can be obtained using Leibniz' rule for parametric integrals (see \citeappendix[Section~8.1]{ProttM1985}):
Given an integral $I(s) := \int_{a(s)}^{b(s)} h(s,t) \dt$ defined on an interval, where $h \colon \mathbb{R}^2 \to \mathbb{R}$ as well as $\partial_s h$ are continuous and $a, b \colon \mathbb{R} \to \mathbb{R}$ are continuously differentiable, the derivative of $I$ is given by partial differentiation under the integral sign:
\[
    I'(s) = h(s,b(s)) \cdot b'(s) - h(s,a(s)) \cdot a'(s) + \int_{a(s)}^{b(s)} \partial_s h(s,t) \dt
    \text{.}
\]

When applying this rule to the integrals \eqref{eq:int1} to \eqref{eq:int4} and summing the results, one notices that most terms cancel.
We are only left with the partial differentiation terms for the two integrals \eqref{eq:int3} and \eqref{eq:int4}, so for all $s \in [0,10] \setminus \{2,5\}$ we have
\begin{align*}
    \mathcal{C}'(s) =
    &- \frac{1}{5} \cdot \int_{s}^{s/2 + 5} \frac{t - s}{\;\sqrt{t^2 - (2/5 \cdot s + 8) \cdot t + (s^2 / 5 + 20)}\;} \dt
    \tag{I5}\label{eq:int5}\\[3pt]
    &- \frac{1}{2} \cdot \int_{s/2 + 5}^{s+6} \frac{t - s - 6}{\;\sqrt{t^2 - (s + 14) \cdot t + (s^2/2 + 6s + 50)}\;} \dt
    \text{.}\tag{I6}\label{eq:int6}
\end{align*}

The exclusion of $s = 2$ and $s = 5$ is because for \eqref{eq:int3} we have that $\partial_s h$ is not defined at $(s,t) = (5,5)$ and for \eqref{eq:int4} it is not defined at $(s,t) = (2,8)$.
The reason for this is the fact that $\gamma_5(5) = (5,5)^{\mathsf{T}}$ and $\gamma_2(8) = (7,1)^{\mathsf{T}}$ are respective bending points of $\gamma_5$ and $\gamma_2$ corresponding to intersections of $P,Q$ (cf.~\autoref{fig:transcendence}), which cause an integrand denominator in \eqref{eq:int5} or in \eqref{eq:int6} to be $0$.
Still, the second claim asserts that $\mathcal{C}'$ exists there as well.
To show this, we simplify the integrals.

\begin{claim}[B2]
    The function $\mathcal{C}$ is differentiable on $[0,10]$, and its derivative at~$s$ is
    $
        \mathcal{C}'(s) =
        - \frac{1}{5} \cdot \int_0^{5/4 \cdot (2 - \kappa_s)} \frac{t}{\;\sqrt{(t + 2\kappa_s)^2 + \kappa_s^2}\;} \dt
        - \frac{1}{2} \cdot \int_{-(\nu_s + 2)}^{0} \frac{t}{\;\sqrt{(t + \nu_s)^2 + \nu_s^2}\;} \dt
    $
    for $\kappa_s := 2/5 \cdot (s - 5)$ and $\nu_s := 1/2 \cdot (s - 2)$.
    It is $\mathcal{C}'(0) < 0$ and $\mathcal{C}'(10) > 0$.
\end{claim}

\begin{proof}
    First, we use integration by substitution on the integrals \eqref{eq:int5} and \eqref{eq:int6}, so that we get $t$ by itself in each numerator.
    For $s \in [0,10] \setminus \{2,5\}$ this gives
    \begin{align*}
        \mathcal{C}'(s) =
        &- \frac{1}{5} \cdot \int_{0}^{5 - s/2} \frac{t}{\;\sqrt{t^2 + 8 \cdot (s/5 - 1) \cdot t + (4/5 \cdot s^2 - 8s + 20)}\;} \dt
        \tag{I7}\label{eq:int7}\\[3pt]
        &- \frac{1}{2} \cdot \int_{-(s/2+1)}^{0} \frac{t}{\;\sqrt{t^2 + (s-2) \cdot t + (s^2/2 - 2s + 2)}\;} \dt
        \text{.}\tag{I8}\label{eq:int8}
    \end{align*}
    
    Here, integral \eqref{eq:int7} has denominator $0$ for $(s,t) = (5,0)$ and integral \eqref{eq:int8} has denominator $0$ for $(s,t) = (2,0)$.
    Using $\kappa_s = 2/5 \cdot (s - 5)$ and $\nu_s = 1/2 \cdot (s - 2)$, we obtain the claimed form, where equality can be verified via expansion:
    \begin{align*}
        \mathcal{C}'(s) =
        &- \frac{1}{5} \cdot \int_0^{5/4 \cdot (2 - \kappa_s)} \frac{t}{\;\sqrt{(t + 2\kappa_s)^2 + \kappa_s^2}\;} \dt
        \tag{I9}\label{eq:int9}\\[3pt]
        &- \frac{1}{2} \cdot \int_{-(\nu_s + 2)}^{0} \frac{t}{\;\sqrt{(t + \nu_s)^2 + \nu_s^2}\;} \dt
        \text{.}\tag{I10}\label{eq:int10}
    \end{align*}
    
    It is now easy to see that in both relevant cases, i.e.\ $\kappa_5 = 0$ and $\nu_2 = 0$, the denominators of \eqref{eq:int9} and \eqref{eq:int10} are $\sqrt{t^2} = |t|$, which actually cancels with $t$ and leaves $\mathrm{sgn}(t)$.
    Consequently, \eqref{eq:int9} evaluates to $-1/5 \cdot \int_0^{5\smash{/}2} 1 \dt = -1/5 \cdot 5/2 = -1/2$ for $(s,t) = (5,0)$, while \eqref{eq:int10} similarly evaluates to $1$ for $(s,t) = (2,0)$.
    This does not yet imply that $\mathcal{C}'(2)$ and $\mathcal{C}'(5)$ exist because these cases were not covered by Leibniz' rule, but we argue next that it only remains to show that the integrands of \eqref{eq:int9} and \eqref{eq:int10} are uniformly bounded by some maximum absolute value.
    
    Given such uniform bounds, the Dominated Convergence Theorem (see \cite{Luxem1971}) implies that $\lim_{s \to 2} \mathcal{C}'(s)$ and $\lim_{s \to 5} \mathcal{C}'(s)$ exist, as the integrands converge to the integrable $t \mapsto \mathrm{sgn}(t)$ for $\nu_s \to 0$ and $\kappa_s \to 0$ respectively.
    The existence of these limits then establishes that $\mathcal{C}'(2)$ and $\mathcal{C}'(5)$ correspond to the limits, which is due to a well-known consequence of L'Hospital's rule:
    We know that $\mathcal{C}$ is continuous by the first claim, so $\lim_{\delta \to 0} [\mathcal{C}(s + \delta) - \mathcal{C}(s)] = 0$ holds and the differential limit $\lim_{\delta \to 0} [1/\delta \cdot (\mathcal{C}(s + \delta) - \mathcal{C}(s))]$ of $\mathcal{C}$ at $s \in \{2,5\}$ is equal to~$\lim_{\delta \to 0} \mathcal{C}'(s + \delta)$.
    
    We next show the uniform bounds to complete the argument.
    Both considered integrands are of the form $t \mapsto t / \sqrt{(t + c \lambda)^2 + \lambda^2}$ with $c \in \mathbb{R}$, so that we obtain the generalised bound $\sqrt{c^2 + 1}$ as follows:
    For all $\lambda \in \mathbb{R}$ and $t \in \mathbb{R}$ we have
    \begin{align*}
        &(t + c \lambda)^2 + \lambda^2 - t^2 / (c^2 + 1) \\[1pt]
        = {} &c^2/(c^2 + 1) \cdot t^2 + 2c \lambda t + (c^2 + 1) \cdot \lambda^2 \\[1pt]
        = {} &\big( \sqrt{c^2/(c^2 + 1)} \cdot t + \mathrm{sgn}(c) \cdot \sqrt{c^2 + 1} \cdot \lambda \big)^2 \\[1pt]
        \geq {} &0
        \quad \text{and thus }
        (t + c \lambda)^2 + \lambda^2 \geq t^2/(c^2 + 1)
        \text{.}
    \end{align*}
    Rearranging and extracting the root gives $|t|/\sqrt{(t + c \lambda)^2 + \lambda^2} \leq \sqrt{c^2 + 1}$.
    
    Finally, we bound the values $\mathcal{C}'(0)$ and $\mathcal{C}'(10)$ as claimed.
    For the latter we have $5/4 \cdot (2 - \kappa_{10}) = 0$, so \eqref{eq:int9} attains $0$ at $s = 10$, while \eqref{eq:int10} starts integrating at $t = -(\nu_{10} + 2) = -6$ and its integrand is negative on $(-6,0)$, so the negative coefficient $-1/2$ implies that \eqref{eq:int10} is positive at $s = 10$.
    Thus, $\mathcal{C}'(10) > 0$.
    
    In the former case $s = 0$ we have $(-\nu_0 + 2) = -1$.
    As the integrand of \eqref{eq:int10} is again negative on $(-1,0)$ and is also subject to the uniform bound $\sqrt{c^2 + 1} = \sqrt{2}$ for $c = 1$, the value of \eqref{eq:int10} at $s = 0$ is at most $-1/2 \cdot \int_{-1}^0 - \sqrt{2} \dt = 1/\sqrt{2}$.
    
    Meanwhile, it is $5/4 \cdot (2-\kappa_0) = 5$ and the value of \eqref{eq:int9} at $s = 0$ is thus given by $-1/5 \cdot \int_0^5 t/\sqrt{(t-4)^2 + 4} \dt$. It suffices to show $\int_0^5 t/\sqrt{(t-4)^2 + 4} \dt > 5/\sqrt{2}$ to overall obtain $\mathcal{C}'(0) < 0$.
    Using the quotient rule and simplifying, the derivative of $t \mapsto t/\sqrt{(t-4)^2 + 4}$ turns out to be $t \mapsto 4 \cdot (5-t) / ((t-4)^2+4)^{3/2}$.
    
    The derivative is positive on $(0,5)$, so the integrand increases and is lower bounded by the left endpoint's value on any subinterval of $(0,5)$.
    We get
    \begin{align*}
        \int_0^5 \frac{t}{\;\sqrt{(t-4)^2 + 4}\;} \dt
        > {} &\frac{1}{2} \cdot \frac{2}{\;\sqrt{(2-4)^2 + 4}\;} + \frac{3}{2} \cdot \frac{5/2}{\;\sqrt{(5/2-4)^2 + 4}\;} \\[3pt]
        {} + {} &1 \cdot \frac{4}{\;\sqrt{(4-4)^2 + 4}\;}
        = \frac{1}{\;2 \cdot \sqrt{2}\;} + \frac{7}{2}
        > \frac{5}{\;\sqrt{2}\;}
        \text{,}
    \end{align*}
    by using the three intervals $(2,5/2)$, $(5/2,4)$ and $(4,5)$.
    The final inequality holds due to $(1/(2 \cdot \sqrt{2}) + 7/2)^2 = 99/8 + 7/(2 \cdot \sqrt{2}) > 100/8 = 25/2 = (5/\sqrt{2})^2$.
    \qed
\end{proof}

So far, we have avoided to evaluate the integrals via antiderivatives, but the third claim relies on this to obtain $\mathcal{C}'$ as a linear combination of logarithms, which allows to apply transcendental number theory by means of \autoref{thm:baker}.

\begin{claim}[B3]
    Some algebraic functions $\alpha_1, \alpha_2, \beta_0, \beta_1, \beta_2 \colon [0,10] \setminus \{2,5\} \to \mathbb{R}$ satisfying $\mathcal{C}'(s) = \beta_0(s) + \beta_1(s) \cdot \ln(\alpha_1(s)) + \beta_2(s) \cdot \ln(\alpha_2(s))$ exist.
    We also have $\mathcal{C}'(2) \neq 0 \neq \mathcal{C}'(5)$, and $\beta_0(s^*_0) = 0$ is fulfilled on $(0,10)$ only by~$s^*_0 := 1/929 \cdot \allowbreak \big( 1880 \cdot \sqrt{10} - 3070 \big) + 60 \cdot \big( \sqrt{16 \cdot \sqrt{10} + 61} \big) / \big( 80 \cdot \sqrt{10} + 269 \big) \in (4,4.5)$.
\end{claim}

\begin{proof}
    Let $h \colon \mathbb{R} \to \mathbb{R}$ be a function with $h(s) := \sqrt{s^2 + \lambda_1 s + \lambda_0}$ for $\lambda_0,\lambda_1 \in \mathbb{R}$.
    By \cite[Equation~4.3.4.1.2]{Jeffr2008}, the antiderivative of $t \mapsto t/h(t)$ is given by
    \[
        \int \frac{t}{h(t)} \dt = - \frac{\lambda_1}{2} \cdot \ln(|2 \cdot h(s) + 2s + \lambda_1|) + h(s) + C
        \text{.}
    \]
    
    Using this antiderivative shows that \eqref{eq:int7} is equal to
    \begin{align}
        &\tfrac{4}{5} \cdot \left( \tfrac{s}{5} - 1 \right) \cdot \ln \Big( \Big| 2 \cdot \sqrt{\tfrac{s^2}{4} - s + 5} + \tfrac{3}{5} \cdot s + 2 \Big| \Big) - \tfrac{1}{5} \cdot \sqrt{\tfrac{s^2}{4} - s + 5}
        \tag{L1}\label{eq:log1} \\[1pt]
        {} - {} &\tfrac{4}{5} \cdot \left( \tfrac{s}{5} - 1 \right) \cdot \ln \Big( \Big| 4 \cdot \sqrt{\tfrac{s^2}{5} - 2s + 5} + 8 \cdot \left( \tfrac{s}{5} - 1 \right) \Big| \Big) + \tfrac{2}{5} \cdot \sqrt{\tfrac{s^2}{5} - 2s + 5} 
        \tag{L2}\label{eq:log2}
    \end{align}
    for all $s \in [0,10] \setminus \{2,5\}$, while \eqref{eq:int8} is equal to
    \begin{align}
        &\tfrac{s - 2}{4} \cdot \ln \Big( \Big| 2 \cdot \sqrt{\tfrac{s^2}{2} - 2s + 2} + s - 2 \Big| \Big) - \tfrac{1}{2} \cdot \sqrt{\tfrac{s^2}{2} - 2s + 2}
        \tag{L3}\label{eq:log3} \\[1pt]
        {} - {} &\tfrac{s - 2}{4} \cdot \ln \Big( \Big| 2 \cdot \sqrt{\tfrac{s^2}{4} - s + 5} - 4 \Big| \Big) + \tfrac{1}{2} \cdot \sqrt{\tfrac{s^2}{4} - s + 5}
        \text{,}\tag{L4}\label{eq:log4}
    \end{align}
    so summing and applying logarithm rules yields
    \begin{align*}
        \mathcal{C}'(s)
        = {} &\frac{4}{5} \cdot \left( \frac{s}{5} - 1 \right) \cdot \ln \left( \left| \frac{2 \cdot \sqrt{\tfrac{s^2}{4} - s + 5} + \tfrac{3}{5} \cdot s + 2}{4 \cdot \sqrt{\tfrac{s^2}{5} - 2s + 5} + 8 \cdot \left( \tfrac{s}{5} - 1 \right)} \right| \right) \\[3pt]
        + {} &\frac{s - 2}{4} \cdot \ln \left( \left| \frac{2 \cdot \sqrt{\tfrac{s^2}{2} - 2s + 2} + s - 2}{2 \cdot \sqrt{\tfrac{s^2}{4} - s + 5} - 4} \right| \right)
        + \beta_0(s)
        \text{.}
    \end{align*}
    
    This is the claimed form, where $\beta_0 \colon [0,10] \to \mathbb{R}$ is defined to be the sum of the algebraic terms of \eqref{eq:log1} to \eqref{eq:log4}.
    Note that for~$s \in \{2,5\}$ the second claim's~proof says that one of the two logarithmic terms in this form can respectively be replaced by an algebraic limit.
    The other logarithmic term remains and is non-zero, so \autoref{thm:baker} implies $\mathcal{C}'(2) \neq 0 \neq \mathcal{C}'(5)$.
    We continue with $\beta_0$:
    \begin{align*}
        \beta_0(s)
        := &- \tfrac{1}{5} \cdot \sqrt{\tfrac{s^2}{4} - s + 5} + \tfrac{2}{5} \cdot \sqrt{\tfrac{s^2}{5} - 2s + 5} \\[1pt]
        &- \tfrac{1}{2} \cdot \sqrt{\tfrac{s^2}{2} - 2s + 2} + \tfrac{1}{2} \cdot \sqrt{\tfrac{s^2}{4} - s + 5} \\[1pt]
        = & \; \tfrac{3}{10} \cdot \sqrt{\tfrac{s^2}{4} - s + 5} + \tfrac{2}{\;5 \cdot \sqrt{5}\;} \cdot |s-5| - \tfrac{1}{\;2 \cdot \sqrt{2}\;} \cdot |s-2|
        \text{.}
    \end{align*}

    Setting $\beta_0(s)$ to $0$, isolating the term with the square root term on one side of the equation, and subsequently squaring both sides gives
    \[
        \tfrac{9}{100} \cdot \left( \tfrac{s^2}{4} - s + 5 \right) = \tfrac{4}{125} \cdot (s-5)^2 + \tfrac{1}{8} \cdot (s-2)^2 - \tfrac{2}{\;5 \cdot \sqrt{10}\;} \cdot |s-5| \cdot |s-2|
        \text{,}
    \]
    where every solution to this also solves a quadratic equation
    \[
        \tfrac{9}{100} \cdot \left( \tfrac{s^2}{4} - s + 5 \right) = \tfrac{4}{125} \cdot (s-5)^2 + \tfrac{1}{8} \cdot (s-2)^2 + \eta \cdot \tfrac{2}{\;5 \cdot \sqrt{10}\;} \cdot (s-5) \cdot (s-2)
    \]
    for $\eta \in \{-1,1\}$.
    A solution $s$ for $\eta = -1$ may only satisfy $\beta_0(s) = 0$ if $s \notin (2,5)$ and a solution for $\eta = 1$ may only do so if $s \in [2,5]$.
    Solving the two quadratic equations results in four candidates $s_{\eta,\zeta} \in \mathbb{R}$ with $\eta, \zeta \in \{-1,1\}$:
    \[
        s_{\eta,\zeta} :=
        \frac{\eta \cdot 1880 \cdot \sqrt{10} - 3070}{929} + \zeta \cdot 60 \cdot \frac{\sqrt{\eta \cdot 16 \cdot \sqrt{10} + 61}}{80 \cdot \sqrt{10} + \eta \cdot 269}
        \text{.}
    \]
    
    To check the validity of the candidates, one can use approximations of $\sqrt{10}$ to establish bounds.
    We now show the claim $s^*_0 = s_{1,1} \in (4,4.5)$ in this way.
    It is $25/8 < \sqrt{10} < 32/10$, as $(25/8)^2 = 625/64 < 10 < 1024/100 = (32/10)^2$, so
    \begin{align*}
        s^*_0 &> \frac{1880 \cdot 25/8 - 3070}{929} + 60 \cdot \frac{\sqrt{16 \cdot 25/8 + 61}}{80 \cdot 32/10 + 269} \\
        &= \frac{2805}{929} + 60 \cdot \frac{\sqrt{111}}{525} > 3 + \frac{600}{525} > 4 \quad \text{and} \\[10pt]
        s^*_0 &< \frac{1880 \cdot 32/10 - 3070}{929} + 60 \cdot \frac{\sqrt{16 \cdot 32/10 + 61}}{80 \cdot 25/8 + 269} \\
        &= \frac{2946}{929} + 60 \cdot \frac{\sqrt{561/5}}{519} < \frac{32}{10} + \frac{660}{519} < \frac{32}{10} + \frac{13}{10} = \frac{9}{2} = 4.5
        \text{.}
    \end{align*}
    
    In particular, $s^*_0 = s_{1,1} \in [2,5]$ is a valid candidate.
    Similarly, one can prove that the bounds $s_{1,-1} < 2$ and $s_{-1,-1} \in (2,5)$ hold.
    Thus, those two candidates are not valid, whereas the last candidate $s_{-1,1} < 0$ is valid but not in $(0,10)$.
    \qed
\end{proof}

Finally, it only remains to show the bound from the fourth claim.

\begin{claim}[B4]
    It is $\beta_1(s) \cdot \ln(\alpha_1(s)) + \beta_2(s) \cdot \ln(\alpha_2(s)) > 0$ for all $s \in (4,4.5)$.
\end{claim}

\begin{proof}
    We have that $\beta_1(s) \cdot \ln(\alpha_1(s)) + \beta_2(s) \cdot \ln(\alpha_2(s))$ for $s \in (4,4.5)$ equals the sum of the logarithmic terms of \eqref{eq:log1} to \eqref{eq:log4}.
    Our strategy is to show lower bounds for all four terms individually and combine them later.
    These individual bounds can be obtained by appropriately plugging in the values $4$ and $4.5 = 9/2$ for $s$, where we may need different values at different places of the term.
    
    We now consider the case of \eqref{eq:log1} in detail:
    We have that $s \mapsto 4/5 \cdot \left( s/5 - 1 \right)$ is increasing and less than $0$ for $s < 5$, while $s \mapsto \sqrt{s^2/4 - s + 5}$ is increasing for $s > 2$, so $s \mapsto 2 \cdot \sqrt{s^2/4 - s + 5} + 3/5 \cdot s + 2$ is increasing and also greater than $1$ on the interval $(2,5)$.
    Hence, $s \mapsto \ln(2 \cdot \sqrt{s^2/4 - s + 5} + 3/5 \cdot s + 2)$ is increasing and greater than $0$ on $(2,5)$, as $\ln$ is increasing and attains $0$ at $1$.
    
    When we have functions $f,g \colon [a,b] \to \mathbb{R}$ with $f(t) < 0$ increasing and $g(t) > 0$ increasing for all $t \in [a,b]$, then we get the bound $f(t) \cdot g(t) = - |f(t)| \cdot |g(t)| > - |f(a)| \cdot |g(b)| = f(a) \cdot g(b)$.
    For \eqref{eq:log1} with $[a,b] := [4,4.5] \subseteq (2,5)$:
    \begin{align*}
        &\tfrac{4}{5} \cdot \left( \tfrac{s}{5} - 1 \right) \cdot \ln \Big( 2 \cdot \sqrt{\tfrac{s^2}{4} - s + 5} + \tfrac{3}{5} \cdot s + 2 \Big) \\[3pt]
        >{} &\tfrac{4}{5} \cdot \left( \tfrac{4}{5} - 1 \right) \cdot \ln \Big( 2 \cdot \sqrt{\tfrac{(9/2)^2}{4} - \tfrac{9}{2} + 5} + \tfrac{3}{5} \cdot \tfrac{9}{2} + 2 \Big) \\[3pt]
        ={} &- \tfrac{4}{25} \cdot \ln \Big( 2 \cdot \sqrt{\tfrac{89}{16}} + \tfrac{47}{10} \Big) > - \tfrac{4}{24} \cdot \ln \Big( 2 \cdot \sqrt{\tfrac{100}{16}} + \tfrac{47}{10} \Big) \\[3pt]
        ={} &- \tfrac{1}{6} \cdot \ln \Big( \tfrac{97}{10} \Big)
        > - \tfrac{1}{6} \cdot \ln(10)
        \text{,}
    \end{align*}
    so the logarithmic term of \eqref{eq:log1} is greater than $- 1/6 \cdot \ln(10)$ on $(4,4.5)$.
    
    In a similar way, one can also show that the logarithmic term of \eqref{eq:log2} is greater than $1/6 \cdot \ln(1/15)$ on~$(4,4.5)$, that the one of \eqref{eq:log3} is greater than $1/2 \cdot \ln(24/5)$ on~$(4,4.5)$ as well as that the one of \eqref{eq:log4} is greater than $- 1/2 \cdot \ln(4/5)$ on~$(4,4.5)$.
    Putting these bounds together and applying logarithm rules yields
    \begin{align*}
        &\beta_1(s) \cdot \ln(\alpha_1(s)) + \beta_2(s) \cdot \ln(\alpha_2(s)) \\[3pt]
        > {} &\frac{1}{2} \cdot \left( \ln \left( \frac{24}{5} \right) - \ln \left( \frac{4}{5} \right) \right) + \frac{1}{6} \cdot \left( \ln \left( \frac{1}{15} \right) - \ln(10) \right) \\[3pt]
        = {} &\frac{1}{2} \cdot \ln \left( \frac{24/5}{4/5} \right) + \frac{1}{6} \cdot \ln \left( \frac{1/15}{10} \right) \\[3pt]
        = {} &\frac{1}{2} \cdot \ln(6) + \frac{1}{6} \cdot \ln \left( \frac{1}{150} \right)
        = \ln \left( \frac{\sqrt{6}}{\;\sqrt[6]{150}\;} \right) > 0
        \text{,}
    \end{align*}
    where the final inequality is implied by $(\sqrt{6}/\sqrt[6]{150})^6 = 216/150 > 1$.
    \qed
\end{proof}

\section{Additional Proofs for \autoref{sec:exact-algorithm}}
\label{app:algorithm-proofs}

\autoref{thm:polygonal-norm-properties} forms the basis of our propagation procedure.
First, in (a) we establish that any polygonal norm $\mathcal{G}_{\psi(R_k)}$ is locally linear, which is required to generalise the methods that we build upon \cite{Klare2020,Brank2022}.
In particular, the below proof of (a) states an evaluation formula for $\mathcal{G}_{\psi(R_k)}$.
Note that there exist other ways to evaluate gauges such as the one described in \citeappendix{ChewD1985}, which however does not readily imply the linearity property.
The proofs of (b) and (c) are constructive as well.

To aid understanding of the constructions and cases from the proof of (b), some example arrangements are visualised in \autoref{fig:arrangements}, which is inspired by~\cite[Figures~5.2~\&~5.3]{Klare2020}.
A comparison shows that even for $k = 4$ there can be many more types of optimal paths than in 1D, each of which needs to have (at least) one face in a desired arrangement.
Accordingly, propagations for 2D CDTW are much more complex in the general case than in the simple example of \autoref{subfig:cdtw-1-norm}, which shows the special case of the optimal path travelling only on valleys.

\settheoremcounter{thm:polygonal-norm-properties}
\begin{lemma}
    \begin{alphaenumerate*}
        \item 
        We can evaluate $\mathcal{G}_{\psi(R_k)}$ in $O(1)$ time.
        Its restriction to each~cone $\psi(\{ \lambda v_{r-1} + \lambda' v_{r} \mid \lambda, \lambda' \geq 0 \})$ is linear, where $r \in \{1,\dotsc,k\}$ and $v_0 := v_k$.
        
        \item 
        Let $\Sigma_{\mathcal{A},\mathcal{B}}$ be the propagation space of borders $\mathcal{A},\mathcal{B}$ under $\mathcal{G}_{\psi(R_k)}$.
        There is an arrangement $(V,E,F)$ of $O(k)$ lines with $\Sigma_{\mathcal{A},\mathcal{B}} = \bigcup F'$ for a subset $F' \subseteq F$~of closed faces such that each restriction of $\mathrm{opt}_{\mathcal{A},\mathcal{B}}$ to an $f \in F'$ is quadratic.
        
        \item 
        The function $\mathrm{opt}_{0,\mathcal{B}}$ is piecewise quadratic for every border $\mathcal{B}$ under $\mathcal{G}_{\psi(R_k)}$.
    \end{alphaenumerate*}
\end{lemma}

\begin{proof}
    \begin{alphaenumerate}
        \item 
        We first provide an evaluation formula for $\mathcal{G}_{\psi(R_k)}$ and afterwards show how to derive it.
        Given $z \in \mathbb{R}^2 \setminus \{\mathbf{0}\}$, we define $\rho(z) := \lceil \ngl(\psi^{-1}(z)) / \theta_1 \rceil$, where $\ngl \colon \mathbb{R}^2 \setminus \{\mathbf{0}\} \to (0,2\pi]$ returns (east-counterclockwise) angles.
        It~is $\rho(z) \in \{1,\dotsc,k\}$ since $\theta_1 = 2\pi/k$.
        Our claim is that $\mathcal{G}_{\psi(R_k)}(z)$ equals
        \[
            \frac{\sin(\theta_{\rho(z)}) - \sin(\theta_{\rho(z)-1})}{\sin(\theta_1)} \cdot \psi^{-1}_1(z) + \frac{\cos(\theta_{\rho(z)-1}) - \cos(\theta_{\rho(z)})}{\sin(\theta_1)} \cdot \psi^{-1}_2(z)
            \text{,}
        \]
        where $\theta_r = r \cdot 2\pi/k$ for $r \in \{0,\dotsc,k\}$.
        This is computable in $O(1)$ time\footnote{See \autoref{fn:rounding}.} and yields a linear function for fixed $\rho(z)$, as $\psi$ and its inverse $\psi^{-1}$ are linear.
        
        The definition of $\rho(z)$ is, if $z \neq \psi(v_r) = \psi(\cos(\theta_r), \sin(\theta_r))$ for $r \in \{0,\dotsc,k\}$, equivalently characterised via $\psi^{-1}(z)$ being contained in the interior of the cone $\Delta := \{ \lambda v_{\rho(z)-1} + \lambda' v_{\rho(z)} \mid \lambda, \lambda' \geq 0 \}$ given by successive vertices~of~$R_k$.
        Therefore, $\mathcal{G}_{\psi(R_k)}$ is linear on the transformed cone $\psi(\Delta)$ containing $z$.
        
        Our derivation starts with reducing the evaluation of $\mathcal{G}_{\psi(R_k)}$ to that of $\mathcal{G}_{R_k}$:
        The properties of $\psi$ and the gauge definition from \autoref{subsec:valleys} give
        \begin{align*}
            \mathcal{G}_{\psi(R_k)}(z)
            &= \inf \{ \lambda \geq 0 \mid z \in \lambda \psi(R_k) \}
            = \inf \{ \lambda \geq 0 \mid z \in \psi(\lambda R_k) \} \\
            &= \inf \{ \lambda \geq 0 \mid \psi^{-1}(z) \in \lambda R_k \}
            = \mathcal{G}_{R_k}(\psi^{-1}(z))
            \text{.}
        \end{align*}
        
        Now let $T \in \mathbb{R}^{2 \times 2}$ be the matrix with vertices $v_{\rho(z)-1}$ and $v_{\rho(z)}$ as columns, so that the linear map $v \mapsto Tv$ maps $\mathbb{R}_{\geq 0}^2$ onto the cone $\Delta$ by construction.
        Particularly, it maps $R_4 \cap \mathbb{R}_{\geq 0}^2$ onto $R_k \cap \Delta$, and together with $\psi^{-1}(z) \in \Delta$ we get $\mathcal{G}_{R_k}(\psi^{-1}(z)) = \mathcal{G}_{T R_4}(\psi^{-1}(z)) = \mathcal{G}_{R_4}(T^{-1} \psi^{-1}(z)) = \| T^{-1} \psi^{-1}(z) \|_1$ similar to before, where the final equality is due to $S_{\leq 1}(\tnorm[1]) = R_4$.
        
        Furthermore, $T^{-1} \psi^{-1}(z) \in \mathbb{R}_{\geq 0}^2$ implies that its length under $\tnorm[1]$ is equal to the sum of its two coordinates.
        To complete the derivation, it remains to determine the inverse matrix $T^{-1}$.
        The standard $2 \times 2$ formula says
        \[
            T^{-1} = \frac{1}{\mathrm{det}(T)} \cdot
            \begin{pmatrix*}[l]
                \phantom{-}\sin(\theta_{\rho(z)}) & -\cos(\theta_{\rho(z)}) \\
                -\sin(\theta_{\rho(z)-1}) \:\: & \phantom{-}\cos(\theta_{\rho(z)-1})
            \end{pmatrix*}
            \text{,}
        \]
        and using a trigonometric identity (see \cite[Equation~2.4.1.4.2]{Jeffr2008}) yields
        \begin{align*}
            \mathrm{det}(T) &= \sin(\theta_{\rho(z)}) \cos(\theta_{\rho(z)-1}) - \cos(\theta_{\rho(z)}) \sin(\theta_{\rho(z)-1}) \\
            &= \sin(\theta_{\rho(z)} - \theta_{\rho(z)-1}) = \sin(\theta_1)
            \text{.}
        \end{align*}
        
        \item 
        By \autoref{def:propagation-space}, the boundary of the propagation space $\Sigma_{\mathcal{A},\mathcal{B}}$ is a rectangle if $\mathcal{A} = \mathrm{adj}(\mathcal{B})$ or an isosceles right triangle if $\mathcal{A} = \mathrm{opp}(\mathcal{B})$, as shown in~\autoref{fig:arrangements}.
        In both cases we add a constant number of arrangement lines for the boundary, which ensure that $\Sigma_{\mathcal{A},\mathcal{B}}$ is a union of faces.
        Next, there are two more kinds of arrangement lines that we consider:
        \emph{Grid lines}, which are vertical or~horizontal, and \emph{split lines}, which may be sloped and can only occur if $\mathcal{A} = \mathrm{adj}(\mathcal{B})$.
        E.g., in \autoref{subfig:arrangement-parallel} there are two grid lines and two (coloured) split lines.
        
        Let $\overline{P},\overline{Q}$ be the curve segments associated with the cell of $\mathcal{A},\mathcal{B}$.
        We construct arrangement lines based on a valley $\ell$ of positive slope for $\overline{P},\overline{Q}$ under $\mathcal{G}_{\psi(R_k)}$, which exists by \autoref{thm:valleys-for-polygonal-norms}.
        Assume first that $\mathcal{A} = \mathrm{adj}(\mathcal{B})$ holds.
        If $\mathcal{B}$ is the right border, we add $\ell$ as a split line, and if $\mathcal{B}$ is the top border, we need to swap coordinates and add $\left( \begin{smallmatrix}0 & 1 \\ 1 & 0\end{smallmatrix} \right) \ell$ instead.
        In both cases we check whether~$\ell$ intersects $\mathcal{A}$ or $\mathcal{B}$ and add a grid line for each intersection, which is a vertical or horizontal line respectively.
        E.g., in \autoref{subfig:arrangement-parallel} the (orange) split line for $\ell$ indicates no such intersections and thus no grid lines for $\ell$, while in \autoref{subfig:arrangement-adjacent} there is a horizontal grid line for $\ell$ because of an intersection with $\mathcal{B}$.
        
        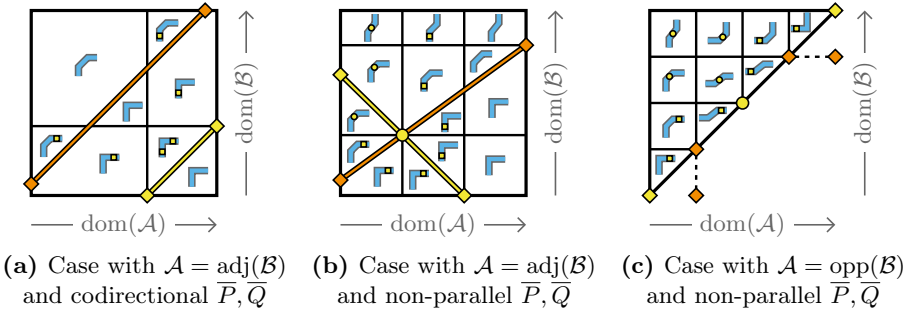
\begin{figure}[H]%
            \centering%
            \begin{subfigure}{0.3\linewidth}
                \centering
                \begin{tikzpicture}[x=2/3*\linewidth,y=2/3*\linewidth,trim left=0pt]
                    \borderaxes
                    \path[main line] (0,0) rectangle (1,1);
                
                    \def\s{0.625}
                    \pgfmathparse{1-1.5*\s}\edef\t{\pgfmathresult}
                    
                    \path[grid line] (\s,0) -- (\s,1);
                    \path[grid line] (0,1-\s) -- (1,1-\s);
                    \path[diagonal split line] (\s,0) -- coordinate[pos=0.25] (ds) coordinate[pos=0.5] (dm) coordinate[pos=0.75] (de) (1,1-\s);
                    \path[valley split line] (0,\t) -- coordinate[pos=1/6] (vs) coordinate[pos=0.5] (vm) coordinate[pos=5/6] (ve) (1-\t,1);
                    
                    \node[diagonal point] at (\s,0) {};
                    \node[diagonal point] at (1,1-\s) {};
                    \node[valley point] at (0,\t) {};
                    \node[valley point] at (1-\t,1) {};

                    \begin{scope}[shift={($(dm)!0.55!(1,0)$)}]
                        \path[matching path shape] (-0.05,-0.0375) |- (0.0375,0.05);
                    \end{scope}

                    \begin{scope}[shift={($(dm)!0.3!(\s,1-\s)$)}]
                        \path[matching path shape] (-0.05,-0.0375) |- (0.0375,0.05);
                        \node[diagonal indicator] at (-0.05,-0.0075) {};
                        \node[diagonal indicator] at (0.0075,0.05) {};
                    \end{scope}

                    \begin{scope}[shift={($(ve)!0.5!(de)$)}]
                        \path[matching path shape] (-0.05,-0.0375) |- (0.0375,0.05);
                        \node[diagonal indicator] at (-0.05,-0.0075) {};
                    \end{scope}
                    
                    \begin{scope}[shift={($(vs)!0.5!(ds)$)}]
                        \path[matching path shape] (-0.05,-0.0375) |- (0.0375,0.05);
                        \node[diagonal indicator] at (0.0075,0.05) {};
                    \end{scope}
                    
                    \begin{scope}[shift={($(vm)!0.6!(\s,1-\s)$)}]
                        \path[matching path shape] (-0.05,-0.0375) |- (0.0375,0.05);
                    \end{scope}
                    
                    \begin{scope}[shift={($(ve)!0.25!(\s,1)$)}]
                        \path[matching path shape] (-0.05,-0.05) -- (-0.05,0.0125) -- (-0.0125,0.05) -- (0.05,0.05);
                        \node[diagonal indicator] at (-0.05,-0.02) {};
                    \end{scope}
                    
                    \begin{scope}[shift={($(vs)!0.25!(0,1-\s)$)}]
                        \path[matching path shape] (-0.05,-0.05) -- (-0.05,0.0125) -- (-0.0125,0.05) -- (0.05,0.05);
                        \node[diagonal indicator] at (0.02,0.05) {};
                    \end{scope}
                    
                    \begin{scope}[shift={($(vm)!0.35!(0,1)$)}]
                        \path[matching path shape] (-0.05,-0.05) -- (-0.05,0.00) -- (0.00,0.05) -- (0.05,0.05);
                    \end{scope}
                \end{tikzpicture}
                \caption{Case with $\mathcal{A} = \mathrm{adj}(\mathcal{B})$ and codirectional $\overline{P},\overline{Q}$}
                \label{subfig:arrangement-parallel}
            \end{subfigure}%
            \hspace*{0.1\linewidth/3}%
            \begin{subfigure}{0.3\linewidth}
                \centering
                \begin{tikzpicture}[x=2/3*\linewidth,y=2/3*\linewidth,trim left=0pt]
                    \borderaxes
                    \path[main line] (0,0) rectangle (1,1);
                    
                    \def\tvl{1/12}
                    \def\tvr{13/16}
                    \def\sd{2/3}
                    \pgfmathparse{2*(\tvl+(\tvr-\tvl)/3)}\edef\td{\pgfmathresult}
                    
                    \path[grid line] (0,\tvr) -- (1,\tvr);
                    \path[grid line] (\sd,0) -- (\sd,1);
                    \path[grid line] (0,\td/2) -- (1,\td/2);
                    \path[grid line] (\sd/2,0) -- (\sd/2,1);
                    \path[diagonal split line] (0,\td) -- coordinate[pos=0.175] (ds1) coordinate[pos=0.3] (ds2) coordinate[pos=0.8] (de) (\sd,0);
                    \path[valley split line] (0,\tvl) -- coordinate[pos=0.145] (vs) coordinate[pos=0.525] (vm) coordinate[pos=0.81] (ve) (1,\tvr);
                    
                    \node[diagonal point] at (0,\td) {};
                    \node[diagonal point] at (\sd,0) {};
                    \node[valley point] at (0,\tvl) {};
                    \node[valley point] at (1,\tvr) {};
                    \node[origin,fill=UmiYellow] at (\sd/2,\td/2) {};

                    \begin{scope}[shift={($(\sd/4,0.5+\tvr/2)$)}]
                        \path[matching path shape] (-0.02,-0.075) -- (-0.02,-0.025) -- (0.02,0.025) -- (0.02,0.075);
                        \node[origin indicator] at (0.00,0.00) {};
                    \end{scope}

                    \begin{scope}[shift={($(0.5,0.5+\tvr/2)$)}]
                        \path[matching path shape] (-0.02,-0.075) -- (-0.02,-0.02) -- (0.02,0.02) -- (0.02,0.075);
                        \node[diagonal indicator] at (-0.02,-0.045) {};
                    \end{scope}
                    
                    \begin{scope}[shift={($(5*\sd/4,0.5+\tvr/2)$)}]
                        \path[matching path shape] (-0.02,-0.075) -- (-0.02,-0.025) -- (0.02,0.025) -- (0.02,0.075);
                    \end{scope}
                    
                    \begin{scope}[shift={($(ds1)!0.45!(\sd/2,\tvr)$)}]
                        \path[matching path shape] (-0.05,-0.05) -- (-0.05,0.0125) -- (-0.0125,0.05) -- (0.05,0.05);
                        \node[origin indicator] at (-0.03125,0.03125) {};
                    \end{scope}
                    
                    \begin{scope}[shift={($(ds2)!0.475!(0,\td/2)$)}]
                        \path[matching path shape] (-0.05,-0.05) -- (-0.05,0.0125) -- (-0.0125,0.05) -- (0.05,0.05);
                        \node[origin indicator] at (-0.03125,0.03125) {};
                    \end{scope}
                    
                    \begin{scope}[shift={($(0.5,15*\td/16)$)}]
                        \path[matching path shape] (-0.05,-0.05) -- (-0.05,0.0125) -- (-0.0125,0.05) -- (0.05,0.05);
                        \node[diagonal indicator] at (-0.05,-0.02) {};
                    \end{scope}
                    
                    \begin{scope}[shift={($(vm)!0.6175!(\sd,\td/2)$)}]
                        \path[matching path shape] (-0.05,-0.0375) |- (0.0375,0.05);
                        \node[diagonal indicator] at (-0.05,-0.0075) {};
                    \end{scope}
                    
                    \begin{scope}[shift={($(ve)!0.3!(\sd,\tvr)$)}]
                        \path[matching path shape] (-0.05,-0.05) -- (-0.05,0.00) -- (0.00,0.05) -- (0.05,0.05);
                    \end{scope}
                
                    \begin{scope}[shift={($(7*\sd/6,\td/2)!0.4!(1,\td)$)}]
                        \path[matching path shape] (-0.05,-0.0375) |- (0.0375,0.05);
                    \end{scope}
                    
                    \begin{scope}[shift={($(vs)!0.275!(0,\td/2)$)}]
                        \path[matching path shape] (-0.05,-0.05) -- (-0.05,0.0125) -- (-0.0125,0.05) -- (0.05,0.05);
                        \node[diagonal indicator] at (0.02,0.05) {};
                    \end{scope}
                    
                    \begin{scope}[shift={($(vs)!0.55!(\sd/2,0)$)}]
                        \path[matching path shape] (-0.05,-0.0375) |- (0.0375,0.05);
                        \node[diagonal indicator] at (0.0075,0.05) {};
                    \end{scope}
                    
                    \begin{scope}[shift={($(de)!0.47!(\sd,\td/2)$)}]
                        \path[matching path shape] (-0.05,-0.0375) |- (0.0375,0.05);
                        \node[diagonal indicator] at (-0.05,-0.0075) {};
                    \end{scope}
                    
                    \begin{scope}[shift={($(de)!0.47!(\sd/2,0)$)}]
                        \path[matching path shape] (-0.05,-0.0375) |- (0.0375,0.05);
                        \node[diagonal indicator] at (0.0075,0.05) {};
                    \end{scope}
                    
                    \begin{scope}[shift={($(\sd,\td/2)!0.55!(1,0)$)}]
                        \path[matching path shape] (-0.05,-0.0375) |- (0.0375,0.05);
                    \end{scope}
                \end{tikzpicture}
                \caption{Case with $\mathcal{A} = \mathrm{adj}(\mathcal{B})$ and non-parallel $\overline{P},\overline{Q}$}
                \label{subfig:arrangement-adjacent}
            \end{subfigure}%
            \hspace*{0.1\linewidth/3}%
            \begin{subfigure}{0.3\linewidth}
                \centering
                \begin{tikzpicture}[x=2/3*\linewidth,y=2/3*\linewidth,trim left=0pt]
                    \borderaxes
                    \path[main line] (0,0) -- (0,1) -- (1,1) -- cycle;
                    
                    \foreach \i in {0.25,0.5,0.75} {
                        \path[grid line] (0,\i) -- (\i,\i);
                        \path[grid line] (\i,1) -- (\i,\i);
                    };
                
                    \path[util dash,dash expand off,grid line] (0.25,0) -- (0.25,0.25);
                    \path[util dash,dash expand off,grid line] (1,0.75) -- (0.75,0.75);
                    \node[valley point] at (0.25,0) {};
                    \node[valley point] at (1,0.75) {};
                    
                    \node[diagonal point] at (0,0) {};
                    \node[valley point] at (0.25,0.25) {};
                    \node[origin,fill=UmiYellow] at (0.5,0.5) {};
                    \node[valley point] at (0.75,0.75) {};
                    \node[diagonal point] at (1,1) {};
                    
                    \begin{scope}[shift={(0.125,0.875)}]
                        \path[matching path shape] (-0.02,-0.075) -- (-0.02,-0.025) -- (0.02,0.025) -- (0.02,0.075);
                        \node[origin indicator] at (0.00,0.00) {};
                    \end{scope}
                    
                    \begin{scope}[shift={($(0.125,0.625) + 0.0125*(1,-1)$)}]
                        \path[matching path shape] (-0.05,-0.05) -- (-0.05,0.0125) -- (-0.0125,0.05) -- (0.05,0.05);
                        \node[origin indicator] at (-0.03125,0.03125) {};
                    \end{scope}
                    
                    \begin{scope}[shift={($(0.125,0.375) + 0.0125*(1,-1)$)}]
                        \path[matching path shape] (-0.05,-0.05) -- (-0.05,0.0125) -- (-0.0125,0.05) -- (0.05,0.05);
                        \node[diagonal indicator] at (0.02,0.05) {};
                    \end{scope}
                    
                    \begin{scope}[shift={(0.1,0.15)}]
                        \path[matching path shape] (-0.05,-0.0375) |- (0.0375,0.05);
                        \node[diagonal indicator] at (0.0075,0.05) {};
                    \end{scope}
                    
                    \begin{scope}[shift={($(0.375,0.875) - 0.0125*(1,-1)$)}]
                        \path[matching path shape] (-0.05,-0.05) -- (0.0125,-0.05) -- (0.05,-0.0125) -- (0.05,0.05);
                        \node[origin indicator] at (0.03125,-0.03125) {};
                    \end{scope}
                    
                    \begin{scope}[shift={(0.375,0.625)}]
                        \path[matching path shape] (-0.075,-0.02) -- (-0.025,-0.02) -- (0.025,0.02) -- (0.075,0.02);
                        \node[origin indicator] at (0.00,0.00) {};
                    \end{scope}
                    
                    \begin{scope}[shift={(0.34,0.44)}]
                        \path[matching path shape] (-0.075,-0.02) -- (-0.025,-0.02) -- (0.025,0.02) -- (0.075,0.02);
                        \node[diagonal indicator] at (0.045,0.02) {};
                    \end{scope}
                    
                    \begin{scope}[shift={($(0.625,0.875) - 0.0125*(1,-1)$)}]
                        \path[matching path shape] (-0.05,-0.05) -- (0.0125,-0.05) -- (0.05,-0.0125) -- (0.05,0.05);
                        \node[diagonal indicator] at (-0.02,-0.05) {};
                    \end{scope}
                    
                    \begin{scope}[shift={(0.59,0.69)}]
                        \path[matching path shape] (-0.075,-0.02) -- (-0.025,-0.02) -- (0.025,0.02) -- (0.075,0.02);
                        \node[diagonal indicator] at (-0.045,-0.02) {};
                    \end{scope}
                    
                    \begin{scope}[shift={(0.798,0.952)}]
                        \path[matching path shape] (-0.0375,-0.05) -| (0.05,0.0375);
                        \node[diagonal indicator] at (-0.0075,-0.05) {};
                    \end{scope}
                \end{tikzpicture}
                \caption{Case with $\mathcal{A} = \mathrm{opp}(\mathcal{B})$ and non-parallel $\overline{P},\overline{Q}$}
                \label{subfig:arrangement-opposing}
            \end{subfigure}%
            \caption{Arrangements on $\Sigma_{\mathcal{A},\mathcal{B}}$, where each face has a type of optimal paths from bottom/left border $\mathcal{A}$ to right border $\mathcal{B}$ of a cell for curve segments $\overline{P},\overline{Q}$ under $\mathcal{G}_{\psi(R_4)}$}%
            \label{fig:arrangements}%
        \end{figure}
        
        Now assume $\mathcal{A} = \mathrm{opp}(\mathcal{B})$.
        In particular, it is $\mathrm{dom}(\mathcal{A}) = \mathrm{dom}(\mathcal{B})$ since $\mathcal{A},\mathcal{B}$ are parametrised in the same coordinate.
        We again identify the intersection points of $\ell$ with $\mathcal{A}$ or $\mathcal{B}$ and project them into the propagation space $\Sigma_{\mathcal{A},\mathcal{B}}$ by duplicating the pertinent coordinate, i.e.\ we use $z \mapsto \left( \begin{smallmatrix}0 & 1 \\ 0 & 1\end{smallmatrix} \right) z$ if $\mathcal{B}$ is the right border and $z \mapsto \left( \begin{smallmatrix}1 & 0 \\ 1 & 0\end{smallmatrix} \right) z$ if $\mathcal{B}$ is the top border, which yields points on the hypotenuse of the right-angled triangle $\Sigma_{\mathcal{A},\mathcal{B}}$.
        We add a vertical and a horizontal grid line at each projected intersection point, see \autoref{subfig:arrangement-opposing} for an instance with four grid lines for $\ell$ due to intersections with $\mathcal{A}$ and $\mathcal{B}$.
        
        So far, we have added a constant number of arrangement lines for $\ell$, which separate different shapes of optimal $(\mathcal{A}(s),\mathcal{B}(t))$-paths induced by \autoref{thm:optimal-paths} for $(s,t)^{\mathsf{T}} \in \Sigma_{\mathcal{A},\mathcal{B}}$.
        To illustrate that, \autoref{fig:arrangements} shows pictograms of the faces' corresponding path shapes.
        The pictograms include another important aspect of evaluating optimal path costs under $\mathcal{G}_{\psi(R_k)}$:
        Breakpoints that are induced by the supporting lines of the cones from (a), namely $L_r := \psi(\{\lambda v_r \mid \lambda \in \mathbb{R}\})$ for $r \in \{1,\dotsc,k/2\}$, each of which corresponds to a diagonal of the polygon $\psi(R_k)$ through two opposite vertices $\psi(v_r)$ and $\psi(v_{r + k/2}) = -\psi(v_r)$.
        
        We first map $L_r$ with $r \in \{1,\dotsc,k/2\}$ into the parameter space cell of $\overline{P},\overline{Q}$ using the affine map $\varphi$ from \autoref{thm:affine}.
        If $\overline{P},\overline{Q}$ are not parallel, then we get the line $\Lambda_r := \varphi^{-1}(L_r)$.
        In the case of parallel $\overline{P},\overline{Q}$ we can proceed in a similar way to the proof of \autoref{thm:valleys}:
        We consider the line $L := \{ \varphi(s,\mp s) \mid s \in \mathbb{R} \}$, where the sign $\mp$ depends on whether $\overline{P},\overline{Q}$ are codirectional or have opposite directions.
        If $L$ is parallel to $L_r$, we omit $\Lambda_r$.
        Otherwise, there exists a unique intersection point $\varphi(s_r,\mp s_r) \in L \cap L_r$ for some $s_r \in \mathbb{R}$, which yields the line $\Lambda_r := \{ (\lambda + s_r, \pm (\lambda - s_r))^{\mathsf{T}} \mid \lambda \in \mathbb{R} \}$ of slope $\pm 1$ in parameter space.
        
        The arrangement lines for a parameter space line $\Lambda_r$ are then obtained in the same way as above for $\ell$.\footnote{Note that $\ell \in \{ \Lambda_1,\dotsc,\Lambda_{k/2} \}$ is possible (cf.\ the proof of \autoref{thm:valleys-for-polygonal-norms}) and that~\autoref{fig:arrangements} assumes this for $k = 4$ in order to keep the number of relevant lines low.}
        Consequently, if $\mathcal{A} = \mathrm{adj}(\mathcal{B})$, we add a split line and up to two grid lines for each $\Lambda_r$, and if $\mathcal{A} = \mathrm{opp}(\mathcal{B})$, we add up to four grid lines for each $\Lambda_r$.
        In case of non-parallel $\overline{P},\overline{Q}$ we also add two grid lines corresponding to the common intersection of $\Lambda_1,\dotsc,\Lambda_{k/2}$ and $\ell$, i.e.\ the polygon centre $\varphi^{-1}(\mathbf{0})$ that is possibly again transformed by $\left( \begin{smallmatrix}0 & 1 \\ 1 & 0\end{smallmatrix} \right)$, $\left( \begin{smallmatrix}0 & 1 \\ 0 & 1\end{smallmatrix} \right)$ or~$\left( \begin{smallmatrix}1 & 0 \\ 1 & 0\end{smallmatrix} \right)$.
        Overall, the constructed arrangement consists of $O(k)$ lines.
        
        We next consider the evaluation of $\mathrm{opt}_{\mathcal{A},\mathcal{B}}$ on an arbitrary face $f \subseteq \Sigma_{\mathcal{A},\mathcal{B}}$ of this arrangement.
        The above already established that the optimal $(\mathcal{A}(s), \mathcal{B}(t))$-path from \autoref{thm:optimal-paths} has the same shape for all $(s,t)^{\mathsf{T}} \in f$, which traces up to three line segments.\footnote{\label{fn:zero-cost}On the boundary of $f$ some line segments may have length $0$. This is covered by the following because it simply means $a(s,t) = b(s,t)$ and thus $\mathrm{cost}(\overline{\gamma}_{s,t}) = 0$.}
        Let $\overline{\gamma}_{s,t} \colon [a(s,t),b(s,t)] \to \mathbb{R}^2$ be a subpath tracing one of these line segments.
        Since the endpoint coordinates of each segment depend linearly on $s$ and $t$, the same is true for $a(s,t)$ and $b(s,t)$ as well as the translational part of the affine map $\overline{\gamma}_{s,t}$.
        The linear part of $\overline{\gamma}_{s,t}$ is vertical or horizontal or in direction of $\ell$, so it does not depend on $s$ and $t$.
        Using the affine map $\varphi$, \autoref{def:path-and-cost} yields the cost of $\overline{\gamma}_{s,t}$ under $\mathcal{G}_{\psi(R_k)}$ through
        \[
            \mathrm{cost}(\overline{\gamma}_{s,t})
            = \int_{a(s,t)}^{b(s,t)} \mathcal{G}_{\psi(R_k)}(\varphi(\overline{\gamma}_{s,t}(\tau))) \dtau
            \text{.}
        \]
        
        Together with (a) it thus follows that the integrand of $\mathrm{cost}(\overline{\gamma}_{s,t})$ is a piecewise linear function on $\mathbb{R}$, where the piece breakpoints are given by $\tau \to \varphi(\overline{\gamma}_{s,t}(\tau))$ switching between cones of the polygon $\psi(R_k)$.
        In the parameter space this corresponds to the subpath $\overline{\gamma}_{s,t}$ crossing some line $\Lambda_r$ with $r \in \{1,\dotsc,k/2\}$ as defined above.
        Therefore, we can partition $\mathrm{cost}(\overline{\gamma}_{s,t})$ at these crossings to obtain a sum of $O(k)$ integrals whose integrands are linear functions in the variable of integration $\tau$.
        This sum is well-defined on the face $f$, in that $\smash{\overline{\gamma}_{s,t}}$ crosses the lines in a consistent way for all $(s,t)^{\mathsf{T}} \in f$ by construction:
        \begin{itemize}
            \item 
            For every line $\Lambda_r$ we have that the number of crossings between $\Lambda_r$ and the optimal $(\mathcal{A}(s),\mathcal{B}(t))$-path is constant on the interior $f^\circ$ of the face $f$, as $f^\circ$ does not intersect the split line for $\Lambda_r$ nor any grid line for $\Lambda_r$.
            
            \item 
            Each crossing between a line $\Lambda_r$ and the optimal $(\mathcal{A}(s),\mathcal{B}(t))$-path on~$f$ has a fixed line segment of the path shape on which it occurs, as $f^\circ$ does not intersect the split line for $\Lambda_r$ nor any grid line for $\Lambda_r$ or $\varphi^{-1}(\mathbf{0})$.
            
            \item 
            Combining the above two statements with the observation that the lines $\Lambda_1,\dotsc,\Lambda_{k/2}$ either are all parallel or all intersect in $\varphi^{-1}(\mathbf{0})$, we get that the ordered sequence in which $\overline{\gamma}_{s,t}$ crosses these lines is fixed.\footnote{On the boundary of $f$ some consecutive crossings may coincide and there may also be additional crossings. Both of these cases can only occur at endpoints of line segments of the optimal path on $f$, so they contribute no cost similar to \autoref{fn:zero-cost}.}
        \end{itemize}
        
        Finally, we evaluate each integral in the sum via a quadratic antiderivative $\tau \mapsto \lambda_2 \tau^2 + \lambda_1(s,t) \tau + C$ of its linear integrand, where $\lambda_1(s,t) \in \mathbb{R}$ depends linearly on $(s,t)^{\mathsf{T}} \in f$ but $\lambda_2 \in \mathbb{R}$ does not, which is inherited by the~translational and the linear part of $\overline{\gamma}_{s,t}$ respectively.
        Plugging $a(s,t)$ and $b(s,t)$ into the antiderivative thus yields a quadratic function in $s$ and $t$.
        By summing up all $O(k)$ quadratics given by the line segments in the optimal path shape of the face $f$, we then obtain a quadratic function rule for $\mathrm{opt}_{\mathcal{A},\mathcal{B}}$ on $f$.
        
        \item 
        We proceed by induction.
        First, consider any border $\mathcal{B}$ on the bottom or on the left side of the parameter space of polygonal curves $P,Q$ under $\mathcal{G}_{\psi(R_k)}$.
        Then for each $t \in \mathrm{dom}(\mathcal{B})$ there is a single and thus optimal $(\mathbf{0},\mathcal{B}(t))$-path, which is horizontal or vertical respectively.
        By \autoref{def:path-and-cost}, the vertical case yields $\mathrm{opt}_{0,\mathcal{B}}(t) = \int_0^t \mathcal{G}_{\psi(R_k)}(P_{\mathcal{G}_{\psi(R_k)}}(0) - Q_{\mathcal{G}_{\psi(R_k)}}(\tau)) \dtau$ for $t \in \mathrm{dom}(\mathcal{B})$; the horizontal one is analogous.
        The integrand composes $\mathcal{G}_{\psi(R_k)}$ and a piecewise affine map, so (a) implies that it is a piecewise linear function.
        Hence, using antiderivatives of the pieces shows that $\mathrm{opt}_{0,\mathcal{B}}$ is piecewise quadratic.
        
        Next, let $\mathcal{B}$ be any other border.
        By our inductive assumption, the optimum functions of $\mathrm{adj}(\mathcal{B}),\mathrm{opp}(\mathcal{B})$ are piecewise quadratic.
        Also, (b) says that every $\mathcal{A} \in \{ \mathrm{adj}(\mathcal{B}), \mathrm{opp}(\mathcal{B}) \}$ has some arrangement of $O(k)$ lines such that $\mathrm{opt}_{\mathcal{A},\mathcal{B}}$ is quadratic on each face $f \subseteq \Sigma_{\mathcal{A},\mathcal{B}}$.
        The overlay of such an arrangement with vertical lines at the breakpoints of $\mathrm{opt}_{0,\mathcal{A}}$ therefore has the property that the function $\smash{(s,t)^{\mathsf{T}}} \mapsto \mathrm{opt}_{0,\mathcal{A}}(s) + \mathrm{opt}_{\mathcal{A},\mathcal{B}}(s,t)$ is quadratic on each face.
        
        Realising \autoeqref{eq:propagation} requires minimising the above sum for all $t \in \mathrm{dom}(\mathcal{B})$.
        Every extremum is attained either on an edge $e$ or on an interior of a face $f$ of the overlayed arrangement, where $\partial_s [\mathrm{opt}_{0,\mathcal{A}}(s) + \mathrm{opt}_{\mathcal{A},\mathcal{B}}(s,t)] = 0$ holds in the latter case by differentiability.
        Given the coefficient representation
        \[
            \mathrm{opt}_{0,\mathcal{A}}(s) + \mathrm{opt}_{\mathcal{A},\mathcal{B}}(s,t)
            = \lambda_{2,0} s^2 + \lambda_{1,1} s t + \lambda_{0,2} t^2 + \lambda_{1,0} s + \lambda_{0,1} t + \lambda_{0,0}
        \]
        on $f$, the partial derivative with respect to $s$ is $2 \lambda_{2,0} s + \lambda_{1,1} t + \lambda_{1,0}$ at~$(s,t)^{\mathsf{T}}$.
        Setting this to $0$ reveals that all extrema on the interior of $f$ are attained on the line segment $e_f := \{ (s,t)^{\mathsf{T}} \in f \mid (2 \lambda_{2,0}, \lambda_{1,1}) \cdot (s,t)^{\mathsf{T}} = - \lambda_{1,0} \}$, which appears in line~11 of \nameref{alg:Propagate}.
        We call it the \emph{extremal edge} of $f$.
        
        Finally, parametrising all (non-horizontal) edges $e$ and extremal edges $e_f$ in the coordinate $t$ gives linear functions $s_e(t)$ and $s_f(t)$.
        Plugging $(s_e(t),t)^{\mathsf{T}}$ and $(s_f(t),t)^{\mathsf{T}}$ into the corresponding quadratic restrictions of the functions $\smash{(s,t)^{\mathsf{T}}} \mapsto \mathrm{opt}_{0,\mathcal{A}}(s) + \mathrm{opt}_{\mathcal{A},\mathcal{B}}(s,t)$ with $\mathcal{A} \in \{ \mathrm{adj}(\mathcal{B}), \mathrm{opp}(\mathcal{B}) \}$ returns a finite number of quadratic functions on subsets of $\mathrm{dom}(\mathcal{B})$.
        By construction, they together cover all minima relevant for \autoeqref{eq:propagation}.
        It thus follows that their lower envelope is a piecewise quadratic function and equal to $\mathrm{opt}_{0,\mathcal{B}}$.
        \qed
    \end{alphaenumerate}
\end{proof}

\autoref{thm:propagation-order} allows faster computations of lower envelopes during propagation and is also crucial for the proof of the 1D polynomial bound \cite{BuchiNW2022}.
The intuition behind it is that one can always choose an optimal $(\mathbf{0},\mathcal{B}(t))$-path and an optimal $(\mathbf{0},\mathcal{B}(t'))$-path that do not cross after diverging at $\mathbf{0}$ or from some shared prefix.
This has been stated by \cite[Lemma~11]{BuchiNW2022} and \cite[Observations~3.2~\&~3.3]{BuchiBW2009} before, but it is in fact not quite enough for proving the correctness of \nameref{alg:Propagate}.

We need to ensure that cost improvements are always pushed onto the stack.
This means avoiding the case that $s$ improves upon $s'$ for $t$ and does not improve upon $s'$ for $t'$, which is higher on the stack and so blocks the improvement~for $t$.
\autoref{thm:propagation-order} implies that propagating $s$ before $s'$ is always possible via a suitable order of $\mathcal{A}$.
Note that we do not require $\mathcal{A}$ to be a border.
It may be any \nolinebreak paramet\-risation \nolinebreak of the same kind, as implicitly used\footnote{More precisely, the relevant part of the proof of \autoref{thm:continuity} constructs a parametrisation $\mathcal{A}$ orthogonal to $\mathcal{B}$ at the point $\mathcal{B}(t_0)$. In fact, there is also a similar implicit use in the proof of \autoref{thm:propagation-correctness}: To compare points $\mathcal{A}_{\mathrm{opp}}(s), \mathcal{A}_{\mathrm{adj}}(s')$ for $\mathcal{A}_{\mathrm{opp}} := \mathrm{opp}(\mathcal{B})$ and $\mathcal{A}_{\mathrm{adj}} := \mathrm{adj}(\mathcal{B})$, one can construct an $\mathcal{A}$ orthogonal to $\mathcal{A}_{\mathrm{opp}}$ at $\mathcal{A}_{\mathrm{opp}}(s)$.} in the proof of \autoref{thm:continuity}.

\settheoremcounter{thm:propagation-order}
\begin{lemma}
    Let $\mathcal{B}$ be a top/right border under $\tnorm$, let $\mathcal{A} \colon [a,b] \to \mathbb{R}^2$ be~a~parametrisation in direction $(1,0)^{\mathsf{T}}$ or $(0,1)^{\mathsf{T}}$, and let $[s\;\!\triangleright\;\!t] := \mathrm{opt}_{0,\mathcal{A}}(s) + \mathrm{opt}_{\mathcal{A},\mathcal{B}}(s,t)$.
    If $[s \triangleright t] < [s' \triangleright t]$ and $[s \triangleright t'] \geq [s' \triangleright t']$ for $t < t'$, then $s < s'$ if and only if $\mathcal{A}$ and $\mathcal{B}$ have the same direction.
    This yields a propagation order for $\mathcal{A} \in \{ \mathrm{adj}(\mathcal{B}),\mathrm{opp}(\mathcal{B}) \}$.
\end{lemma}

\begin{proof}
    We first show that the suffixes of corresponding paths starting respectively at points $\mathcal{A}(s)$ and $\mathcal{A}(s')$ cannot cross:
    Assume for the sake of contradiction that some optimal $(\mathcal{A}(s),\mathcal{B}(t))$-path $\gamma_{t}$ intersects some optimal $(\mathcal{A}(s'),\mathcal{B}(t'))$-path~$\gamma_{t'}$ in a point $z$.
    Let $\gamma_z^t, \gamma_z^{t'}$ denote the prefix paths of $\gamma_t, \gamma_{t'}$ up to $z$, so that we obtain $\mathrm{opt}_{0,\mathcal{A}}(s) + \mathrm{cost}(\gamma_z^t) < \mathrm{opt}_{0,\mathcal{A}}(s') + \mathrm{cost}(\gamma_z^{t'})$ due to $[s \triangleright t] < [s' \triangleright t]$.
    Consequently, concatenating an optimal $(\mathbf{0},\mathcal{A}(s))$-path with $\gamma_z^t$ and with the suffix of $\gamma_{t'}$ starting at $z$ yields $[s \triangleright t'] < [s' \triangleright t']$.
    This is in contradiction to $[s \triangleright t'] \geq [s' \triangleright t']$.
    
    If $\mathcal{A}$ and $\mathcal{B}$ have the same direction (e.g.\ for $\mathcal{A} = \mathrm{opp}(\mathcal{B})$), then the above implies $s < s'$ since in this case $s > s'$ would mean that every $(\mathcal{A}(s),\mathcal{B}(t))$-path crosses every $(\mathcal{A}(s'),\mathcal{B}(t'))$-path by $t < t'$ and continuity.
    Otherwise, $\mathcal{A}$ and $\mathcal{B}$ have orthogonal directions (e.g.\ for $\mathcal{A} = \mathrm{adj}(\mathcal{B})$), so the above now implies $s > s'$ since in that case $s < s'$ would mean that every $(\mathcal{A}(s),\mathcal{B}(t))$-path crosses every $(\mathcal{A}(s'),\mathcal{B}(t'))$-path by $t < t'$ and continuity.
    See \autoref{fig:crossing-paths} for both cases.
    \qed
\end{proof}

\begin{figure}[H]%
    \centering%
    \begin{minipage}[b]{0.45\linewidth}%
        \centering%
        \begin{tikzpicture}[x=0.95\linewidth,y=0.95\linewidth,scale=0.45]
            \path[main line] (0,0) rectangle (1,0);
            \path[main line] (0,1) rectangle (1,1);
            
            \coordinate (s) at (0.5,0);
            \coordinate (s') at (1/6,0);
            \coordinate (t) at (0.5,1);
            \coordinate (t') at (5/6,1);
            
            \path[matching path] (s) -- (t);
            \path[matching path] (s') -- ($(s') + (0,1/3)$) -- ($(t') - (0,1/3)$) -- (t');
            
            \node[matching point] at (s) {};
            \node[matching point] at (s') {};
            \node[matching point,label={above:\color{UmiSkyblue}$\mathcal{B}(t)$}] at (t) {};
            \node[matching point,label={above:\color{UmiSkyblue}$\mathcal{B}(t')$}] at (t') {};
            
            \begin{scope}[shift={(1.25,0)}]
                \path[main line] (0,0) -- (0,1) -- (1,1);
                \coordinate (tr) at (1,1);
                
                \coordinate (s) at (0,1/6);
                \coordinate (s') at (0,0.5);
                \coordinate (t) at (0.5,1);
                \coordinate (t') at (5/6,1);
                
                \path[matching path] (s) -- ($(s) + (1/3,0)$) -- ($(t) - (0,2/3)$) -- (t);
                \path[matching path] (s') -- ($(s') + (2/3,0)$) -- ($(t') - (0,1/3)$) -- (t');
                
                \node[matching point] at (s) {};
                \node[matching point] at (s') {};
                \node[matching point,label={above:\color{UmiSkyblue}$\mathcal{B}(t)$}] at (t) {};
                \node[matching point,label={above:\color{UmiSkyblue}$\mathcal{B}(t')$}] at (t') {};
            \end{scope}
            
            \pgfresetboundingbox
            \useasboundingbox (0,0) rectangle (tr);
        \end{tikzpicture}%
        \captionof{figure}{Crossing paths to border $\mathcal{B}$}%
        \label{fig:crossing-paths}%
    \end{minipage}%
    \hspace*{0.05\linewidth}%
    \begin{minipage}[b]{0.45\linewidth}%
        \centering%
        \begin{tikzpicture}[x=0.95\linewidth,y=0.95\linewidth,scale=0.6]
            \useasboundingbox (0,0) rectangle (1,{sin(60)});
            
            \path[util dash,dash expand off,sublevel set]
                (1,0) coordinate (v) arc[start angle=0,end angle=60,radius=1] coordinate (v');
            \path[sublevel set,draw=none]
                (0,0) -- coordinate[pos=0.2] (a) (v) -- (v') -- (0,0);
            \path[draw,semithick]
                (0,0) -- (a) arc[start angle=0,end angle=30,radius=0.2] -- cycle;
            
            \coordinate (z*) at ($(v)!0.5!(v')$);
            \path[draw,semithick,rotate=30]
                ($(z*) + 0.1*(-1,0)$) |- ($(z*) + 0.1*(0,-1)$);
            \path[sublevel set] (v) -- (v');
            
            \path[main line] (0,0) -- node[pos=0.5,above,inner sep=2pt] {$1$} (v);
            \path[main line] (0,0) -- node[pos=0.55,above,sloped,inner sep=1pt] {$\cos(\pi/k)$} (z*);
            
            \node[origin,label={[label distance=0.5pt]left:$\mathbf{0}$}] at (0,0) {};
            \node[point,draw=UmiBlue,label={[label distance=0.5pt]right:$v$}] at (v) {};
            \node[point,UmiBlue] at (v') {};
            
            \path[draw,thin] (z*) -- ++(30:30pt) node[anchor=mid west,inner sep=0.5pt] {$z^*$};
            \node[point,draw=UmiBlue] at (z*) {};
        \end{tikzpicture}%
        \captionof{figure}{Polygonal approximation}%
        \label{fig:polygonal-approximation}%
    \end{minipage}%
\end{figure}

The following proof summarises the algorithm under $\mathcal{G}_{\psi(R_k)}$, and it derives a value for $k$ such that CDTW under the $2$-norm is $(1+\varepsilon)$-approximated.

\settheoremcounter{thm:algorithm-results}
\begin{corollary}
    We can compute CDTW exactly under $\mathcal{G}_{\psi(R_k)}$, yielding a $(1+\varepsilon)$-approximation for CDTW under $\tnorm[2]$ using $\psi := \mathrm{id}$ and some $k \in O(\varepsilon^{-1/2})$.
\end{corollary}

\begin{proof}
    To reiterate the exact CDTW algorithm under $\mathcal{G}_{\psi(R_k)}$:
    We first compute the optimum functions of all the borders on the parameter space's bottom and left side, where for every point $z$ there is a single  $(\mathbf{0},z)$-path that is horizontal or vertical respectively.
    The costs of these paths are integrals of piecewise linear functions due to \autoref{thm:polygonal-norm-properties}a.
    Then we use \nameref{alg:Propagate} to consecutively obtain all the remaining borders' optimum functions, which are piecewise quadratic by \autoref{thm:polygonal-norm-properties}c and correctly computed by \autoref{thm:propagation-correctness}.
    Finally, we return the value at the end of the optimum function of the final cell's top or right border, which is equal to the CDTW value under $\mathcal{G}_{\psi(R_k)}$ by \autoref{def:borders-and-optimum-function} and \autoref{def:path-and-cost}.
    
    Moreover, this value is an approximation for the CDTW value under $\tnorm[2]$.
    The $2$-norm is the gauge $\mathcal{G}_D$ of the unit disk $D := S_{\leq 1}(\tnorm[2])$, as demonstrated in \autoref{subsec:valleys}.
    Because the vertices of the convex regular $k$-gon $R_k$ all lie on the boundary of $D$ by construction in \autoref{sec:exact-algorithm}, it is $\psi(R_k) = R_k \subseteq D$ using~$\psi = \mathrm{id}$.
    Hence, it follows $\| z \|_2 = \mathcal{G}_D(z) \leq \mathcal{G}_{R_k}(z)$ for all $z \in \mathbb{R}^2$.
    We proceed to prove that we further have the bound $\mathcal{G}_{R_k}(z) \leq 1/\cos(\pi/k) \cdot \| z \|_2$:
    The ratio $\mathcal{G}_{R_k}(z) / \| z \|_2$ is maximised at each midpoint $z^* := (v + v')/2$ of two adjacent vertices $v,v'$ of~$R_k$ since the midpoints deviate strongest from $D$.
    Consider the right-angled triangle of a midpoint $z^*$, a related vertex $v$ and the origin $\mathbf{0}$, see \autoref{fig:polygonal-approximation}.
    The interior angle at $\mathbf{0}$ is $\pi/k$ and the hypotenuse length is $\| v \|_2 = 1$, so it is $\| z^* \|_2 = \cos(\pi/k)$.
    This means $\mathcal{G}_{R_k}(z^*) / \| z^* \|_2 = 1 / \cos(\pi/k)$ and establishes the claimed bound.
    
    Given the above relations between the norms, \autoref{def:cdtw} now implies
    \[
        \mathrm{cdtw}_{\| \cdot \|_2}(P,Q) \leq \mathrm{cdtw}_{\mathcal{G}_{R_k}}(P,Q) \leq 1/\cos(\pi/k)^2 \cdot \mathrm{cdtw}_{\| \cdot \|_2}(P,Q)
        \text{,}
    \]
    where the squared factor is due to the norm $\tnorm$ appearing in the distance term as well as in the speed regularisation term of CDTW under~$\tnorm$.
    Now the goal is to find some $k \in \{4,6,\dotsc\}$ with $1/\cos(\pi/k)^2 \leq 1+\varepsilon$ for a given $\varepsilon > 0$.
    If $\varepsilon \geq 1$, then $k := 4$ yields $1/\cos(\pi/4)^2 = 1/(1/\sqrt{2})^2 = 2 \leq 1+\varepsilon$.
    Assume $\varepsilon \leq 1$ next.
    We utilise the power series expansion of cosine (see \cite[Equation~2.8.1.1.2]{Jeffr2008}):
    \[
        \cos(t) = \sum_{r=0}^{\infty} (-1)^r \cdot \frac{t^{2r}}{(2r)!} = 1 - \frac{t^2}{2!} + \frac{t^4}{4!} - \frac{t^6}{6!} + \dotsb
        \quad
        \text{for }
        t \in \mathbb{R}
        \text{.}
    \]
    
    It is $\cos(t) \geq 1 - t^2/2$ for $t \in [0,1]$ because then 
    $t^{2r} / (2r)! \geq t^{2(r+1)} / (2(r+1))!$ holds for $r \in \mathbb{N}$ such that the remaining positive terms have no smaller absolute values than their negative successors.
    As a consequence, $\cos(t)^2 \geq (1 - t^2/2)^2 = 1 - t^2 + t^4/4 \geq 1 - t^2$ for $t \in [0,1]$.
    We now choose $k := 2 \lceil \pi \varepsilon^{-1/2} \rceil \in O(\varepsilon^{-1/2})$.
    In particular, we have an even $k \in \{8,10,\dotsc\}$ with $\pi/k \in [0,1]$.
    This gives
    \begin{align*}
        \frac{1}{\cos(\pi/k)^2}
        &\leq \frac{1}{1 - \pi^2/k^2}
        = \frac{1}{1 - \pi^2 / (2 \lceil \pi \varepsilon^{-1/2} \rceil)^2}
        \leq \frac{1}{1 - \varepsilon/4} \\[5pt]
        &= \frac{1 - \varepsilon/4 + \varepsilon/4}{1 - \varepsilon/4}
        = 1 + \frac{\varepsilon}{4 - \varepsilon}
        \leq 1 + \frac{\varepsilon}{3}
        \leq 1 + \varepsilon
        \text{,}
    \end{align*}
    where the penultimate inequality follows from the above assumption $\varepsilon \leq 1$.
    \qed
\end{proof}

Finally, \autoref{thm:running-time} reduces the algorithm's running time analysis to the combinatorial problem of bounding the total number $N$ of propagated quadratic pieces.
Like in 1D \cite[Lemma~12]{BuchiNW2022}, we achieve a linear dependency on $N$ thanks to the generalised stack-based propagation procedure.
Of course, we also get a dependency on the parameter $k$.
Note that $N$ itself likely depends on $k$ too.

\settheoremcounter{thm:running-time}
\begin{proposition}
    The running time of our algorithm for CDTW under $\mathcal{G}_{\psi(R_k)}$ is bounded by $O(N \cdot k^2 \log(k) \alpha(k))$, where $N$ denotes the total number of quadratic pieces over all optimum functions and $\alpha$ is the inverse Ackermann function.
\end{proposition}

\begin{proof}
    Let $N_\mathcal{B}$ denote the number of quadratic pieces of the optimum function $\mathrm{opt}_{0,\mathcal{B}}$ for a cell border $\mathcal{B}$.
    The base case of the dynamic program takes time proportional to the sum of all $N_\mathcal{A}$ for borders $\mathcal{A}$ on the bottom and on the left side of the parameter space.
    Line~1 of \nameref{alg:Propagate} takes $O(1)$ time per cell by \autoref{thm:valleys-for-polygonal-norms}.
    In the following, we show that the running time of the main loop is bounded by $O((N_{\mathrm{adj}(\mathcal{B})} + N_{\mathrm{opp}(\mathcal{B})}) \cdot k^2 \log(k) \alpha(k))$.
    Because each border $\mathcal{A}$, apart from the final cell's top and right border, is equal to $\mathrm{adj}(\mathcal{B})$ for exactly one call to \nameref{alg:Propagate} and is also equal to $\mathrm{opp}(\mathcal{B})$ for exactly one call to \nameref{alg:Propagate}, summing over all calls yields a total running time in $O(N \cdot k^2 \log(k) \alpha(k))$.
    
    For each $\mathcal{A} \in \{ \mathrm{adj}(\mathcal{B}), \mathrm{opp}(\mathcal{B}) \}$ we first compute the $O(k)$ arrangement lines as described in the proof of \autoref{thm:polygonal-norm-properties}b.
    We store the different kinds of arrangement lines separately, that means split lines, vertical grid lines and horizontal grid lines.
    They are based on the valley $\ell$ and the other parameter space lines $\Lambda_1,\dotsc,\Lambda_{k/2}$ from the proof of \autoref{thm:polygonal-norm-properties}b, which are all parallel if the cell's curve segments are parallel, otherwise they all share a common intersection.
    Except for the two potential grid lines for the common intersection and the at most four lines for the boundary of $\Sigma_{\mathcal{A},\mathcal{B}}$, each arrangement line is associated with one of $\ell,\Lambda_1,\dotsc,\Lambda_{k/2}$.
    Iterating through $\Lambda_1,\dotsc,\Lambda_{k/2}$ in order yields their intersection points with the borders $\mathcal{A}$ and $\mathcal{B}$, which induce their grid lines, in cyclic order.
    Thus, it is possible to compute a sorted list of each kind of arrangement lines within $O(k)$ time.
    
    As the arrangement $(V,E,F)$ from line~3 of \nameref{alg:Propagate} consists of $O(k)$~lines, we have $|V|, |E|, |F| \in O(k^2)$ and that $(V,E,F)$ can be computed in $O(k^2)$ time using an incremental algorithm (see \citeappendix[Section~8.3]{deBerCKO2008}).
    We also precompute the quadratic restriction of $\mathrm{opt}_{\mathcal{A},\mathcal{B}}$ to each face $f \in F$ with $f \subseteq \Sigma_{\mathcal{A},\mathcal{B}}$.
    Doing this individually for all faces would require $O(k^3)$ time since the proof of \autoref{thm:polygonal-norm-properties}b constructs each such restriction from a sum of $O(k)$ quadratic functions.
    
    In order to prevent the additional factor, we utilise a traversal of $(V,E,F)$ that starts at some face $f_0 \subseteq \Sigma_{\mathcal{A},\mathcal{B}}$.
    We fully evaluate the sum of quadratic functions for $f_0$ to get a representation of $\mathrm{opt}_{\mathcal{A},\mathcal{B}}$ on $f_0$ with $O(1)$ coefficients.~%
    After a face has been processed, we add its yet unprocessed neighbours to a queue.
    
    Processing a face $f$ from the queue works as follows:
    Take an already processed neighbour $f'$ of $f$ and compute the terms that differ in the sums for $\mathrm{opt}_{\mathcal{A},\mathcal{B}}$ on $f$ and $f'$.
    Then evaluate the difference of these terms and add it to the coefficient representation for $f'$ to receive the one for $f$.
    This is possible in time proportional to the number of differing terms.
    If $f$ and $f'$ are separated by a split line or a grid line for $\ell,\Lambda_1,\dotsc,\Lambda_{k/2}$, then that number is in $O(1)$ by construction.
    
    The grid lines for their common intersection can yield $O(k)$ differing terms, but there are at most two of them, so they need to be traversed only $O(1)$ times in order to process all faces.
    We thus omit additions to the queue that would unnecessarily traverse one such grid line again.
    Overall, the traversal of the arrangement computes representations of $\mathrm{opt}_{\mathcal{A},\mathcal{B}}$ on all faces in $O(k^2)$ time.
    
    Next up are the overlay and interval computations in lines~4--6 of \nameref{alg:Propagate}.
    Merging the new vertical grid lines at the $N_{\mathcal{A}} - 1$ breakpoints of $\mathrm{opt}_{0,\mathcal{A}}$ into the sorted list of the existing $O(k)$ vertical grid lines and potentially reversing the resulting list yields the desired sorted list $\mathcal{I}$ of intervals in $O(N_{\mathcal{A}} + k)$ time.
    
    Because the newly added lines are all vertical and hence parallel, the overlayed arrangement then has $O(N_{\mathcal{A}} \cdot k^2)$ faces.
    In particular, we can update $(V,E,F)$ within the same time bound using the abovementioned incremental algorithm.
    For every interval $I \in \mathcal{I}$ let now $X_I := | \{ f \in F \mid f \subseteq (I \times \mathbb{R}) \cap \Sigma_{\mathcal{A},\mathcal{B}} \} |$ be the number of faces that are iterated over in line~10 of the algorithm.
    We have
    \[
        X_I \in O(k^2) \text{ for each } I \in \mathcal{I} \text{ and } \sum_{I \in \mathcal{I}} X_I \in O(N_{\mathcal{A}} \cdot k^2) \text{ overall.}
    \]
    
    Lines~8--12 take $O(X_I)$ time thanks to the above precomputation of $\mathrm{opt}_{\mathcal{A},\mathcal{B}}$, and because every closed face $f \in F$ is incident to only $O(1)$ edges from $E$ that are contained in its boundary.
    More precisely, $f$ may be bounded by at most two split lines, as either they are all parallel or they all share a common intersection, by at most two vertical grid lines, and by at most two horizontal grid lines.
    
    In particular, $H$ now contains $O(X_I)$ quadratic functions.
    Hence, computing its lower envelope in line~13 is possible within $O(X_I  \log(X_I) \alpha(X_I))$ time (see \citeappendix{Hersh1989}).
    Line~14 then takes $O(X_I \log(X_I) \alpha(X_I) + Y_I)$ time, where $Y_I$ denotes the number of quadratic pieces popped from the stack $\mathcal{S}$.
    The sum of $O(X_I \log(X_I) \alpha(X_I))$ over all intervals $I \in \mathcal{I}$ iterated over in line~7 of \nameref{alg:Propagate} is bounded by
    \begin{align*}
        &\sum_{I \in \mathcal{I}} c \cdot X_I \log(X_I) \alpha(X_I)
        \leq c \cdot \sum_{I \in \mathcal{I}} X_I \log(c \cdot k^2) \alpha(c \cdot k^2) \\[3pt]
        \leq{} &c \cdot \sum_{I \in \mathcal{I}} X_I \cdot c' \log(k) \alpha(k)
        \leq c' c^2 \cdot N_\mathcal{A} \cdot k^2 \log(k) \alpha(k)
    \end{align*}
	for a suitable constant $c \in \mathbb{N}$ with $c' := (\log(c) + 2)^2$.
    Thus, over both considered borders $\mathcal{A} \in \{\mathrm{adj}(\mathcal{B},\mathrm{opp}(\mathcal{B}))\}$ no more than $O((N_{\mathrm{adj}(\mathcal{B})} + N_{\mathrm{opp}(\mathcal{B})}) \cdot k^2 \log(k) \alpha(k))$ quadratic pieces are created and subsequently pushed onto the stack $\mathcal{S}$.
    
    The sum of the second term $O(Y_I)$ over all intervals of both borders has the same asymptotic bound, which is revealed through a simple amortised argument:
    If a quadratic piece is popped completely from the stack $\mathcal{S}$, then we charge this to the popped piece itself.
    If a piece is popped only partially from~$\mathcal{S}$, then we charge this to the newly pushed piece.
    It follows that every created piece is charged at most twice to cover all needed pop operations.
    This completes the proof.
    \qed
\end{proof}

    \bibliographyappendix{literature.bib}
\end{document}